\documentclass[11pt,letterpaper]{article}

\newif\ifcom
\comtrue 

\newcommand{\blind}{0}

\usepackage{latexsym,multirow}
\usepackage{amssymb,amsmath, bm}
\usepackage{graphicx}
\usepackage{verbatim}

\usepackage{mathtools}

\usepackage{enumitem}

\usepackage{makecell}

\usepackage{natbib}
\usepackage[pdftex, bookmarksopen=true, bookmarksnumbered=true,
pdfstartview=FitH, breaklinks=true, urlbordercolor={0 1 0},
citebordercolor={0 0 1}, hidelinks]{hyperref}
\usepackage{colortbl}
\usepackage{subfigure}

\usepackage{dcolumn}
\newcolumntype{.}{D{.}{.}{-1}}
\newcolumntype{d}[1]{D{.}{.}{#1}}

\usepackage{theorem}
\theoremstyle{plain}

\newtheorem{theorem}{Theorem}

\newtheorem{assumption}{Assumption}
\newtheorem{definition}{Definition}
\newtheorem{lemma}{Lemma}
\newcommand{\qed}{\hfill \ensuremath{\Box}}
\newcommand{\indep}{\mbox{$\perp\!\!\!\perp$}}

\usepackage{rotating}

\usepackage[compact]{titlesec}

\allowdisplaybreaks 

\begin{document}

\newcommand\ud{\mathrm{d}}
\newcommand\dist{\buildrel\rm d\over\sim}
\newcommand\ind{\stackrel{\rm indep.}{\sim}}
\newcommand\iid{\stackrel{\rm i.i.d.}{\sim}}
\newcommand\logit{{\rm logit}}
\renewcommand\r{\right}
\renewcommand\l{\left}
\newcommand\pre{{(t-1)}}
\newcommand\cur{{(t)}}
\newcommand\cA{\mathcal{A}}
\newcommand\cB{\mathcal{B}}
\newcommand\bone{\mathbf{1}}
\newcommand\E{\mathbb{E}}
\newcommand\Var{{\rm Var}}
\newcommand\cC{\mathcal{C}}
\newcommand\cD{\mathcal{D}}
\newcommand\cF{\mathcal{F}}
\newcommand\cJ{\mathcal{J}}
\newcommand\cK{\mathcal{K}}
\newcommand\cP{\mathcal{P}}
\newcommand\cT{\mathcal{T}}
\newcommand\cX{\mathcal{X}}
\newcommand\cXR{\mathcal{X,R}}
\newcommand\wX{\widetilde{X}}
\newcommand\wT{\widetilde{T}}
\newcommand\wY{\widetilde{Y}}
\newcommand\wZ{\widetilde{Z}}
\newcommand\bX{\mathbf{X}}
\newcommand\bY{\mathbf{Y}}
\newcommand\bT{\mathbf{T}}
\newcommand\bt{\mathbf{t}}
\newcommand\bwT{\widetilde{\mathbf{T}}}
\newcommand\bwt{\tilde{\mathbf{t}}}
\newcommand\bbT{\overline{\mathbf{T}}}
\newcommand\bbt{\overline{\mathbf{t}}}
\newcommand\ubT{\underline{\mathbf{T}}}
\newcommand\ubt{\underline{\mathbf{t}}}
\newcommand\bhT{\widehat{\mathbf{T}}}
\newcommand\bht{\hat{\mathbf{t}}}

\newcommand\bV{\mathbf{V}}
\newcommand\bv{\mathbf{v}}

\newcommand\bU{\mathbf{U}}
\newcommand\bu{\mathbf{u}}
\newcommand\bg{\mathbf{g}}

\newcommand\pgamma{\pmb{\gamma}}

\newcommand\bA{\mathbf{A}}

\newcommand\I{{\bf I}}

\newcommand\Pra{\mbox{Pr}^\ast}
\newcommand\Prd{\mbox{Pr}^\dagger}
\newcommand\Pru{\mbox{Pr}^{\texttt{U}}}
\newcommand\Prr{\mbox{Pr}^{\texttt{R}}}
\newcommand{\1}{\mathbf{1}}
\newcommand\Cov{\mbox{\normalfont Cov}}
\newcommand\bTm{\bT^{\mathcal{M}}}
\newcommand\bTc{\bT^{\mathcal{C}}}

\newcommand\Tm{T^{\mathcal{M}}}
\newcommand\Tc{T^{\mathcal{C}}}
\newcommand\tm{\widetilde{m}}
\newcommand\cY{\mathcal{Y}}
\newcommand\cZ{\mathcal{Z}}

\newcommand\cG{\mathcal{G}}
\newcommand\cN{\mathcal{N}}
\newcommand\tcN{\widetilde{\cN}}

\newcommand\bL{\mathbf{L}}
\newcommand\bell{\pmb{\ell}}
\newcommand\bZ{\mathbf{Z}}
\newcommand\bz{\mathbf{z}}
\newcommand\ba{\mathbf{a}}
\newcommand\bD{\mathbf{D}}
\newcommand\bd{\mathbf{d}}
\newcommand\R{\mathbb{R}}
\newcommand\bC{\mathbf{C}}
\newcommand\bc{\mathbf{c}}
\newcommand\bG{\mathbf{G}}
\newcommand\bx{\mathbf{x}}
\newcommand\cS{\mathcal{S}}

\newcommand{\argmax}{\operatornamewithlimits{argmax}}
\newcommand{\argmin}{\operatornamewithlimits{argmin}}

\newcommand\spacingset[1]{\renewcommand{\baselinestretch}%
  {#1}\small\normalsize}

\newcommand\sd{s^\dagger}

\newcommand{\etc}[1]{\textbf{\textcolor{blue}{(ETT: #1)}}}
\newcommand{\nec}[1]{\ifcom \textbf{\textcolor{blue}{(NE: #1)}} \fi}

\makeatletter
\newcommand*{\centernot}{%
  \mathpalette\@centernot
}
\def\@centernot#1#2{%
  \mathrel{%
    \rlap{%
      \settowidth\dimen@{$\m@th#1{#2}$}%
      \kern.5\dimen@
      \settowidth\dimen@{$\m@th#1=$}%
      \kern-.5\dimen@
      $\m@th#1\not$%
    }%
    {#2}%
  }%
}
\makeatother

\spacingset{1.25}

\newcommand{\tit}{Identification and Estimation of Causal Peer Effects
  Using Double Negative Controls for Unmeasured Network Confounding}


\if0\blind

{\title{{\tit}} 
  \author{\hspace{0.7in} Naoki Egami\thanks{Assistant Professor, Department of
      Political Science, Columbia University, New York NY 10027. Email:
      \href{mailto:naoki.egami@columbia.edu}{naoki.egami@columbia.edu}, URL:
      \url{https://naokiegami.com}\vspace{0.1in}} \hspace{0.7in} Eric
    J. Tchetgen
    Tchetgen\thanks{Luddy Family President’s Distinguished Professor,
      Department of Statistics and Data Science, the Wharton School, University of
      Pennsylvania, Philadelphia PA 19104.
      Email: \href{mailto:ett@wharton.upenn.edu}{ett@wharton.upenn.edu}, URL:
      \url{https://statistics.wharton.upenn.edu/profile/ett}}}}

\date{\today}

\fi

\if1\blind
\title{\tit}

\fi

\maketitle

\begin{abstract}
  Scientists have been interested in estimating causal peer effects to understand how people's behaviors are affected by their
  network peers. However, it is well known that identification and
  estimation of causal peer effects are challenging in observational
  studies for two reasons. The first is the identification challenge due to unmeasured
  network confounding, for example, homophily
  bias and contextual confounding. The second issue is network
  dependence of observations, which one must take into account for valid
  statistical inference. Negative control variables, also
  known as placebo variables, have been widely used in observational
  studies including peer effect analysis over networks, although they have been used primarily for bias detection.
  In this article, we establish a formal framework which leverages a pair of negative control outcome
  and exposure variables (double negative controls) to
  nonparametrically identify
  causal peer effects in the presence of unmeasured network
  confounding. We then propose a generalized method of moments
  estimator for causal peer effects, and establish its consistency and
  asymptotic normality under an assumption about $\psi$-network
  dependence. Finally, we provide a network heteroskedasticity
  and autocorrelation consistent variance estimator. Our methods are
  illustrated with an application to peer effects in education.
\end{abstract}

\clearpage
\spacingset{1.4}

\section{Introduction}
Social and biomedical scientists have long been interested in how people's behaviors
are affected by peer behaviors. For example, scholars
have studied peer effects on voting behaviors
\citep{sinclair2012social, jones2017social}, educational outcomes \citep{epple2011peer, sacerdote2011peer}, criminal behaviors \citep{glaeser1996crime}, and job
opportunities \citep{granovetter1973strength}. Epidemiologists
and researchers in public health have studied related concepts such as
``contagion'' effects of infectious disease \citep{halloran1995causal,
  crawford2018risk} and health behaviors \citep{fowler2013review}.

Despite its importance, identification and estimation of causal
peer effects have been challenging for two reasons. The first issue is
that it is often difficult to identify causal peer effects in observational studies due to unmeasured network
confounding, such as \textit{homophily bias} and \textit{contextual
  confounding} \citep{manski1993identification,
  vanderweele2013social, ogburn2018}. Homophily bias arises when people
become connected due to unobserved characteristics. Contextual confounding exists when peers share some
unobserved contextual factors. Highlighting concerns about these potential biases, influential papers
across disciplines have criticized prior peer effect analyses from observational studies \citep[e.g.,][]{lyons2011spread,
  angrist2014perils}. \cite{shalizi2011homophily} argue that it is nearly impossible to
credibly estimate causal peer effects in observational studies using direct confounding adjustment methods (e.g. regression based adjustment) due
to pervasive concerns about unmeasured network confounding. 

In addition to unmeasured network confounding, another important
challenge is that one needs to account for network dependence of
observations in order to obtain valid statistical inference. When
such network dependence is ignored, standard errors may be underestimated,
and confidence intervals may be anti-conservative
\citep{lee2020network}. While recent studies have allowed for some extent of network
dependence across units in observational causal inference \citep{van2014causal, ogburn2017causal,
  fora2020, ogburn2020causal, tchetgen2020auto, leung2021network}, they have largely
relied on an assumption of no uncontrolled network confounding.

In this paper, we propose to resolve these challenges by
using a pair of negative control outcome and exposure variables
(\textit{double negative controls}). A negative control outcome (also known as a placebo outcome) is an outcome variable that is
known not to be causally affected by the treatment of
interest. Likewise, a negative control exposure (also known as a
placebo treatment) is a treatment variable that does not
causally affect the outcome of interest \citep{tchetgen2010negative}. There is a long-standing tradition in
biomedical and social sciences of using negative controls to detect
unmeasured confounding. A non-null effect of the treatment on the
negative control outcome or a non-null effect of the negative control
exposure on the outcome of interest amounts to compelling evidence of
unmeasured confounding. In the literature of causal peer effects, \cite{egami2018diffusion}
exploits a negative control outcome, and \cite{liu2020regression} use
a negative control exposure to address unmeasured network confounding. While they require relatively weak
assumptions to detect unmeasured network confounding, both works
require much stronger assumptions for identification of causal
peer effects as they each use only one type of negative control
variable but not both. Recently, a series of papers \citep{kuroki2014,
  miao2018confounding, miao2018identifying,
  deaner2018proxy, shi2020dnc, tchetgen2020proximal, kallus2021causal} propose to use
double negative controls for identification of causal effects, but
they have focused on i.i.d or panel data settings, and to date none of these papers has considered network data. 

Our contribution is to propose a general framework for using double
negative controls for identification and estimation of causal peer
effects in the presence of uncontrolled network confounding, while taking into
account network dependence. We first derive nonparametric
identification of causal peer effects in the presence of
unmeasured network confounding by exploiting double negative controls
that are associated with unmeasured confounders. In particular, we
incorporate double negative control variables via a network
outcome confounding bridge function, a network version of the outcome
confounding bridge function studied in \citet{miao2018confounding,
  miao2018identifying, shi2020dnc, tchetgen2020proximal}. We discuss general approaches for
selecting negative controls from network data in practice. We then propose a generalized method of moments (GMM) estimator
\citep{hansen1982gmm} for the causal peer effect, and we establish consistency and asymptotic normality of the resulting
estimator under correct specification of the network confounding
bridge function, and an assumption about $\psi$-network dependence \citep{kojevnikov2020},
which expresses the degree of stochastic dependence between variables
in terms of network distance. This assumption of $\psi$-network dependence restricts
the speed by which network dependence decays as network
distance increases, and the speed by which the density of the network
changes as sample size increases. Finally, we provide a network heteroskedasticity and autocorrelation
consistent (network HAC) variance estimator, with which researchers can
construct asymptotic confidence intervals of causal peer effects.

The paper is organized as follows. In Section~\ref{sec:dyad}, we
consider dyadic data to focus on the use of double negative controls
for identification of causal peer effects. A corresponding framework for estimation and
inference via GMMs is relatively straightforward in this setting assuming a sample of independent and identically distributed dyads is available. In Section~\ref{sec:net}, we examine a more general setting where one
observes data from a single network. We study identification as
well as estimation and inference by accounting for $\psi$-network dependence. In Section~\ref{sec:sim}, we assess the
finite sample performance of our proposed estimators via extensive
simulations. Also, we illustrate our methods by applying them to the Add
Health network data to infer the extent of causal peer effects in education in
Section~\ref{sec:app}. We extend our results in Section~\ref{sec:ext} to settings where researchers are interested in causal peer effects from
higher-order peers (i.e., those not directly connected to a
given focal unit). Section~\ref{sec:con} concludes the paper.

\subsection*{Related Literature}
This article builds on a growing literature on causal peer effects. Various methods have been proposed to address concerns about uncontrolled
network confounding in observational studies. A popular approach for identification
of causal peer effects is the so-called instrumental variable method. \cite{bramoulle2009identification} use instrumental
variables to deal with simultaneity in the linear-in-mean models
\citep{manski1993identification, imbens2013social}.
\cite{omalley2014diff} propose to use genes as instruments to study
causal peer effects of body mass index among friends. In addition to the well known
exclusion restriction, both methods assume (conditional) exogeneity of
instrumental variables. However, this assumption may be untenable in a
wide range of applications because it is violated as long
as instrumental variables are associated with unmeasured variables
at the source of homophily or contextual confounding. In contrast, our methods
allow for and in fact leverage such association between negative controls and
unobserved confounders. \cite{shalizi2021} propose a consistent
estimator of causal peer effects, which adjusts for
estimated latent homophilous attributes in settings where the data
generating process is linear and the network grows according to
either a stochastic block model or a continuous space model. In
contrast, we establish nonparametric identification of causal peer
effects by using double negative controls, and our results accommodate
both latent homophily and contextual confounding. Also, our asymptotic
results make an alternative assumption of $\psi$-network dependence
\citep{kojevnikov2020} instead of assuming specific network models, and we allow for asymptotic normality and construction of asymptotic
confidence intervals in addition to consistent estimation of causal peer effects.

There is also a literature focusing on a different research goal, such as testing
\citep[e.g.,][]{anagnostopoulos2008influence, vanderweele2012why},  partial
identification \citep{versteeg2010difftest,
  versteeg2013statistical}, and sensitivity analysis
\citep{vanderweele2011sen}, rather than (point) identification of
causal peer effects. Finally, our methods addressing unmeasured
network confounding in observational studies are complementary to
approaches based on randomized experiments or natural experiments
\citep{sacerdote2001peer, duflo2011peer, sinclair2012diffIV, eckles2017Online,
  basse2019peer, li2019randomization}. 

\section{Double Negative Controls for Dyadic Data}
\label{sec:dyad}

\setcounter{assumption}{0}
\renewcommand {\theassumption} {1.\arabic{assumption}}
In this section, in order to ground ideas, we focus on identification and estimation of causal peer
effects from dyadic data. In general, even in this simple setting, causal peer effects are not identified based on
standard covariate adjustment in the presence of unmeasured
network confounding, such as latent homophily and contextual confounding. We propose an alternative approach based on negative
controls. We consider a more general network setting in Section~\ref{sec:net}.

\subsection{Notation and Definitions}
We consider data on dyads, i.e., pairs of two individuals. For each
dyad $i$, let $S_i = 1$ when the two units are connected and $S_i = 0$
when the two units have no tie between them. For example, in dyadic
data based on a students' friendship survey, $S_i = 1$ encodes the two
students being friends with each other and $S_i = 0$ otherwise.  

Suppose one has observed $n$ independent and identically distributed
samples of connected dyads ($S_i = 1$) where each
dyad is labeled $i \in \{1, \ldots, n\}$. For each unit within dyads,
we observe focal behavior $Y$ at two time points,
 baseline and a single follow-up. Define $Y_{kt}$ to be the focal
behavior of unit $k \in \{1, 2\}$ at time $t \in  \{1,2\}$ where $t = 1, 2$
denotes baseline and follow-up, respectively. Without loss of
generality, define $k=1$ to be the \textit{ego}  --- a unit on whom
we estimate a causal peer effect --- and define $k=2$ to be the
\textit{peer} --- a unit whose effect on the ego we estimate. 

The outcome of interest is ego’s behavior at follow-up
$Y_{12}$. The treatment variable of interest is peer's behavior at
the baseline $Y_{21}$. Using the potential outcomes framework
\citep{neyman1923, rubin1974causal, robins1986causal}, define
$Y_{12}(y_{21})$ to be the potential outcome had possibly contrary to fact, the
treatment variable been set to $Y_{21} = y_{21}$, which we assume to exist and to be well defined. Throughout the paper, we make
the standard consistency assumption linking observed and potential
outcomes: 
\begin{equation}
  Y_{12} = Y_{12}(Y_{21}), \ \ \ a.s. \label{eq:con}
\end{equation}

Our goal is to estimate the Average Causal Peer Effect (ACPE), defined as
\begin{equation*}
  \tau (y_{21}, y_{21}^\prime) \coloneqq \E\{Y_{12}(y_{21}) - Y_{12}(y_{21}^\prime) \mid S = 1\} 
\end{equation*}
where $y_{21}, y_{21}^\prime \in \mathcal{Y}_{21}$ where $\mathcal{Y}_{21}$ is the support
of $Y_{21}$. We condition on $S = 1$ because we only consider connected dyads.

\subsection{Identification Challenge}
\label{subsec:challenge}
Covariate adjustment based on conditional ignorability is the  most common 
approach to identification of causal effects in observational studies \citep{rosenbaum1983central,
  robins1986causal}. In causal peer analysis, conditional ignorability entails assuming 
\begin{equation}
  Y_{12}(y_{21}) \ \indep \ Y_{21} \mid Y_{11}, \bX, S = 1 
\end{equation}
where $\bX$ represent observed pre-treatment covariates for the dyad. However, such approach is, in general,
not plausible when estimating the ACPE due to unmeasured homophily
\citep{shalizi2011homophily} and contextual confounding
\citep{vanderweele2013social}. Homophily bias arises when people
become connected due to unobserved characteristics. Contextual
confounding exists when peers share some unobserved contextual factors. 
In this paper, we use the term \textit{unmeasured network confounding}
to refer to confounding that violates conditional ignorability in causal peer analysis, and thus it contains unmeasured homophily and contextual confounding as
special cases.

\begin{figure}[!t]
  \begin{minipage}[t]{.45\textwidth}
    \begin{center}
      \includegraphics[width=\textwidth]{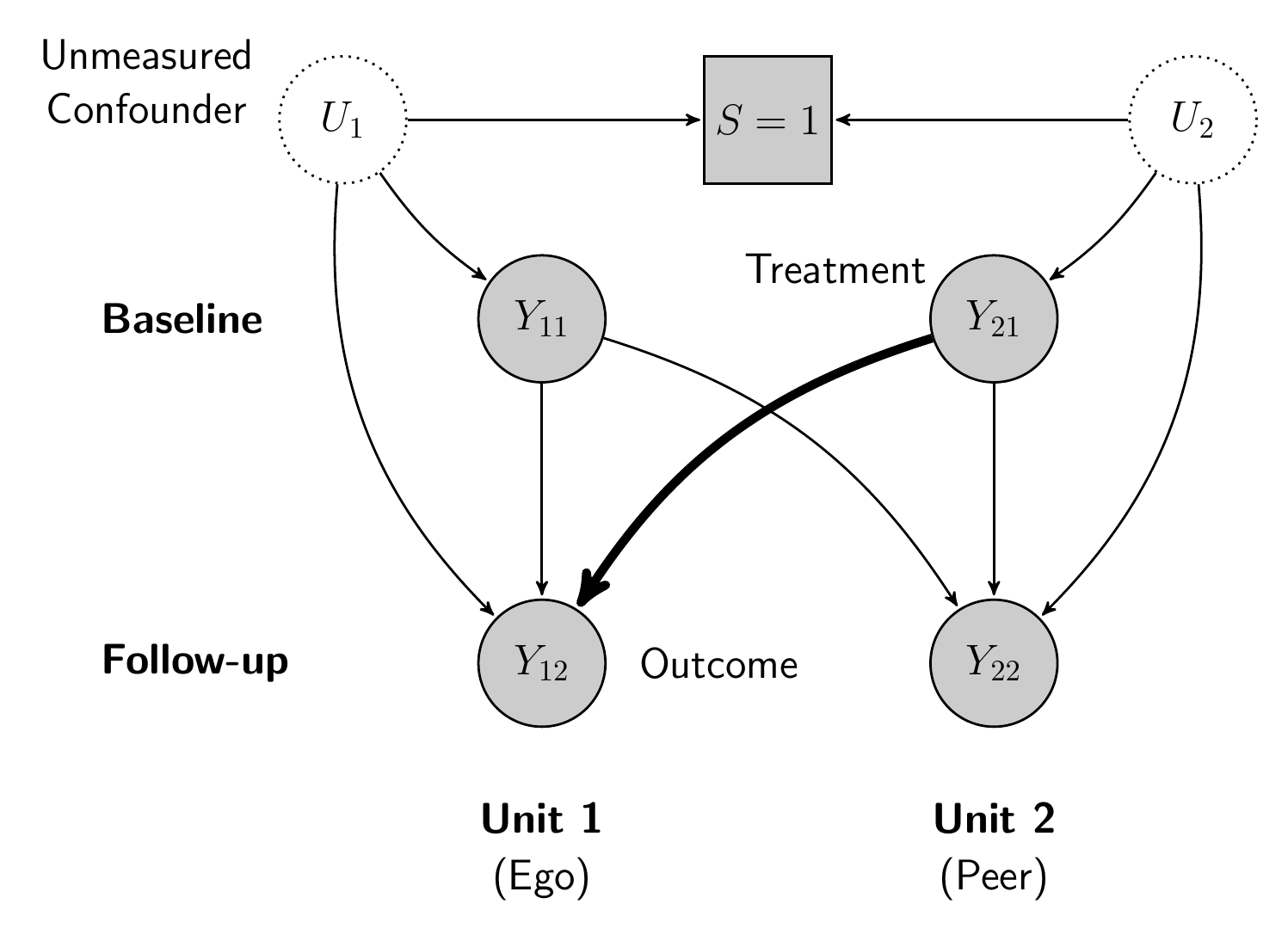}\\
      \hspace{0.3in} (i) Unmeasured Homophily
    \end{center}   
  \end{minipage}
  \hspace{0.3in}
  \begin{minipage}[t]{.5\textwidth}
    \begin{center}
      \includegraphics[width=\textwidth]{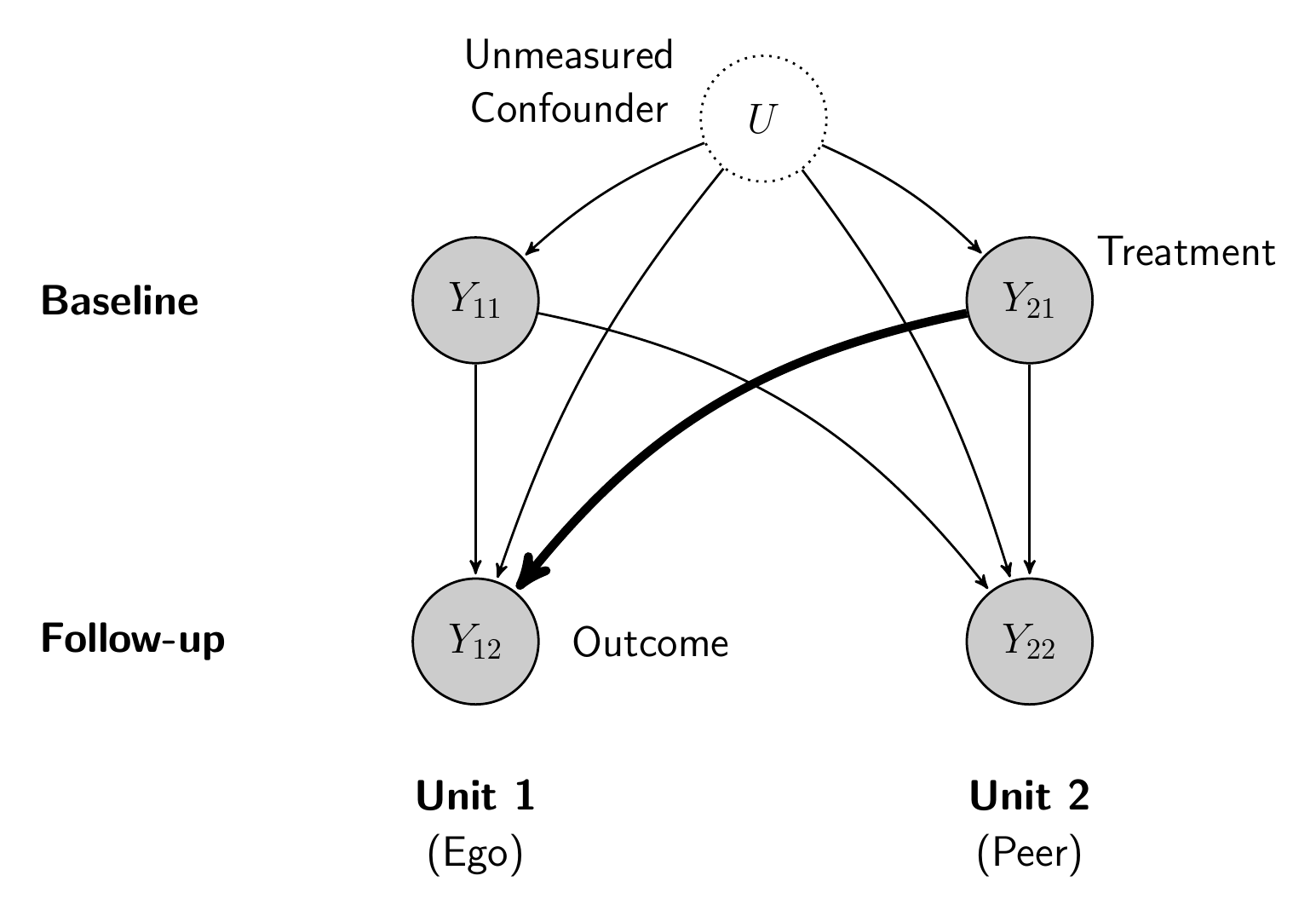}\\
      \hspace{0.3in} (ii) Contextual Confounding
    \end{center}    
  \end{minipage}
  \spacingset{1}{\caption{A DAG for dyadic data in the presence of
      unmeasured network confounding. \textit{Note:} The thick arrow
      from $Y_{21}$ to $Y_{12}$ indicates the causal peer effect of
      interest. We use shaded (dotted) nodes to denote observed
      (unobserved) variables. For simplicity, we have suppressed
      observed covariates $\bX$. In Figure~\ref{fig:dag-dyad}.(i), the square box around $S = 1$ represents
      that we observe dyads conditional on $S = 1$.}\label{fig:dag-dyad}}
\end{figure}

Figure~\ref{fig:dag-dyad}.(i) represents a causal directed acyclic graph
(DAG) illustrating latent homophily. We use $U_k$ with $k
\in \{1, 2\}$ to represent unobserved characteristics that affect a tie
relationship $S$ as well as focal behavior
$Y$. Because the tie relationship $S$ is affected by $U_1$ and $U_2$,
variable $S$ is a collider in the terminology of graph
theory \citep{pearl2000}. To estimate the ACPE, we condition on $S = 1$, and
therefore, there is an unblocked backdoor path $Y_{21} \leftarrow U_2
\rightarrow \fbox{$S=1$} \leftarrow U_1 \rightarrow Y_{12}$ where
$U_1$ and $U_2$ are unobserved, and the square box around $S=1$ means
that we observe dyads conditional on $S = 1$ \citep{shalizi2011homophily}. Thus, even in the simple setting of dyadic
data, identification of the ACPE is not possible without additional
assumptions.

Figure~\ref{fig:dag-dyad}.(ii) represents a causal DAG illustrating
contextual confounding. Here, we use $U$ to represent an unmeasured
shared context that affects both the ego and peer. Due to the
unblocked back-door path $Y_{21} \leftarrow U \rightarrow Y_{12}$, conditional ignorability is violated. 

\subsection{Identification with Double Negative Controls}
\label{subsec:iden}
In this section, we consider an alternative approach for identification and estimation by exploiting
auxiliary variables called negative controls. In particular, we will
use negative control outcome (NCO) and negative control exposure (NCE)
variables, which we define below.  

We first make the latent ignorability assumption, which states that conditional ignorability holds if we could measure all factors at the source of network confounding.
\begin{assumption}[Latent Ignorability]
  \label{l-ig}
  For all $y_{21} \in \cY_{21}$, 
  \begin{equation*}
    Y_{12}(y_{21}) \ \indep \ Y_{21} \mid U, \bX, S = 1.
  \end{equation*}
\end{assumption}
Assumption~\ref{l-ig} states that $U$, $\bX$, and $S$ suffice to account
for confounding of the relationship between $Y_{21}$ and $Y_{12}(y_{21})$, whereas
$\bX$ and $S$ alone may not. This assumptions is often plausible as there is no direct
restriction on the nature of the latent characteristic $U$. However,
this assumption alone is not sufficient for identification given
that we do not observe the latent characteristic $U$. 

The key to the proposed approach is to suppose that one can
measure two auxiliary variables, negative control outcome $W$ and
negative control exposure $Z$ that satisfy the following conditions. 
\begin{assumption}[Negative Controls]
  \label{nc} $ $ \\
  \textnormal{1. Negative Control Outcome (NCO):} 
  \begin{equation}
    W \ \indep \ Y_{21} \mid U, \bX, S = 1    
  \end{equation}
  \textnormal{2. Negative Control Exposure (NCE):} 
  \begin{equation}
    Z \ \indep \ Y_{12} \mid Y_{21}, U, \bX, S = 1 \ \ \mbox{and} \ \ \  Z \ \indep
    \ W \mid Y_{21}, U, \bX, S = 1.
  \end{equation}
\end{assumption}
Assumption~\ref{nc}.1 states that $W$ is an auxiliary variable that is conditionally independent of the
treatment $Y_{21}$ given the latent confounder $U$, observed
pre-treatment covariates $\bX$, and the dyadic type
$S$. Assumption~\ref{nc}.2 means that $Z$ is an
auxiliary variable that is conditionally independent of the
outcome $Y_{12}$  and NCO $W$ given the treatment $Y_{21}$, the latent confounder
$U$, observed pre-treatment
covariates $\bX$. 

\begin{figure}[!t]
  \begin{minipage}[t]{.45\textwidth}
    \begin{center}
      \includegraphics[width=\textwidth]{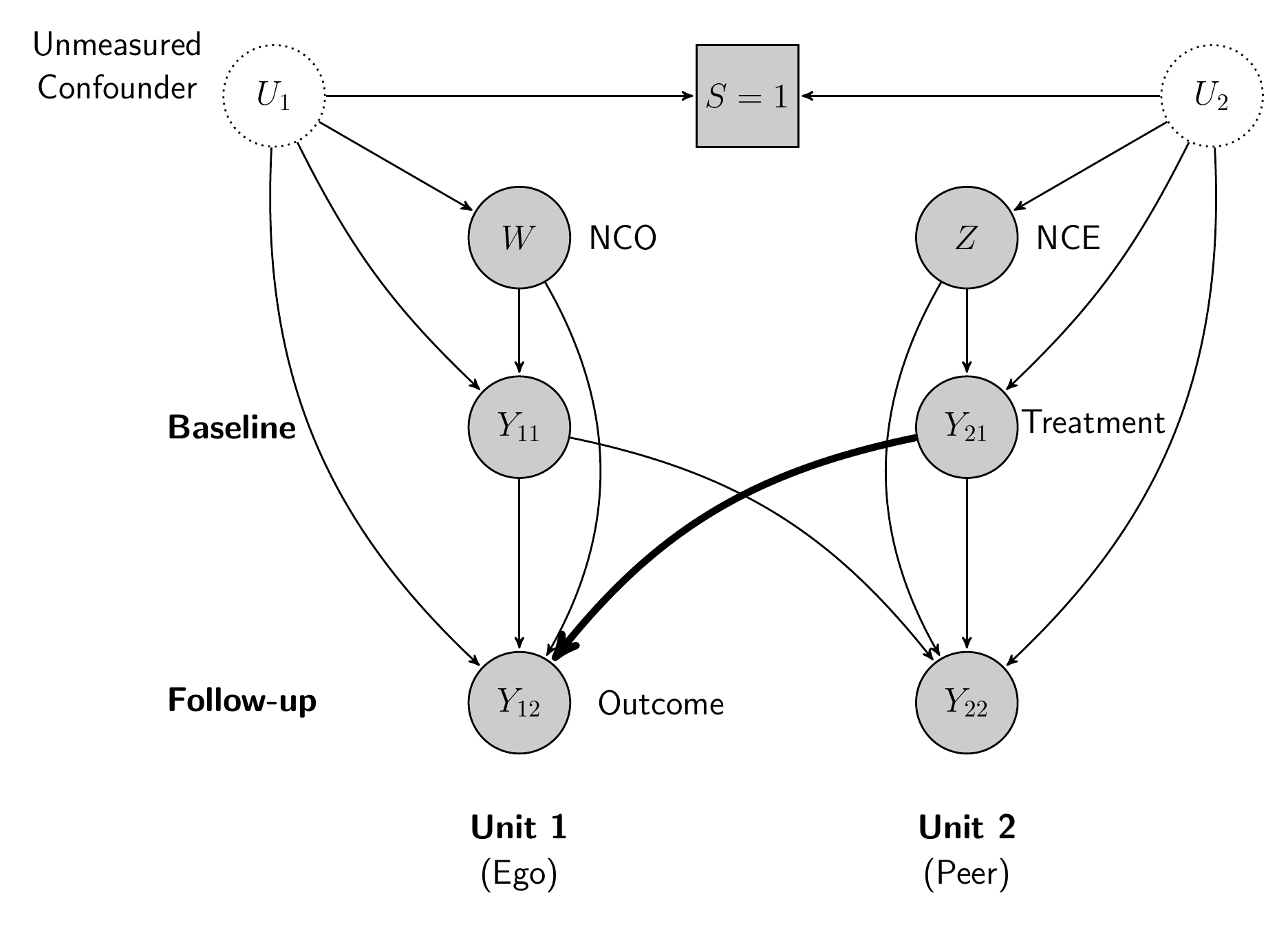}    
      \\ \hspace{0.3in}  (i)
    \end{center}
  \end{minipage}
  \hspace{0.4in}
  \begin{minipage}[t]{.45\textwidth}
    \begin{center}
      \includegraphics[width=\textwidth]{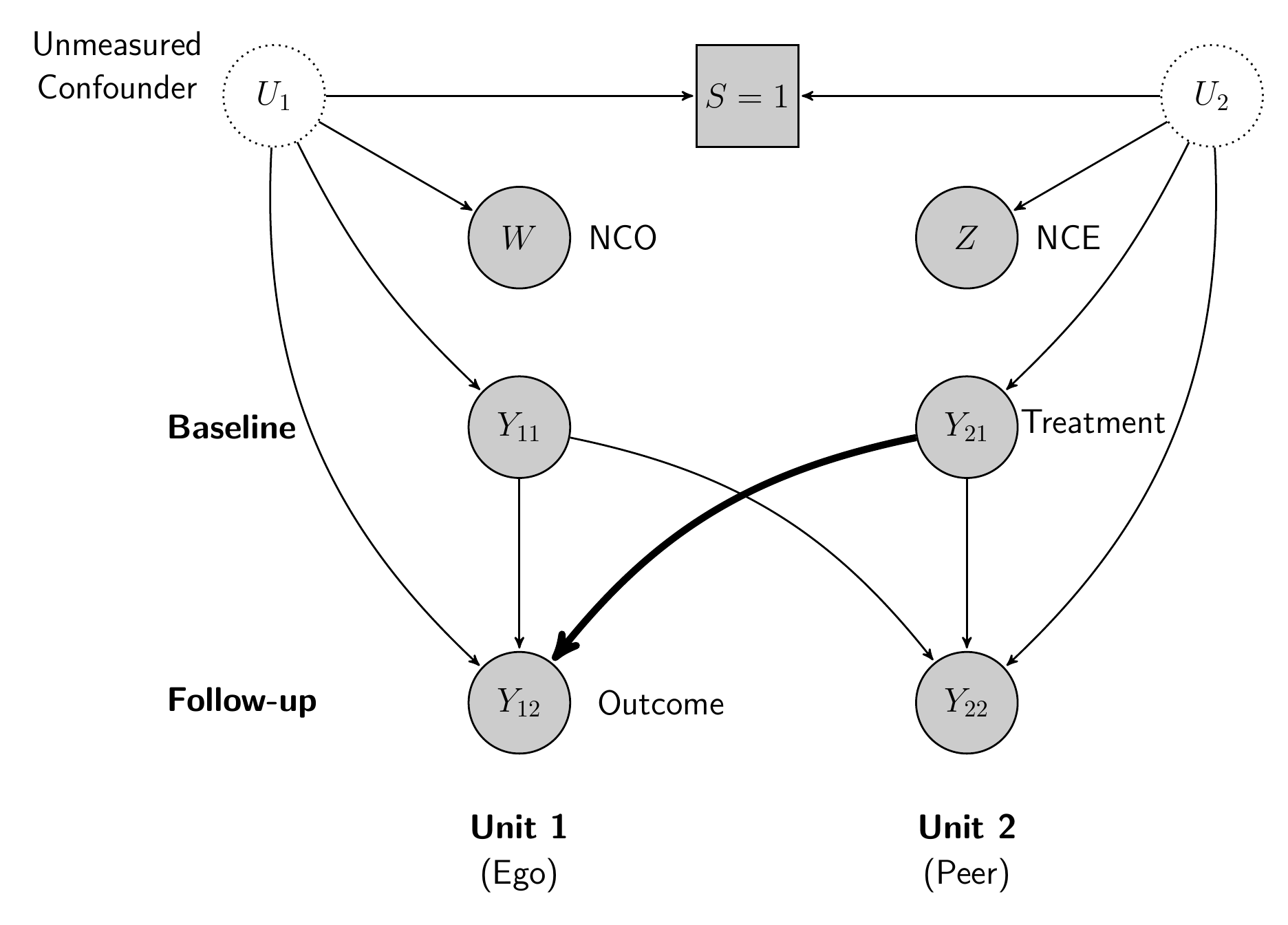}
       \\ \hspace{0.3in} (ii) 
    \end{center}
  \end{minipage}
  \vspace{0.3in} \\
  \hspace{0.03in}
  \begin{minipage}[t]{.45\textwidth}
    \begin{center}
      \includegraphics[width=\textwidth]{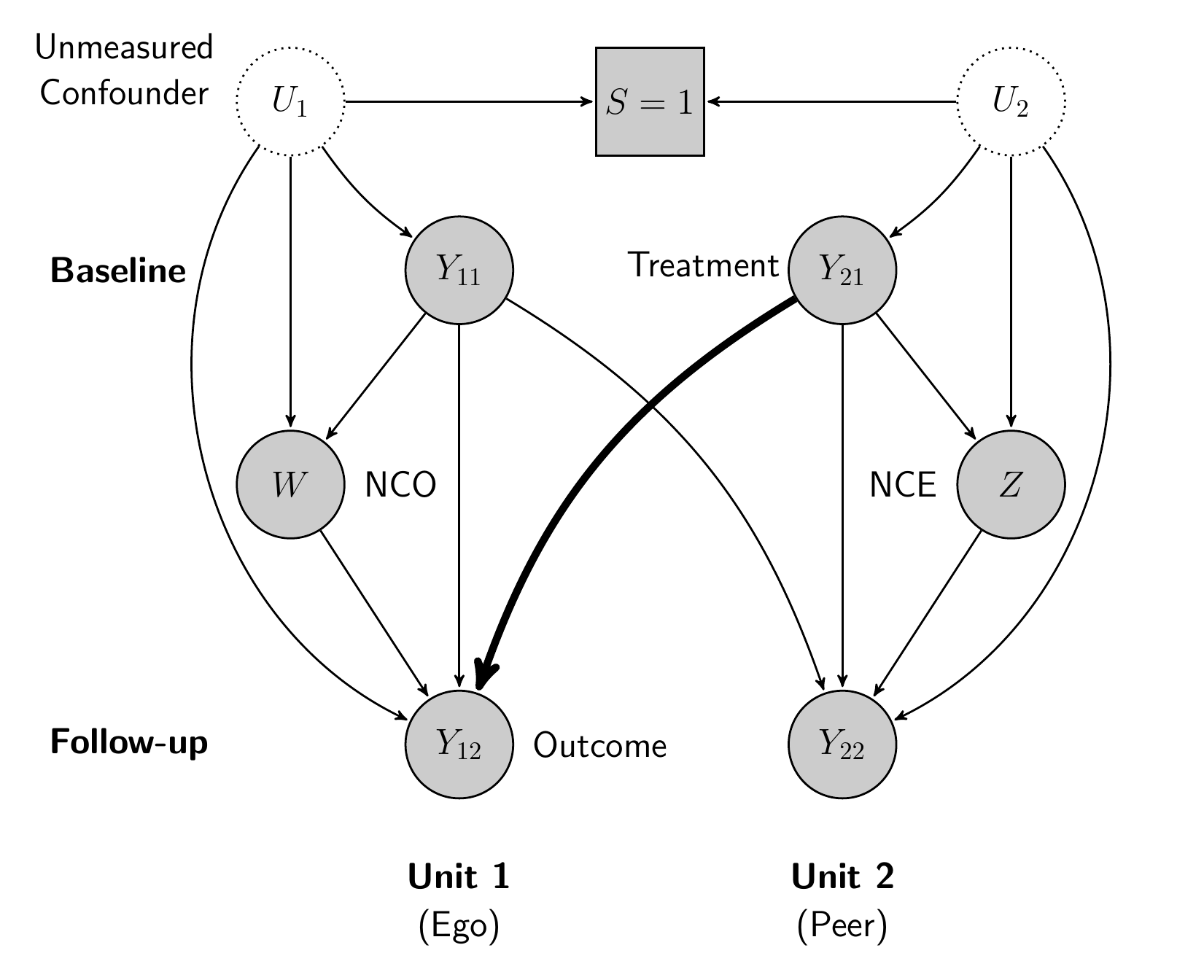}
       \\ \hspace{0.2in} (iii)
    \end{center}
  \end{minipage}
  \hspace{0.4in}
  \begin{minipage}[t]{.45\textwidth}
    \begin{center}
      \includegraphics[width=\textwidth]{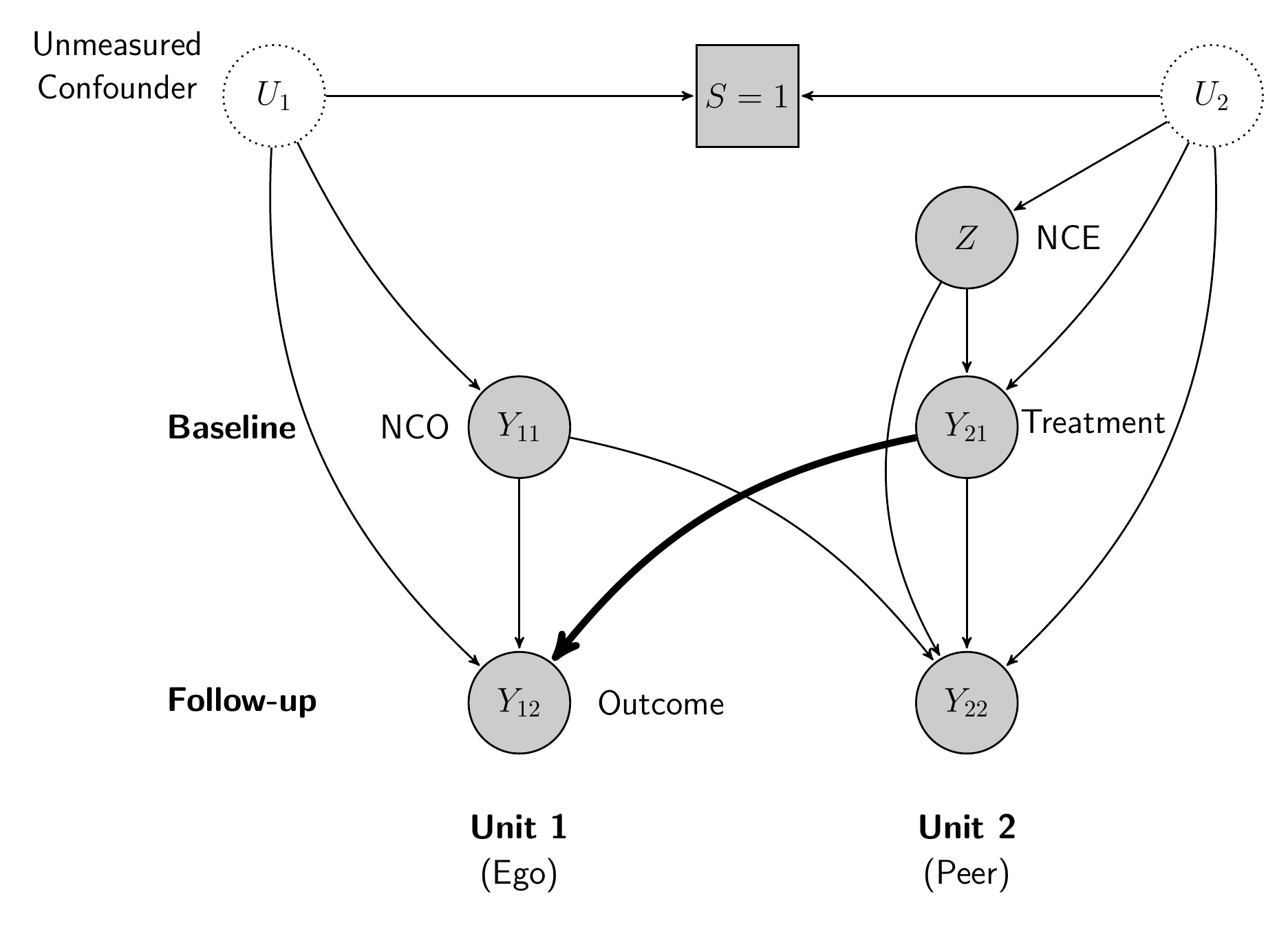}    
       \\ \hspace{0.3in} (iv)
    \end{center}
  \end{minipage}
  \spacingset{1}{\caption{Examples of causal DAGs with Double Negative
      Controls. \textit{Note:} Auxiliary variables $W$ and $Z$ are
      added to Figure~\ref{fig:dag-dyad}.(i). The causal relationships between focal behaviors $Y_{kt}$, unmeasured confounder $U_k$, and the
      tie relationship $S$ are the same as those in Figure~\ref{fig:dag-dyad}.(i).}\label{fig:dag-dyad-nc}}
\end{figure}

In practice, plausible candidates for negative controls are auxiliary
variables that (a) do not affect network relationships and (b) do not affect variables of other units. 
Figure~\ref{fig:dag-dyad-nc} represents examples of causal DAGs that
extend Figure~\ref{fig:dag-dyad}.(i) and encode Assumption~\ref{nc}. In Figures~\ref{fig:dag-dyad-nc}.(i)--(iii), $W$ and $Z$ are auxiliary variables that do not affect the dyadic
relationship $S$ or do not affect variables of the other
unit. A variety of relationships between the focal behaviors and
negative controls can be accommodated. In Figure~\ref{fig:dag-dyad-nc}.(i), $W$ and $Z$ are pre-treatment variables affecting
the focal behaviors, while in Figure~\ref{fig:dag-dyad-nc}.(ii),  $W$ and $Z$ have no causal
relationship with the focal behaviors. In Figure~\ref{fig:dag-dyad-nc}.(iii), $W$ and $Z$ are
intermediate variables between focal behavior at baseline $Y_{k1}$
and focal behavior at follow-up $Y_{k2}$. Figure~\ref{fig:dag-dyad-nc}.(iv) shows that focal behavior at
baseline $Y_{11}$ can also serve as NCO in the scenario represented
by Figure~\ref{fig:dag-dyad-nc}.(i). Indeed, in all 
Figures~\ref{fig:dag-dyad-nc}.(i)--(iii), $Y_{11}$ may also serve as
NCO.

\paragraph{Example (Negative Controls).}
In the context of Add Health data, \cite{cohen2008} estimated
peer effects on three health outcomes --- acne, height, and headaches --- known ex ante not to
exhibit peer effects. The purpose of \cite{cohen2008}'s analysis was to
investigate the validity of popular statistical approaches that assume the absence of unmeasured
homophily and contextual confounding \citep[e.g.,][]{fowler2007spread}. By finding implausible peer
effects on the three health outcomes mentioned above, \cite{cohen2008}
warn that unmeasured homophily and contextual confounding may be operating
in studies of peer effects based on Add Health data.

In the proposed double negative control approach, the three health
outcomes can be used not only for detecting confounding but also as
negative controls to potentially correct for such confounding. For example, whether a student has headaches is
unlikely (a) to causally affect whether students
are friends to each other and (b) to causally affect peers' headaches. If
these conditions are plausible in applied contexts, an ego's headache
and a peer's headache can be used as the NCO and NCE, respectively. \qed

\vspace{0.1in}
Upon selecting valid NCO and NCE variables, one approach supposes that there exists an outcome confounding bridge
\citep{miao2018confounding} that relates the confounders' effects on
negative control outcome $W$ to the  confounders' effects on outcome of primary interest $Y_{12}.$
\begin{assumption}[Outcome Confounding Bridge]
  \label{cb}
  There exists a function $h(W, Y_{21}, \bX)$ such that for all $y_{21} \in \cY_{21}$,
  \begin{equation}
    \E(Y_{12} \mid Y_{21} = y_{21}, U, \bX, S = 1)  = \E\{h(W,
    y_{21}, \bX) \mid  Y_{21} = y_{21}, U, \bX, S = 1\}. \label{eq:cb}
  \end{equation}
\end{assumption}
Assumption~\ref{cb} states that the confounding effect of $U$ on
outcome $Y_{12}$ is equal to the confounding effect of $U$ on $h(W,
y_{21}, \bX)$, a transformation of $W$. One simple yet important
implication of this assumption is that $W$ should be associated
with $U$ conditional on the treatment, observed covariates, and the
dyadic relationship. 

This assumption formally connects the confounding effect on the outcome $Y_{12}$ and the
confounding effect on the negative control outcome $W.$ Instead
of assuming complete knowledge of its relationship, the proposed double negative
control approach will use negative control exposure $Z$ to identify it, as described below in Theorem~\ref{iden}.

Formally, equation~\eqref{eq:cb} is a Fredholm integral equation of the first
kind \citep{kress1989, carrasco2007linear}. Existence of a solution to this equation can be established under regularity
conditions regarding the NCO relevance and a certain singular value
decomposition of the operator defining the integral equation. These conditions are somewhat technical and so we
reserve their details to Section~\ref{subsec:proof-existence} in the
supplementary material, where we provide further
discussion and a proof of the following lemma. 
\begin{lemma}
  \label{existence}
  Under Assumptions~\ref{com-wu} and~\ref{assum-exis} $($defined in Section~\ref{subsec:proof-existence}$)$, there exists a function $h(W, Y_{21}, \bX)$ such that for all $y_{21} \in \cY_{21}$, equation~\eqref{eq:cb} holds. 
\end{lemma}

\paragraph{Example (Linear Confounding Bridge).}
While the bridge function $h$ can take any functional form, we illustrate the assumption with a linear confounding bridge.
Suppose that $\E(Y_{12} \mid Y_{21}, U, \bX, S = 1) = (1, Y_{21},
U, \bX)^\top \beta$ and that $\E(W \mid U, \bX, S = 1)$ is linear in
$U$ and $\bX$, then equation~\eqref{eq:cb} holds with $h(W, Y_{21}, \bX;
\gamma) = (1, W, Y_{21}, \bX)^\top \gamma$, with an appropriate value of
$\gamma$. We do not need to assume the value of $\gamma$. Rather, as shown below, we can use an appropriate choice of negative control
exposure $Z$ to identify and estimate $\gamma$. \qed

\paragraph{Example (Categorical Variables).}
Suppose $W, U, Z, Y_{21}$ are all categorical variables. We define
possible values for $W$ and $U$ as $w_i$ and $u_j$ for $i = 1, \ldots,
|W|,  j=1, \ldots, |U|$ where $|\cdot|$ denotes the cardinality of a
categorical variable. $|Y_{21}|$ and $|Z|$ are also similarly
defined. Define also $\Pr(W \mid y_{21}, U, \bx, S = 1)$ to be a $|W| \times |U|$ matrix with $\Pr(W \mid y_{21}, U, \bx, S
= 1)_{ij} = \Pr(W = w_i \mid y_{21}, U = u_j, \bx, S = 1).$ As long
as $\Pr(W \mid y_{21}, U, \bx, S = 1)$ have full column rank (which implies
that $|W| \geq |U|$), the confounding bridge $h(W, y_{21}, \bx)$
exists as a solution to $\E(Y_{12} \mid y_{21}, U, \bx, S = 1) =
h(W, y_{21}, \bx) \Pr(W \mid y_{21}, U, \bx, S = 1)$ where $h(W, y_{21}, \bx)$ is a $1
\times |W|$ vector. When the number of categories for $U$ is equal to that of $W$, the
confounding bridge function is unique and given by $h(W, y_{21}, \bx)
= \E(Y_{12} \mid y_{21}, U, \bx, S = 1) \Pr(W \mid y_{21}, U, \bx, S = 1)^{-1}$. \qed

\vspace{0.1in}
Finally, we make the following assumption about negative control relevance. 
\begin{assumption}[Negative Control Relevance]
  \label{com}
  For any square integrable function $f$ and any $y_{21}$ and $\bx$, 
  if $\E\{f(W) \mid Z = z, Y_{21} = y_{21}, \bX = \bx, S = 1\} = 0$
  for almost all $z$, then $f(W) = 0$ almost surely. 
\end{assumption}
This assumption states that $Z$ is sufficiently informative about
$W$, which is essential to ensure identification of the outcome bridge
function $h$. This condition is formally known as a completeness condition, a well known technical condition central to the study of sufficiency
in statistical inference \citep{casella2001statistical}. Many commonly-used parametric and
semiparametric models, such as semi-parametric
exponential family \citep{newey2003} and semiparametric location-scale
family \citep{hu2018}, satisfy the completeness condition.

The completeness condition has been widely used to achieve identification in
non-parametric instrumental variable models \citep[e.g.,][]{d2011complete,
  darolles2011}. In the nonparametric instrumental variable
literature, completeness is an instrumental variable
relevance condition, which generalizes the rank condition of
linear instrumental variable models \citep{newey2003}. Thus Assumption~\ref{com}
essentially means that $Z$ is a relevant variable for $W$ conditional on $(Y_{21}, \bX, S)$. 
\paragraph{Examples.}
To gain further intuition,  we consider implications of the completeness condition. 
\begin{itemize}[leftmargin=0in]
\item[] (Categorical Variables). Consider a special case of categorical NCO and NCE. In this case, Assumption~\ref{com}
  requires that the number of levels in NCE must be at least as large as
  the number of levels in NCO. 
\item[] (Continuous Variables).
  When both NCO and NCE are continuous variables, Assumption~\ref{com}
  requires that the number of NCEs must be at least as large as
  the number of NCOs. 
\item[] (Parametric or Semiparametric Confounding Bridge). While Assumption~\ref{com} is important for accommodating a nonparametric
  confounding bridge function, we can relax the completeness condition when a bridge function belongs to a parametric or
  semiparametric model $h(W, Y_{21}, \bX; \gamma)$ indexed by a finite
  or infinite dimensional parameter $\gamma$. Under such a model, the
  completeness condition only requires that, for all $y_{21}$ and $\bx$, $\E\{h(W,
  y_{21}, \bx; \gamma) - h(W, y_{21}, \bx; \gamma^\prime) \mid Z, \bX
  = \bx, Y_{21} = y_{21}, S = 1\} \neq 0$ with a positive probability for
  any $\gamma \neq \gamma^\prime$ \citep[see][for further
  details]{miao2018confounding}. \qed
\end{itemize}

\paragraph{Remark.}
We note that alternative completeness conditions
may also be sufficient for identification of the ACPE. While 
completeness condition (Assumption~\ref{com}) is analogous to the one used
in \cite{miao2018confounding}, alternative completeness conditions
have also been considered in related studies
\citep[e.g.,][]{deaner2018proxy, miao2018identifying, shi2020dnc,
  kallus2021causal}, all in the more tractable i.i.d. or panel data
settings and not in network settings. In addition to Theorem~\ref{iden} given below, for the sake of completeness, we also establish nonparametric identification of the
ACPE under alternative identifying conditions in
Section~\ref{subsec:prove-iden-2} of the supplementary material. We
further discuss completeness condition in Section~\ref{subsec:more-com} of the supplementary material. Interested readers can also see
\cite{chen2014local} and \cite{andrews2017examples}, and references
therein for an overview of the role of completeness in nonparametric causal inference. \qed  

The following theorem demonstrates nonparametric identification of the
ACPE under the stated assumptions. 
\begin{theorem}
  \label{iden}
  Under Assumptions~\ref{l-ig}-\ref{com}, the confounding bridge
  function is identified as the unique solution to the following
  equation.
  \begin{equation*}
    \E(Y_{12} \mid Z, Y_{21}, \bX, S = 1) = \E\{h(W, Y_{21}, \bX) \mid
    Z, Y_{21}, \bX, S = 1\}.
  \end{equation*}
  Using the identified confounding bridge function $h(W, Y_{21}, \bX)$,
  the ACPE is identified by 
  \begin{equation*}
    \tau(y_{21}, y_{21}^\prime) = \E\{h(W, y_{21}, \bX) - h(W,
    y_{21}^\prime, \bX) \mid S = 1\}.
  \end{equation*}
\end{theorem}
Importantly, under Assumptions~\ref{l-ig}-\ref{com}, we have
identified the ACPE without imposing any parametric restriction on the
confounding bridge or negative controls. We provide a proof in
Section~\ref{subsec:prove-iden} of the supplementary material.

\paragraph{Example (Identification under Linear Confounding Bridge).}
While Theorem~\ref{iden} establishes nonparametric identification of
the ACPE, here we consider a simple case with a binary treatment $Y_{21}$
and a binary NCE $Z$ without any pre-treatment covariates, which admits a closed-form solution. Suppose
the confounding bridge function is linear: $h(W, Y_{21}; \gamma) =
\gamma_0 + \gamma_1 W + \gamma_2 Y_{21}$. Then, the ACPE is
identified as 
\begin{equation*}
  \tau(1, 0) \ = \ \E( \mbox{OD}_{Y_{12} Y_{21} \mid Z}) - \E(
  \mbox{OD}_{W Y_{21} \mid Z}) \times
  \cfrac{\E(\mbox{OD}_{Y_{12} Z \mid Y_{21}})}{\E(\mbox{OD}_{W Z \mid Y_{21}} )} 
\end{equation*}
where $\mbox{OD}_{V_{1} V_{2} \mid V_{3}} = \E(V_{1} \ | \ V_{2} = 1,
V_{3}, S = 1) - \E(V_{1} \ | \ V_{2} = 0, V_{3}, S = 1).$

The first term $\E( \mbox{OD}_{Y_{12} Y_{21} \mid Z})$ corresponds to
a biased estimator of the ACPE; a regression of outcome $Y_{12}$ on treatment $Y_{21}$ conditional on
$Z$, which is equal to the ACPE only in the absence of unmeasured
confounder. The second term $\E(
\mbox{OD}_{W Y_{21} \mid Z})$ corresponds to an estimator of the
confounding effect on $W$, which should be zero in the absence of
unmeasured confounding. This captures the amount of confounding
to be corrected. Finally, the third term
$\E(\mbox{OD}_{Y_{12} Z \mid Y_{21}})/\E(\mbox{OD}_{W Z \mid
    Y_{21}} )$ represents a ratio of the confounding effects
on the outcome $Y_{12}$ and on the negative control outcome $W$,
which is estimated by using NCE $Z$. It is clear here that the
negative control relevance (Assumption~\ref{com}) is essential to
guarantee $\E(\mbox{OD}_{W Z \mid Y_{21}}) \neq 0.$ Intuitively, the double negative
control approach subtracts an estimated bias (the second term) scaled
by the differential confounding effects
on the outcome and on the NCO (the third term) from the original biased estimator (the first term). 
  
This explicit form contains two important special cases: (1)
conditional ignorability (i.e., no unmeasured network confounding), and (2) the
well-known difference-in-differences (DID) design
\citep{angrist2008mostly}. First, when conditional ignorability
holds and there exists no unmeasured network confounding, $\E(
\mbox{OD}_{W Y_{21} \mid Z}) = 0$ and $\tau(1, 0) = \E(
\mbox{OD}_{Y_{12} Y_{21} \mid Z})$, which reduces to the standard
identification formula under conditional ignorability \citep{rosenbaum1983central,
  robins1986causal}. Second, the formula reduces to DID when we
assume the confounding effect on the outcome is equal to the
confounding effect on the NCO. \cite{tchetgen2016negative} show that
the DID uses a pre-treatment outcome ($Y_{11}$ in our setting) as $W$ and assumes the entire bridge
function is known, i.e., not only a functional form but also the value of
coefficients $\gamma$, without using any NCE. In particular, the widely-used assumption of
parallel trends assumes that the confounding effects on the outcome
$Y_{12}$ and on pre-treatment outcome $Y_{11}$ (used as the NCO) are
the same, i.e., assuming the third term equal to one. 

In contrast, the double negative control approach can use any valid $W$
(including pre-treatment outcome $Y_{11}$ as a special case; see
Figure~\ref{fig:dag-dyad-nc}.(iv)). Most importantly, we use NCE $Z$ to estimate the confounding bridge function --- the
differential confounding effects
on the outcome $Y_{12}$ and on the negative control outcome $W$
--- as the ratio $\E(\mbox{OD}_{Y_{12}Z \mid
  Y_{21}})/\E(\mbox{OD}_{W Z \mid Y_{21}} )$ in the third term. It is
important to emphasize that while we use this closed-form solution in the linear confounding bridge case
to illustrate the intuition behind the double negative control approach, our
identification results do not impose any parametric restriction on the
confounding bridge or negative controls. \qed

\subsection{Estimation and Inference}
We now propose a strategy for estimation and inference of the
ACPE. Because we observe $n$ independent and identically distributed
samples of dyads, we observe
independent and identically distributed samples on $(Y_{12}, Y_{21},
W, Z, \bX)$ given $S = 1$ where $Y_{12}, Y_{21}, W, Z, \bX$ are the outcome of interest, 
treatment, NCO, NCE, and observed pre-treatment
covariates, respectively. 

Suppose that an analyst has specified a parametric or semiparametric model for the
confounding bridge $h(W, Y_{21}, \bX; \gamma)$ with parameter
$\gamma$. Then, based on Theorem~\ref{iden}, we can
estimate $\gamma$ by solving the following empirical moment equations.
\begin{equation*}
  \frac{1}{n} \sum_{i=1}^n \{Y_{i12} - h(W_{i}, Y_{i21}, \bX_{i}; \gamma)\}
  \times \eta(Z_{i}, Y_{i21}, \bX_i) = 0,   
\end{equation*}
where $\eta$ is a user-specified vector function with dimension equal to that of $\gamma$. For example, if a linear confounding
bridge function is used, i.e., $h(W, Y_{21}, \bX; \gamma) = (1,
W, Y_{21}, \bX)^\top\gamma$, we can use $\eta(Z, Y_{21}, \bX) = (1,Z, Y_{21}, \bX)^\top$. 

Once the bridge function $h$ is estimated,
we can estimate the ACPE by
\begin{equation*}
  \frac{1}{n} \sum_{i=1}^n \{h(W_{i}, y_{21}, \bX_i; \widehat{\gamma})
  - h(W_{i}, y_{21}^\prime, \bX_i; \widehat{\gamma})\}.
\end{equation*}
To appropriately account for uncertainty of the estimated bridge function and for the possibility that dimension of $\eta$ might be larger than that of $\gamma$, we
combine the two moments into generalized method of moments (GMM) with parameter
$\theta = (\tau, \gamma)$ \citep{hansen1982gmm}. We define a moment for dyad $i$ to be
\begin{equation*}
  m(Y_{i12}, Y_{i21}, W_{i}, Z_{i}, \bX_{i}; \theta) = \left\{
    \begin{array}{l}
      \tau - \{h(W_{i}, y_{21}, \bX_i; \gamma) - h(W_{i},
      y_{21}^\prime, \bX_i; \gamma)\}\\
      \{Y_{i12} - h(W_{i}, Y_{i21}, \bX_i; \gamma)\} \times \eta(Z_i,
      Y_{i21}, \bX_i)     
    \end{array}
  \right\}. 
\end{equation*}
Then, the GMM estimator is
\begin{equation}
  \widehat{\theta} = \argmin_{\theta} \overline{m}(\theta)^\top \Omega
  \ \overline{m}(\theta) \label{eq:dyad-gmm}
\end{equation}
where $\overline{m}(\theta) = \frac{1}{n} \sum_{i=1}^n m(Y_{i12},
Y_{i21}, W_{i}, Z_{i}, \bX_{i}; \theta)$ and $\Omega$ is a
user-specified positive-definite weight matrix. Asymptotic properties
described below hold for any positive-definite weight matrix $\Omega$. 

The proposed double negative control (DNC) estimator
$\widehat{\tau}(y_{21}, y_{21}^\prime)$ for $\tau(y_{21},
  y_{21}^\prime)$ is the first element of $\widehat{\theta}$ defined
  in equation~\eqref{eq:dyad-gmm}. Because we consider i.i.d samples of dyads in this section, the moment
$m(Y_{i12}, Y_{i21}, W_{i}, Z_{i}, \bX_{i}; \theta)$ is also i.i.d., and thus, under the standard
regularity conditions for GMM \citep{hansen1982gmm, newey1994}, the DNC estimator is consistent: 
\begin{equation*}
  \widehat{\tau}(y_{21}, y_{21}^\prime) \xrightarrow{p} \tau(y_{21}, y_{21}^\prime), 
\end{equation*}
and asymptotically normal:  
\begin{equation*}
  \cfrac{\widehat{\tau}(y_{21}, y_{21}^\prime) - \tau(y_{21},
    y_{21}^\prime)}{\sqrt{\sigma^2/n}} \xrightarrow{d}
  \textsf{Normal}(0, 1),
\end{equation*}
where $\xrightarrow{p}$ denotes convergence in probability, and
$\xrightarrow{d}$ denotes convergence in
distribution. Moreover, the asymptotic variance $\sigma^2$ can be
consistently estimated by $\widehat{\sigma}^2 = (\widehat{\Gamma}
    \widehat{\Lambda} \widehat{\Gamma}^\top)_{11}$, which is the $(1,1)$ th
element of matrix $\widehat{\Gamma}
\widehat{\Lambda} \widehat{\Gamma}^\top$, and 
\begin{eqnarray*}
  \widehat{\Lambda} & = & \frac{1}{n}
    \sum_{i=1}^n m(Y_{i12}, Y_{i21}, W_{i}, Z_{i}, \bX_{i}; \widehat{\theta}) \ m(Y_{i12}, Y_{i21}, W_{i}, Z_{i}, \bX_{i}; \widehat{\theta})^\top, \\
\widehat{\Gamma} & = & (\widehat{M}^\top \Omega
\widehat{M})^{-1} \widehat{M}^\top \Omega, \ \ \mbox{and} \ \ \widehat{M} = \frac{1}{n} \sum_{i=1}^n \frac{\partial}{\partial
  \theta}  m(Y_{i12}, Y_{i21}, W_{i}, Z_{i}, \bX_{i}; \widehat{\theta}).
\end{eqnarray*}
Therefore, an asymptotically valid $(1-\alpha)$
confidence interval for $\tau(y_{21}, y_{21}^\prime)$ is given by
$[\widehat{\tau}(y_{21}, y_{21}^\prime) - \Phi(1-\alpha/2) \times 
\widehat{\sigma}/\sqrt{n}, \ \widehat{\tau}(y_{21}, y_{21}^\prime) +
\Phi(1-\alpha/2) \times \widehat{\sigma}/\sqrt{n}]$ where $\Phi(\cdot)$
denotes the quantile function for the standard normal distribution.

To minimize the asymptotic variance within the GMM class, we can use
the two-step GMM to estimate the optimal
$\widehat{\Omega}.$ In the first step, we choose an identity matrix as
$\Omega$ or some other positive-definite matrix, and compute
preliminary GMM estimate $\widehat{\theta}_{(1)}$. This estimator is
consistent, but not efficient. In the second step, we compute
$\widehat{\Lambda}$ based on $\widehat{\theta}_{(1)}$, which is
denoted by $\widehat{\Lambda}_{(1)}.$ Then, we can get the final
estimate by solving equation~\eqref{eq:dyad-gmm} with $\Omega =
\widehat{\Lambda}_{(1)}^{-1}.$ The resulting estimator
$\widehat{\theta}$ is consistent and asymptotically normal, and is
asymptotically efficient within the GMM class \citep{hansen1982gmm}. The asymptotic variance
also simplifies to $(\widehat{M}^\top \widehat{\Lambda}^{-1}
\widehat{M})^{-1}.$ To further improve finite sample performance,
researchers can consider alternative GMM estimators, such as
continuously updating GMM \citep{hansen1996finite}.

\section{Double Negative Controls for Network Data}
\label{sec:net}
\setcounter{assumption}{1}
\renewcommand {\theassumption} {\arabic{assumption}}
In this section, we consider a sample of interconnected units in a single
network. Extending results in Section~\ref{sec:dyad}, we propose the
double negative control approach to identification of the ACPE in the presence of unmeasured network confounding, which includes latent
homophily and contextual confounding as special cases. We then examine
estimation and inference while accounting for both
unmeasured network confounding and network-dependent observations. 

\subsection{Notation and Definitions}
Suppose one has observed data on a population of $n$ units
interconnected by a network. We let $N_n = \{1, \ldots, n\}$ be the
set of unit indices. We consider an undirected network $\cG_n$
where ties or links between units are mutual, and connected units can
affect each other. Formally, we define $\cG_n = (N_n, \bG)$, i.e., a set of units
$N_n$ connected by mutual ties, represented by a network adjacency
matrix $\bG$. The network adjacency
matrix $\bG$ depends on sample size $n$, but the index will be
suppressed in the following discussion as it eases the exposition
without confusion.  The entry $G_{ij}$ takes the value of one if unit $i$
and $j$ are connected and takes the value of zero otherwise. We follow
the convention that $G_{ii} = 0$ for $i \in N_n$, and we call units $i$ and $j$ \textit{peers} if $G_{ij} = 1.$
We also define two network notations useful throughout the paper. We
define network distance $d_n(i,j)$ to be the length of the shortest
path between nodes $i$ and $j$ on network $\cG_n$. We define $\cN_n(i;s)$ to be a set of nodes that are
at distance $s$ from node $i$:
\begin{equation*}
  \cN_n(i;s) = \{j \in N_n: d_n(i, j) = s\}.
\end{equation*}

For each $i \in N_n$, one observes $(Y_{i1}, Y_{i2}, \bX_i)$, where $Y_{it}$ denotes the focal behavior
of unit $i$ at time $t \in \{1, 2\}$, and $\bX_i$ are covariates of unit $i$
measured prior to $Y_{i1}$. $\bX_i$ can include
network-characteristics, such as the network degree of unit $i$. We
call $t = 1$ baseline and $t=2$ follow-up.  

For the sake of clarity in the exposition, we restrict presentation of all
main results to the causal effect from peers. It is important to
emphasize that results in this section, however, do not assume the absence of the causal effects
from higher-order peers (e.g., peers-of-peers); we only consider such
higher-order peer effects as nuisance. In Section~\ref{sec:ext},
we discuss similar results for a general case where higher-order peer effects (e.g., the
causal effect from peers-of-peers) is the main causal estimand of interest. 

As the outcome variable, we focus on focal behavior at
follow-up $Y_{i2}$. In principle, it is possible to perform causal inference by defining a
multivariate treatment variable based on focal behaviors of peers at baseline $\{Y_{j1}: j \in \cN(i;1)\}.$ However, in practice, researchers may need to make a dimension-reducing
assumption, known as an exposure mapping
\citep{aronow2012interference}, to define the treatment variable, $A_i
= \phi(\{Y_{j1}: j \in \cN(i;1)\}) \in \mathbb{R}$ where function $\phi$
is specified by a researcher based on subject matter knowledge. For example,
the most common choice is $A_i = \sum_{j=1}^n G_{ij}
Y_{j1}/\sum_{j=1}^n G_{ij}$, while our results can accommodate any
choice of $\phi$.  The potential outcome $Y_{i2}(a)$ is defined as the
outcome that would realize when the treatment variable is set to $A_i
= a$.  We make the standard
consistency assumption linking observed and potential outcomes, 
$Y_{i2} = Y_{i2}(A_i)$, throughout the paper. Our goal is to estimate the Average Causal Peer Effect (ACPE), defined as
\begin{equation}
  \tau (a, a^\prime) \coloneqq \frac{1}{n} \sum_{i =1}^n
  \E \left\{Y_{i2}(a) - Y_{i2}(a^\prime)\right\} \label{eq:acpe} 
\end{equation}
where $a, a^\prime \in \mathcal{A}$ where $\mathcal{A}$ is the support
of $A$. We define the expectation conditional on the observed network
$\cG_n,$ while we omit its conditioning for notational simplicity.
Unlike typical causal parameters in i.i.d settings, units' potential outcome
may not share a common expectation, i.e., $\E \left\{Y_{i2}(a) -
  Y_{i2}(a^\prime)\right\} \neq \E \left\{Y_{j2}(a) -
  Y_{j2}(a^\prime)\right\}$ for $i \neq j$. Thus,
the causal estimand is explicitly written as the empirical mean.   

To identify the ACPE, existing works rely upon the assumption that
the observed variables are sufficient to
account for confounding of the relationship between $A_i$ and $Y_{i2}(a)$, i.e., 
\begin{equation*}
  Y_{i2}(a) \ \indep \ A_i \mid Y_{i1}, \bX_{i}, \label{eq:obs}
\end{equation*}
where observed pre-treatment covariates $\bX_{i}$ can include observed
covariates of peers of unit $i$ or covariates of other units who are indirectly connected to unit $i$. Even though some recent methods allow for
network dependence across units \citep[e.g.,][]{ogburn2017causal, tchetgen2020auto},
they assume such latent network dependence does not confound the main
outcome-treatment relationship. However, such assumption is, in general, untenable in many
applications due to unmeasured network confounding, including 
unmeasured homophily \citep{shalizi2011homophily} and contextual
confounding \citep{vanderweele2013social}, as discussed in
Section~\ref{subsec:challenge}. 

\subsection{Identification Assumptions with Double Negative Controls}
We propose an alternative approach based on double negative controls.  We generalize Assumptions~\ref{l-ig}--\ref{com} in Section~\ref{sec:dyad} to the
network setting.

\begin{assumption}
  \label{net-assum}
  \ \  \vspace{-0.3in} \\
  \begin{itemize}
  \item[] \textnormal{1. (Latent Ignorability).} For all $a \in \cA$ and all $i \in N_n$, 
    \begin{equation*}
      Y_{i2}(a) \ \indep \ A_i \mid U_i, \bX_i.
    \end{equation*}
  \item[] \textnormal{2. (Negative Controls).} For all $i \in N_n$,
    \begin{eqnarray*}
      && \mbox{\textnormal{(Negative Control Outcome)}} \ \ \ W_i \
         \indep \ A_{i} \mid U_i, \bX_i,\\
      && \mbox{\textnormal{(Negative Control Exposure)}} \ \ \ \bZ_i \ \indep \ Y_{i2} \mid A_i, U_i, \bX_i, \ \ \mbox{and} \ \ \  \bZ_i \ \indep
         \ W_i \mid A_i, U_i, \bX_i, 
    \end{eqnarray*}
    where $\bZ_i$ is a vector of negative control exposures. 
  \item[]  \textnormal{3. (Outcome Confounding Bridge).}   There exists an outcome confounding bridge function $h(W_{i}, A_{i}, \bX_i)$ such
    that for all $a \in \cA$, and all $i \in N_n,$
    \begin{equation}
      \E(Y_{i2} \mid A_{i} = a, U_i, \bX_i)  = \E\{h(W_{i}, a, \bX_i) \mid  A_{i} = a, U_i, \bX_i\}. \label{eq:cb-net}
    \end{equation}
  \item[]  \textnormal{4. (Negative Control Relevance).} For any square integrable
    function $f$ and any $a$ and $\bx$, if $\E\{f(W_{i}) \mid \bZ_{i} = \bz, A_i = a, \bX_i =
    \bx\} = 0$ for almost all $\bz$, then $f(W_{i}) = 0$ almost surely.
  \end{itemize}
\end{assumption}

\subsection{Selecting Negative Controls in Network Settings}
\label{subsubsec:select-nc}
In the proposed double negative control approach, selection of negative
control outcome and exposure is essential in practice. While, in principle any
negative controls satisfying Assumption~\ref{net-assum} can be used,
we discuss two convenient strategies to select negative controls in the network
setting.

\subsubsection{Using Auxiliary Variables}
\label{subsub:auxi}
First, as in Section~\ref{sec:dyad}, a plausible candidate is an
auxiliary variable $C_i$ that (a) does not affect network
relationships $\cG_n$ and (b) does not affect variables of other units. For example, in the context of Add Health data, whether a student experiences
headaches is likely to satisfy this condition (See ``Example (Negative
Controls)'' in Section~\ref{subsec:iden}).

Figure~\ref{fig:dag-net-genNC} represents
examples of causal graphs where Assumption~\ref{net-assum}.2 holds. We
view unit 2 as an ego, who has two peers (units 1 and 3) and one peer-of-peers (unit
4). We use a fully connected chain graph \citep{lauritzen2002} to
denote general network dependence of latent confounders $U$ across
units. This chain graph representation is one general way to capture
unmeasured network confounding, which can accommodate both unmeasured
homophily and contextual confounding. 

In Figure~\ref{fig:dag-net-genNC}.(i), $C_2$ satisfies  NCO conditions and three variables $\{C_1, C_3, C_4\}$ satisfy NCE conditions. Importantly, not only auxiliary variables of peers but
also those of peers-of-peers also satisfy the conditions of the NCE. 

Finally, we provide primitive sufficient conditions that imply the
negative control conditions. In particular, the following conditions
capture a general approach for using auxiliary variables as negative controls. 
\begin{eqnarray}
  C_{i} & \indep & \{(C_j, Y_{j1}): j \neq i\} \ \mid \ U_i, \bX_i, \label{eq:a-1}\\ 
  Y_{i2} & \indep & \{C_j: j \neq i\} \ \mid \ A_i, U_i, \bX_i, \label{eq:a-2}
\end{eqnarray}
Equations~\eqref{eq:a-1} and~\eqref{eq:a-2} formalize the notion 
that $C_i$ should not affect network relationships and should
not affect peers' variables. Lemma~\ref{nc-aux} below shows that auxiliary variable $C_i$ can serve as a valid negative control if it
satisfies the stated conditions (Figure~\ref{fig:dag-net-genNC}.(i) is an example).
\begin{lemma}
  \label{nc-aux}
  Suppose auxiliary variable $C_i$ satisfies the two conditions
  $($equations~\eqref{eq:a-1} and~\eqref{eq:a-2}$)$. Then, Assumption~\ref{net-assum}.2 holds
  with $W_i = C_i$ and $\bZ_i = \{C_{j}: j \neq i\}.$
\end{lemma}
Several points are worth noting. First, these are sufficient
conditions, not necessary conditions for Assumption~\ref{net-assum}.2. Therefore, 
any negative controls that satisfy Assumption~\ref{net-assum} can in principle be
used for identification. Second, Lemma~\ref{nc-aux} suggests that in practice, there may be multiple NCEs because one can use
auxiliary variables of all other units $\{C_{j}: j \neq i\}.$ Therefore, it
may be possible to enhance identification and increase estimation efficiency by exploiting a large number of
NCEs. We plan to examine optimal selection and specification of NCEs in future work. 

\subsubsection{Using the Focal Behaviors}
\label{subsub:focal}
\begin{figure}[!p]
  \begin{center}
    \includegraphics[width=0.9\textwidth]{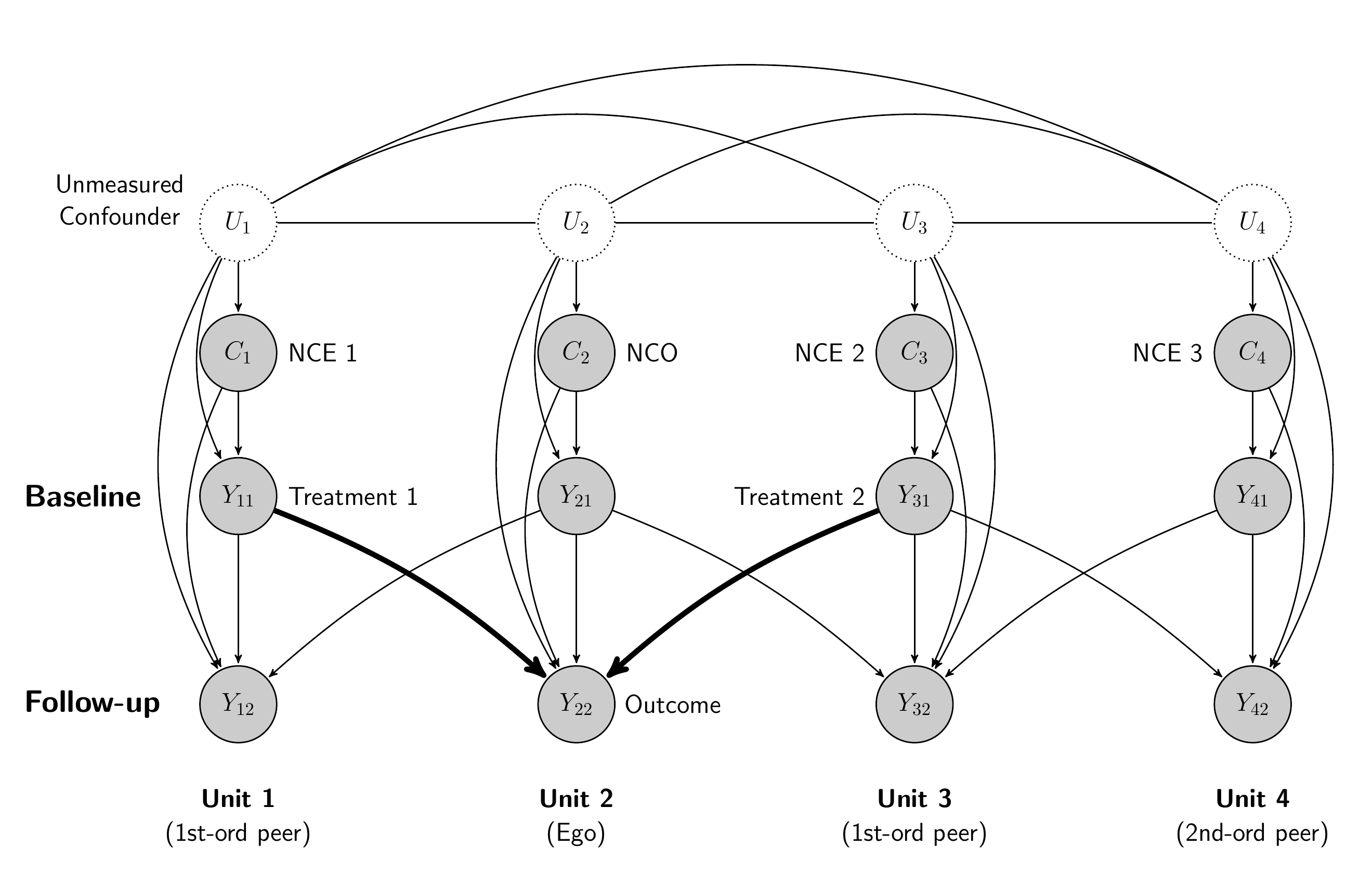}
    \\ \hspace{0.4in} (i) 
  \end{center}
  \ 
  \begin{center}
    \includegraphics[width=0.9\textwidth]{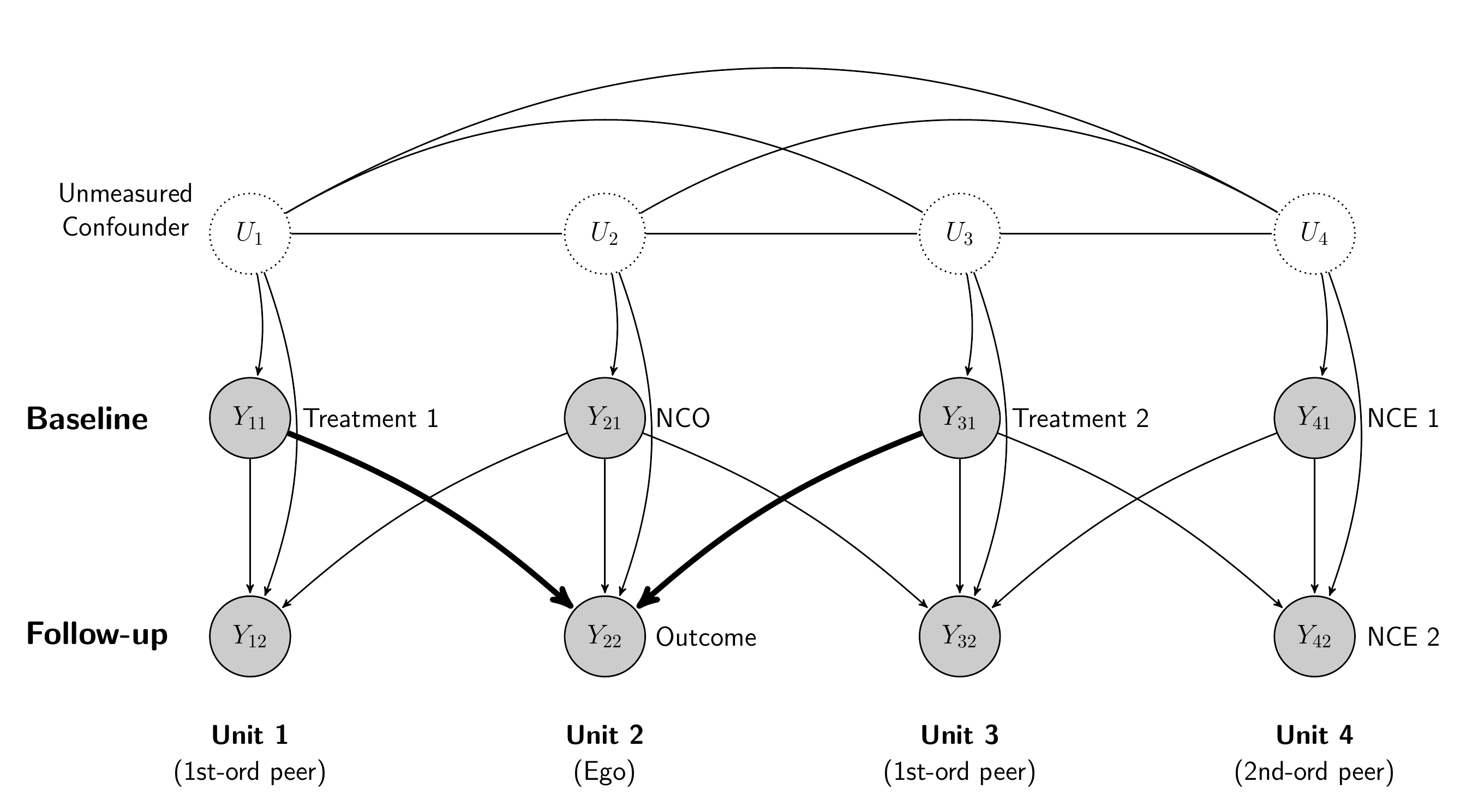}
    \\ \hspace{0.4in} (ii)
  \end{center}
  \spacingset{1}{\caption{Examples of chain graphs with Double
      Negative Controls. \textit{Note:} We use fully connected chain
      graphs to denote general network dependence of
      latent confounders $U$ across units. For concreteness, we show Unit 2 as the ego. The thick arrows
      from $Y_{11}$ to $Y_{22}$ and from $Y_{31}$ to $Y_{22}$ indicate the causal peer effects of
      interest. We use shaded (dotted) nodes to denote observed
      (unobserved) variables. }\label{fig:dag-net-genNC}}
\end{figure}

Second, in certain settings,  we may also use focal behaviors $Y_{it}$ of peers and
those measured at baseline as plausible candidates for negative
controls. In particular, ego's focal behavior at baseline may serve as valid NCO and focal
behaviors of peers-of-peers $\{Y_{jt}: j \in \cN(i; 2), t \in \{1, 2\}\}$ may constitute valid
NCEs. Figure~\ref{fig:dag-net-genNC}.(ii) represents a causal graph illustrating
an instance of the causal model where $Y_{21}$ qualifies as NCO and variables $\{Y_{41}, Y_{42}\}$ qualify as NCE. More generally, when focal behaviors of peers-of-peers constitute valid negative control exposures,  focal behaviors of units at least of network distance $2$ from node
$i$ may be credible negative control exposures. A hybrid approach might entail combining the auxiliary variables and focal behaviors as negative
controls. In Figure~\ref{fig:dag-net-genNC}.(i), we define focal
behavior measured at baseline $Y_{21}$ as NCO (instead of $C_2$)
and auxiliary variables of peers $\{C_1, C_3, C_4\}$ as NCEs.

This selection of negative controls is particularly plausible when focal
behaviors do not have direct causal relationships with peers'
focal behaviors measured concurrently. In Figure~\ref{fig:dag-net-genNC}.(ii), while peers' focal behaviors
measured at baseline affect egos' focal behaviors measured at
follow-up (e.g., $Y_{11}$ and $Y_{31}$ affect $Y_{22}$), peers' focal
behaviors cannot causally affect egos' focal behaviors measured concurrently (e.g., $Y_{11}$ and $Y_{31}$ do not affect $Y_{21}$; $Y_{12}$
and $Y_{32}$ do not affect $Y_{22}$). This absence of causal
simultaneity has previously been assumed in the literature of causal peer effects
\citep{shalizi2011homophily, ogburn2014DAG, egami2018diffusion,
  liu2020regression, shalizi2021}.

In practice, this assumption is most credible when researchers a priori know
that the focal behaviors of units are indeed measured
concurrently. For example, in Add Health data, students' GPA are
likely to be measured at the same time for students within a school, and
thus, a student's GPA cannot be affected by peers' GPA in the same
semester. Importantly, a student's GPA can be affected by peers' GPA in
the last semester, and a student's study habit might be affected by
peer's study habits within the same semester. These, however, do not
invalidate the use of GPA of peers-of-peers as NCEs as long
as students' GPA within the same semester do not have direct causal
relationships with each other. In some applications, analysts can directly measure focal behaviors of interest. In such cases, by virtue of survey/study design,
researchers can ensure that the focal behaviors measured at each wave 
do not affect peers' focal behaviors within the same wave by
conducting surveys concurrently. This
assumption is less credible when measurements of focal behaviors are
aggregated over long periods of time, such as the number of political tweets
over a year, which is likely to be affected by peers' tweets
within the same year. This is often called the temporal
aggregation problem, which invalidates not only peer effect analysis
but also a large class of panel data analyses
\citep{granger1988some}.

As in Section~\ref{subsub:auxi}, we provide primitive sufficient conditions for valid
negative controls. In particular, the following conditions
capture a general approach for leveraging focal behaviors as negative controls. 
\begin{eqnarray}
  Y_{i1} & \indep & \{Y_{j1}: j \neq i\} \ \mid \ U_i, \bX_i, \label{eq:f-1}\\ 
  Y_{i2} & \indep & \{Y_{j1}: j \in \cN(i; s), s \geq \tilde{s} \} \ \mid \ A_i, U_i,
                    \bX_i, \ \ \ \ \mbox{with some integer $\tilde{s}$},\label{eq:f-2}
\end{eqnarray}
Equation~\eqref{eq:f-1} formalizes the notion that focal behaviors do not have direct causal relationships with the peers'
focal behaviors measured at the same time. Equation~\eqref{eq:f-2}
captures the assumption that only peers closer than distance $s$ can have
causal peer effects on an ego. While we focus on the causal effect from peers as the causal estimand, we
do not necessarily need to assume the absence of higher-order peer effects.

Lemma~\ref{nc-focal} below establishes that focal behaviors can serve as
valid negative controls when they satisfy the stated conditions
(Figure~\ref{fig:dag-net-genNC}.(ii) is an example).
\begin{lemma}
  \label{nc-focal}
  Suppose that focal behaviors satisfy conditions
  ~\eqref{eq:f-1} and ~\eqref{eq:f-2}. Then, Assumption~\ref{net-assum}.2 holds
  with $W_i = Y_{i1}$ and $\bZ_i = \{Y_{j1}: j \in \cN(i;s), s \geq \tilde{s} \}.$
\end{lemma}
Again, we emphasize that these conditions are sufficient for Assumption~\ref{net-assum}.2, and thus, there
may be other ways to justify negative control conditions. Any negative controls that satisfy Assumption~\ref{net-assum} may be
used for identification of ACPE.

This particular selection strategy of negative controls has two advantages when
valid. First, when the NCO entails focal behaviors measured at baseline, the confounding bridge assumption
(Assumption~\ref{net-assum}.3) is often more likely to hold because the
NCO and the main outcome are measured on the same scale \citep{tchetgen2016negative}. Second,
if researchers can leverage focal behaviors of peers and
those measured at baseline as negative controls, researchers do
not need to collect additional auxiliary variables, which lowers data
collection requirements and improves applicability of the double
negative control approach. However, selection of valid negative
control variables must always be based on reliable domain
knowledge because Assumption~\ref{net-assum}.1 --
Assumption~\ref{net-assum}.4 must be met.  

\subsection{Nonparametric Identification}
Analogous to Theorem~\ref{iden}, we now establish nonparametric identification of the ACPE under Assumption~\ref{net-assum}. 
\begin{theorem}
  \label{iden-net}
  Under Assumption~\ref{net-assum}, the confounding bridge
  function is identified as the unique solution to the following
  equation.
  \begin{equation*}
    \E(Y_{i2} \mid \bZ_{i}, A_{i}, \bX_i) = \E\{h(W_{i},
    A_{i}, \bX_i) \mid \bZ_{i}, A_{i}, \bX_i\},
  \end{equation*}
 and ACPE is identified by 
  \begin{equation*}
    \tau(a, a^\prime) = \frac{1}{n} \sum_{i=1}^n \E\{h(W_{i}, a,
    \bX_i)- h(W_{i}, a^\prime, \bX_i)\}.
  \end{equation*}
\end{theorem}
We provide a proof in Section~\ref{subsec:prove-iden} of the supplementary material.

Despite the complexity of network data, the identification
assumptions and formula of the ACPE are remarkably similar to those in the dyadic study design. Clearly, selection possibilities of negative
controls are more flexible in the network
setting. As discussed in Section~\ref{subsubsec:select-nc}, focal
behaviors of peers may be used as NCEs in some network applications in
addition to auxiliary variables. Thus, from a perspective of causal
identification, network data present richer opportunities for NC adjustment than dyadic
data in that they offer more options of credible negative
controls. An important difference from the dyadic case emerges in estimation and
inference where one must appropriately account for network-dependence, a challenging task we consider next.

\subsection{Estimation and Inference}
\label{subsec:est-net}
In this section, we consider estimation and inference for the ACPE
while allowing for network-dependent observations. 

We define a triangular array of $\R^v$-valued random vector, $\bL_{n,
  i} = (Y_{i2}, A_i, W_{i}, \bZ_i, \bX_i)$ for $i
\in N_n$, adapted to a network $\cG_n$ where $v$ is the length
of the vector $\bL_{n, i}$. Similar to Section~\ref{sec:dyad}, we define a moment estimating function with parameter $\theta = (\tau, \gamma)$.
\begin{equation*}
  m(\bL_{n,i}; \theta) = \left\{
    \begin{array}{l}
      \tau - \{h(W_{i}, a, \bX_i; \gamma) - h(W_{i}, a^\prime, \bX_i; \gamma)\}\\
      \{Y_{i2} - h(W_{i}, A_i, \bX_i; \gamma)\} \times \eta(A_i, \bZ_i, \bX_i)     
    \end{array}
  \right\}.
\end{equation*}
The GMM estimator for $\theta$ is
\begin{equation}
  \widehat{\theta} = \argmin_{\theta} \overline{m}(\theta)^\top \Omega
  \ \overline{m}(\theta), \label{eq:net-gmm}
\end{equation}
where $\overline{m}(\theta) = \frac{1}{n} \sum_{i=1}^n m(\bL_{n,i};
\theta)$ and $\Omega$ is a user-specified positive-definite weight
matrix. Therefore, the proposed DNC estimator $\widehat{\tau}(a, a^\prime)$ for $\tau(a,
  a^\prime)$ is the first element of $\widehat{\theta}$ defined in
  equation~\eqref{eq:net-gmm}. Asymptotic results we derive below
  hold for the two-step GMM or other alternative GMM estimators
  \citep{hansen1996finite}, too.
  
Since we consider a sample of interconnected units in a network, the
assumption that $m(\bL_{n,i}; \theta)$ is independently and
identically distributed is unrealistic. Below, we consider assumptions on
the observed data law that are considerably weaker, but still allow for valid
inferences about the ACPE in network settings. For ease of exposition, we consider a setting in which the expected causal peer effect, $\E\{Y_{i2}(a) -  Y_{i2}(a^\prime) \}$, is constant across units, while otherwise allowing for network-dependent (i.e., non-independent) errors. In the supplementary material, we extend our results to more general settings of heterogeneous ACPE.  

We define $\psi$-network dependence \citep{kojevnikov2020},
which encodes the degree of stochastic dependence between variables
in terms of network distance. 

\begin{definition}[$\psi-$Network Dependence \citep{kojevnikov2020}]
  The triangular array $\{\bV_{n,i}\}_{i \in N_n}, n \geq
  1, \bV_{n,i} \in \mathbf{R}^v$, is called conditionally $\psi$-weakly
  dependent given $\cG_n$, if for each $n \in \mathbb{N},$ there exist
  a $\cG_n$-measurable sequence
  $\beta_n = \{\beta_{n,s}\}, s \geq 0, \beta_{n,0} = 1$, and a collection of nonrandom functions
  $(\psi_{q_1,q_2})_{q_1,q_2\in\mathbb{N}}, \ \psi_{q_1,q_2}: \mathcal{L}_{v,q_1}
  \times \mathcal{L}_{v,q_2} \rightarrow [0, \infty)$, such that for all
  $(Q_1, Q_2) \in \cP_n(q_1, q_2;s)$\footnote{$\cP_n(q_1, q_2;s)$ denotes the
    collection of two sets of nodes of size $q_1$ and $q_2$ with distance
    between each other of at least $s$. Formally, $\cP(q_1,q_2;s) = \{(Q_1,
    Q_2): Q_1, Q_2 \subset N_n, |Q_1| =q_1, |Q_2|=q_2, \mbox{ and  } d_n(Q_1,Q_2) \geq s\}.$} with $s > 0$ and all $f_1 \in
  \mathcal{L}_{v,q_1}$ and $f_2 \in
  \mathcal{L}_{v,q_2},$\footnote{$\mathcal{L}_{v,q_1}$ and $\mathcal{L}_{v,q_2}$ denote the
    collection of bounded Lipschitz real functions on $\mathbf{R}^{v
      \times q_1}$ and $\mathbf{R}^{v
      \times q_2},$ respectively.}
  \begin{equation}
    |\Cov(f_1(\bV_{n,Q_1}), f_2(\bV_{n,Q_2}) \mid \cG_n)| \leq \psi_{q_1,q_2}(f_1,f_2) \beta_{n,s}  \ \ \  \mbox{a.s.}
  \end{equation}
  In this case, we call the sequence $\beta_n =
  \{\beta_{n,s} \}_{s=1}^{\infty}$ the weak dependent coefficients of $\{\bV_{n,i}\}_{i \in N_n}$.
\end{definition}
Coefficient $\beta_{n,s}$ captures network dependence between units
that are at network distance greater than or equal to $s$ in network $\cG_n$
by the covariance of nonlinearly transformed variables. Thus, a
sequence of coefficients $\beta_n = \{\beta_{n,s}
\}_{s=1}^{\infty}$ captures how fast network dependence between units
decays as network distance $s$ increases. Assumption~\ref{net1},
which we will introduce next, 
restricts the rate by which this network dependence decays.
Importantly, the $\psi-$network dependence permits dependence between units $i$
and $j$ that are only indirectly connected in the network, and thus, any two units can be dependent as long as there is a network path
between them. This is in contrast to two other popular approaches; (1) 
dependency graphs, which can allow units to be dependent only when
they are adjacent in a given network, and (2) Markov random fields, which
impose conditional independence restrictions based on the network
structure (e.g., a given unit's observed data are independent of data observed for all units in the network  conditional on observed data for its first-order network peers). 

Using the notion of $\psi-$network dependence, we make the following
assumptions on the observed data distribution that permit network
dependent error, but still allow for making inferences about the ACPE.
\begin{assumption}
  \label{net1}
  The triangular arrays $\{m(\bL_{n,i}; \theta)\}_{i \in N_n, n \geq 1}$ and $\left\{\frac{\partial}{\partial
      \theta} m(\bL_{n,i}; \theta)\right\}_{i \in N_n, n \geq 1}$ are
  conditionally $\psi$-weakly dependent given $\cG_n$, respectively, for all $\theta
  \in \Theta$ with the weak dependent coefficients $\beta_n$ that
  satisfy the following conditions. $\sup_{n \geq 1} \max_{s \geq 1}
  \beta_{n,s} < \infty$, and for some constant $\lambda$, $\psi_{q_1,q_2}(f_1,f_2)
  \leq \lambda \times q_1q_2 (||f_1||_{\infty} + \mbox{\normalfont
    Lip}(f_1))
  (||f_2||_{\infty} + \mbox{\normalfont
    Lip}(f_2)).$\footnote{$\mbox{\normalfont
    Lip}(f)$ represents Lipschitz constant of $f$, and $||\cdot||_{\infty}$ denotes the sup norm,
  i.e., $||f||_{\infty} = \sup |f(x)|.$} There exist $p > 4$ and a sequence $\xi_n \rightarrow \infty$ such
  that
  \begin{itemize}
  \item[] \textnormal{1.} $\beta_{n, \xi_n}^{(p-1)/p} = o_{a.s.}(n^{-3/2})$, 
  \item[] \textnormal{2.} for each $k \in \{1, 2\},$
    \begin{equation*}
      \cfrac{1}{n^{k/2}} \sum_{s \geq 0} r_n(s, \xi_n; k) \beta_{n,s}^{1
        - (k+2)/p} = o_{a.s.}(1)
    \end{equation*}
    where we define the neighborhood shell: 
    \begin{equation}
      r_n (s, \xi_n; k) = \inf_{\alpha > 1} \biggl\{\frac{1}{n} \sum_{i \in N_n} \max_{j \in
        \cN_n(i; s)} |\tcN_n(i; \xi_n)
      \setminus \tcN_n(j;
      s-1)|^{k\alpha}\biggr\}^{\frac{1}{\alpha}}
      \times \biggl\{ \frac{1}{n} \sum_{i \in
        N_n} |\cN_n(i;s)|^{\frac{\alpha}{\alpha-1}}\biggr\}^{1- \frac{1}{\alpha}},
    \end{equation}
    and the within-$s$ peers $\tcN_n(i;s) = \{j \in N_n: d_n(i, j)
    \leq s\}.$ We use $|\cdot|$ to denote the cardinality of a set.
  \end{itemize}  
\end{assumption}
This is an adaptation of Condition ND in \cite{kojevnikov2020} to our
setup. Assumption~\ref{net1}.1 restricts the speed by which weak dependent
coefficients $\beta_{n, s}$ decay as network distance $s$
increases. Assumption~\ref{net1}.2 restricts the speed by which the density of
the network changes as sample size increases. When network
dependence $\beta_{n, s}$ decays faster with network distance $s$,
it can accommodate denser networks. See \cite{kojevnikov2020} for
further discussion on these conditions.

We are now ready to state asymptotic properties of the proposed DNC estimator. 
\begin{theorem}
  \label{est-net}
  Under the conditions given in Theorem~\ref{iden-net}, Assumption~\ref{net1} and standard GMM
  regularity conditions,\footnote{In the supplementary material, we provide the regularity conditions
    widely used in the GMM framework \citep{hansen1982gmm,
      newey1994}.} as $n$ goes to infinity, the DNC estimator is consistent:
  \begin{equation*}
    \widehat{\tau}(a, a^\prime)  \xrightarrow{p} \tau(a, a^\prime),
  \end{equation*}
  and asymptotically normal: 
  \begin{equation*}
    \cfrac{\widehat{\tau}(a, a^\prime) - \tau(a, a^\prime)}{\sqrt{\sigma^2/n}} \xrightarrow{d}
    {\normalfont \textsf{Normal}}(0,1).
  \end{equation*}
  The asymptotic variance $\sigma^2$ is the $(1,1)$ th element of
  matrix $\Sigma$ where 
  \begin{eqnarray*}
    && \Sigma  = \Gamma_{0}
    \Lambda_0 \Gamma_{0}^\top, \hspace{0.1in} \Lambda_0 = \Var\left(\frac{1}{\sqrt{n}} \sum_{i=1}^n m(\bL_{n,i};
       \theta_{0}) \right)\\
&& \Gamma_{0} = (M_{0}^\top \Omega M_{0})^{-1} M_{0}^\top
\Omega, \hspace{0.1in} M_{0} = \frac{1}{n} \sum_{i=1}^n \E \left\{\frac{\partial}{\partial \theta}  m (\bL_{n,i};
   \theta_{0}) \right\},
  \end{eqnarray*}
  and we define $\theta_{0}$ to be the true parameter such that, for
  all units $i \in N_n$, $\E \left\{m(\bL_{n,i}; \theta) \right\} = 0$
  only when $\theta = \theta_{0}$.
\end{theorem}
We provide a proof in Section~\ref{appsec:asym} of the supplementary material.

To estimate the standard error of the DNC estimator, the key is to
estimate $\Lambda_0.$ To account for network-dependent errors, we rely
on the network HAC variance estimator \citep{kojevnikov2020} adapted
to our setting:  
\begin{equation}
  \widehat{\Lambda} = \sum_{s \geq 0} \omega(s/b_n) \left\{\frac{1}{n}
    \sum_{i \in N_n} \sum_{j \in \cN_n(i;s)} m (\bL_{n,i};
    \widehat{\theta}) m (\bL_{n,j}; \widehat{\theta})^\top \right\}, \label{eq:hac0}
\end{equation}
where a kernel function $\omega(\cdot)$ is defined as follows:
$\omega: \overline{\mathbb{R}} \rightarrow [-1, 1]$ such that
$\omega(0) = 1, \omega(c) = 0$ for $|c| \geq 1$ and $\omega(c) =
\omega(-c)$ for all $c \in \overline{\mathbb{R}}$. Examples include the truncated,
Parzen, and Tukey–Hanning kernels. $b_n$ denotes a bandwidth of the
network HAC variance estimator. This bandwidth
determines how far $\widehat{\Lambda}$ takes
into account network dependence; kernel weight $\omega(s/b_n) > 0$ for $s < b_n$ and
$\omega(s/b_n) = 0$ for $s \geq b_n.$

The variance estimator of the DNC estimator can be computed as
the $(1,1)$ th element of matrix $\widehat{\Sigma}$ defined as 
\begin{equation}
  \widehat{\Sigma} = \widehat{\Gamma} \widehat{\Lambda}
  \widehat{\Gamma}^\top \label{eq:hac}
\end{equation}
where $\widehat{\Gamma} = (\widehat{M}^\top
\Omega \widehat{M})^{-1} \widehat{M}^\top \Omega$, and
$\widehat{M} = \frac{1}{n} \sum_{i=1}^n
\frac{\partial}{\partial \theta}  m (\bL_{n,i}; \widehat{\theta})$.

We now establish consistency of this network HAC variance
estimator. The following result formally restricts the speed by which
bandwidth $b_n$ should increase as sample size increases. When
network dependence $\beta_{n,s}$ decays slower and the average number
of network peers at distance $s$ increases faster with network
distance, bandwidth $b_n$ should increase faster as sample size increases.
\begin{theorem}
  \label{hac}
  Under the conditions given in Theorem~\ref{est-net}, suppose the choice
  of bandwidth and kernel satisfies the following condition with $p$ that
  satisfies Assumption~\ref{net1}.  
  \begin{equation}
    \lim_{n \rightarrow \infty} \sum_{s \geq 0}
    |\omega(s/b_n) - 1 | \rho_n(s) \beta_{n,s}^{1 -
      2/p} = 0 \ \ \mbox{a.s.,}  \label{eq:reg-band}
  \end{equation}
  where $\rho_n(s)$ is the average number of network peers at
  the distance $s$, $\rho_n(s) =
  \frac{1}{n} \sum_{i =1}^n |\cN_n(i; s)|$. Then, $\widehat{\Sigma} \xrightarrow{p} \Sigma.$
\end{theorem}
We provide a proof in Section~\ref{appsec:asym} of the supplementary material.

In practice, we recommend using the default choice of bandwidth $b_n$
provided  by \cite{kojevnikov2020}, i.e.,
\begin{equation}
  b_n = \mbox{constant} \times \cfrac{\log(n)}{\log\{\max(\mbox{average degree}, 1.05)\}}. \label{eq:band}
\end{equation}
In our simulation studies (Section~\ref{sec:sim}), we set the constant
in equation~\eqref{eq:band} to $1.0$ and find this default choice performs well across various settings. 

Finally, under conditions given in Theorem~\ref{hac}, we obtain an asymptotically valid $(1-\alpha)$
confidence interval for $\tau(a, a^\prime)$ by
$[\widehat{\tau}(a, a^\prime) - \Phi(1-\alpha/2) \times 
\widehat{\sigma}/\sqrt{n}, \ \widehat{\tau}(a, a^\prime) +
\Phi(1-\alpha/2) \times \widehat{\sigma}/\sqrt{n}]$ where
$\widehat{\sigma}$ is the $(1,1)$th element of matrix
$\widehat{\Sigma}$ defined in equation~\eqref{eq:hac}. 

\subsection{Linear Double Negative Control Estimator}
\label{subsec:l-dnc}
Here, we discuss an important special case under a linear
specification for the confounding bridge function, which admits a closed form solution. Suppose we assume a linear confounding bridge:
\begin{equation*}
  h(W_{i}, A_i, \bX_i; \gamma) = \gamma_\alpha + \gamma_A A_i
  + \gamma_W W_{i} + \gamma_X^\top \bX_i.
\end{equation*}
Under this linear model, $\tau(a, a^\prime) = \gamma_A \times (a -
a^\prime).$ We can estimate coefficients $\gamma = (\gamma_\alpha,
\gamma_A, \gamma_W, \gamma_X^\top)^\top$ by fitting the linear GMM
estimator:
\begin{equation}
  \widehat{\gamma} = (\bV_W^\top \bV_Z \Omega \bV_Z^\top
  \bV_W)^{-1}\bV_W^\top \bV_Z \Omega \bV_Z^\top \bY \label{eq:lin-gmm}
\end{equation}
where $\bV_W$ is a matrix with $n$ rows with $i$th row $\bV_{iW} =
(1, A_i, W_{i}, \bX_i^\top)^\top$, $\bV_Z$ is a matrix with $n$ rows with
$i$th row $\bV_{iZ} = (1, A_i, \bZ_{i}^\top, \bX^\top_i)^\top,$ and
$\bY$ is a $n$-dimensional vector with $i$th element equal to
$Y_{i2}.$

To account for network dependence of samples, we adopt the
network HAC variance estimator in Section~\ref{subsec:est-net}. The key is to
estimate 
\begin{equation}
  \Lambda ^{\textsf{lin}}_0 = \Var\left(\frac{1}{\sqrt{n}}
    \sum_{i=1}^n  e_{i2} \bV_{iZ}\right),  \ \ \ \mbox{and} \ \ \ e_{i2} = Y_{i2} -
  \gamma^\top \bV_{iW}.
\end{equation}
Using the network HAC variance estimator, we can estimate the
variance of $\widehat{\gamma}$ as 
\begin{equation*}
  \widehat{\Var}(\widehat{\gamma}) = n (\bV_W^\top \bV_Z \Omega \bV_Z^\top
  \bV_W)^{-1}\bV_W^\top \bV_Z \Omega  \widehat{\Lambda}^{\textsf{lin}}
  \Omega \bV_Z^\top \bV_W (\bV_W^\top \bV_Z \Omega \bV_Z^\top \bV_W)^{-1},
\end{equation*}
where 
\begin{equation*}
  \widehat{\Lambda}^{\textsf{lin}} = \sum_{s \geq 0}
  \omega(s/b_n) \left\{\frac{1}{n} \sum_{i \in N_n} \sum_{j \in \cN_n(i;s)}
    \widehat{e}_{i2}\widehat{e}_{j2} \bV_{iZ} \bV_{jZ}^\top
  \right\}, \ \ \ \mbox{and} \ \ \ \widehat{e}_{i2} = Y_{i2} -
  \widehat{\gamma}^\top \bV_{iW}.
\end{equation*}
Researchers can use the two-step GMM to minimize the asymptotic
variance within this class. In the first step, we set $\Omega =
(\bV_Z^\top \bV_Z)^{-1}$, and compute a preliminary GMM estimate
$\widehat{\gamma}_{(1)}$. Importantly, one may estimate
$\widehat{\gamma}_{(1)}$ using any off-the-shelf software package for two-stage
least squares, such as \texttt{ivreg} in \texttt{R}, by viewing $W_i$
as the endogenous treatment, $\bZ_i$ as the instrument, and $(A_i,
\bX_i^\top)$ as covariates \citep[see][]{tchetgen2020proximal}. Here we use the two-stage least
squares as a convenient way to compute this preliminary GMM estimate,
and thus, we do not make
any assumptions required for standard instrumental variable analysis. See
\cite{miao2018confounding} for relationships between the double
negative control approach and the instrumental variable approach in
general. In the second step, we compute
$\widehat{\Lambda}^{\textsf{lin}}$ based on $\widehat{\gamma}_{(1)}$,
which is denoted by
$\widehat{\Lambda}^{\textsf{lin}}_{(1)}.$ Then, we obtain the final
estimator by solving equation~\eqref{eq:lin-gmm} with $\Omega = (\widehat{\Lambda}^{\textsf{lin}}_{(1)})^{-1}.$ The resulting estimator
$\widehat{\gamma}$ is consistent and asymptotically normal, and is
asymptotically efficient within the GMM class under the
conditions given in Theorems~\ref{est-net} and~\ref{hac}. The variance also simplifies to
\begin{equation*}
  \widehat{\Var}(\widehat{\gamma}) = n (\bV_W^\top \bV_Z (\widehat{\Lambda}^{\textsf{lin}})^{-1} 
  \bV_Z^\top \bV_W)^{-1}. 
\end{equation*}

\subsubsection*{Choice of Bandwidth}
In general settings of network-dependent errors
(Section~\ref{subsec:est-net}), one must estimate a bandwidth $b_n$
for the network HAC variance estimator \citep[e.g.,][]{kojevnikov2020} because how far network dependence
persists is a priori unknown. However, when the following assumption holds, we can analytically select the bandwidth.
\begin{assumption}\label{ana-var}
  \ 
  \begin{itemize}
  \item[\textnormal{1.}] The ACPE is equal to a linear function of
    parameters $\gamma$ in the confounding bridge function. 
  \item[\textnormal{2.}] There exists integer $s^\ast$ such that for units $i,
    j$ with distance $d_n(i,j) \geq s^\ast$,
    \begin{equation*}
      \bL_{n,j} \ \indep \ \bL_{n,i} \mid A_i, \bZ_i, \bX_i, U_i.
    \end{equation*}
  \end{itemize}
\end{assumption}
Assumption~\ref{ana-var}.1 holds for a linear confounding bridge function as we
consider in this section. Assumption~\ref{ana-var}.2 requires that observed data for unit $j$,
$\bL_{n,j}$, is conditionally independent of observed data for unit $i$,
$\bL_{n,i}$, given unit $i$'s treatment, NCEs, observed pre-treatment
covariates, and the unmeasured confounder. This
conditional independence is required only upon conditioning on latent
confounder $U_i$, and thus, it does not restrict network dependence of
the observed data law itself. 

Importantly, we emphasize that Assumption~\ref{ana-var}.2 holds under
many relevant scenarios. Figure~\ref{fig:dag-net-genNC} provides
examples of causal graphs where Assumption~\ref{ana-var}.2 is
satisfied. In Figure~\ref{fig:dag-net-genNC}.(i), suppose one uses
$C_i$ as the NCO and $\bZ_i = \{C_{j}: j \in \cN_n(i; 1)\}$ as the
NCEs. Then, Assumption~\ref{ana-var}.2 holds with $s^\ast = 2$. If one
uses auxiliary variables of both peers and peers-of-peers,
$\bZ_i = \{C_{j}: j \in \{\cN_n(i; 1), \cN_n(i; 2)\}\}$,
Assumption~\ref{ana-var}.2 holds with $s^\ast = 3.$
Figure~\ref{fig:dag-net-genNC}.(ii) represents another example. Suppose
one exploits $Y_{i1}$ as the NCO and $\bZ_i = \{Y_{jt}: j \in \cN_n(i;2), \ t \in \{1, 2\}\}$ as the
NCEs. Then, Assumption~\ref{ana-var}.2 holds with $s^\ast = 4$.

Under Assumption~\ref{ana-var}, Lemma~\ref{hac-choice} below shows
that one can analytically select the bandwidth for the network HAC variance estimator. 
\begin{lemma}
  \label{hac-choice}
  Suppose the conditions given in Theorem~\ref{est-net} hold. Under
  Assumption~\ref{ana-var}.1, we can simplify the moment function to
   $\tm(\bL_{n,i}; \gamma) =  \{Y_{i2} - h(W_{i}, A_i, \bX_i;
  \gamma)\} \times \eta(A_i, \bZ_i, \bX_i)$ as our target parameter is a
  linear function of $\gamma.$ Then, under Assumption~\ref{ana-var}.2 with integer $s^\ast$,
  we can use the following network HAC variance estimator for
  $\widehat{\gamma}$, which is the GMM estimator
  with moment function $\tm (\bL_{n,i}; \gamma)$.
\begin{equation}
  \widehat{\Var}(\widehat{\gamma}) = \frac{1}{n} \widehat{\Gamma}_{\gamma} \widehat{\Lambda}_{s^\ast}
  \widehat{\Gamma}_{\gamma}^\top \label{eq:hac-2}
\end{equation}
where 
\begin{equation*}
  \widehat{\Lambda}_{s^\ast} = \sum_{s = 0}^{s^\ast - 1} \omega(s/b_n) \left\{\frac{1}{n}
    \sum_{i \in N_n} \sum_{j \in \cN_n(i;s)} \tm (\bL_{n,i};
    \widehat{\gamma}) \tm (\bL_{n,j}; \widehat{\gamma})^\top \right\},
\end{equation*}
$\widehat{\Gamma}_{\gamma} = (\widehat{M}_{\gamma}^\top
\Omega \widehat{M}_{\gamma})^{-1} \widehat{M}_{\gamma}^\top \Omega$, and
$\widehat{M}_{\gamma} = \frac{1}{n} \sum_{i=1}^n
\frac{\partial}{\partial \gamma} \tm (\bL_{n,i}; \widehat{\gamma})$.
\end{lemma}
The key to this result is that, to compute the variance of a sum of products of moments, one only needs to
consider moments of units with distance less than $s^\ast$ (i.e., we
added $s \in \{0, 1, \ldots, s^\ast - 1\}$), as the remaining contributions are null. This is in contrast to
the default network HAC variance estimator (equation~\eqref{eq:hac0})
where we have to incorporate all products of moments of units with distance less than $b_n$, which is
in general larger than $s^\ast$. We provide a proof in Section~\ref{appsec:asym} of the supplementary material.

For the linear DNC estimator, Assumption~\ref{ana-var}.1
automatically holds, and thus, as long as Assumption~\ref{ana-var}.2
holds, we can rely on this analytical choice of bandwidth. We
evaluate both analytical and default bandwidth
selections (equation~\eqref{eq:band}) in a simulation study (Section~\ref{sec:sim}).

\section{Simulation Study}
\label{sec:sim}

We investigate the finite sample performance of the proposed DNC
estimator of the ACPE using networks of varying density and size. In
the supplementary material, we also examine the performance of the
proposed estimator in settings where key identification assumptions
are violated; in Section~\ref{subsec:vio-nc}, we consider violation of
the negative control assumption (Assumption~\ref{net-assum}.2), and
in Section~\ref{subsec:vio-comp}, we examine violation of the outcome
confounding bridge assumption (Assumption~\ref{net-assum}.3) due to violation of the underlying completeness condition. 

\paragraph{Setup.}
To investigate the performance of the proposed estimator, we consider
two different types of networks: the small world network and the real-world
network from Add Health data. To generate the small world
network, we use \texttt{sample\_smallworld} with the rewiring
probability of $0.15$ based on \textsf{R} package \texttt{igraph}. We consider
two levels of densities: low (the average degree of four) and high
(the average degree of eight). Add Health project collected
detailed information about friendship networks by an in-school
survey. We define friendships as symmetric relationships: the pair of
students $i$ and $j$ in the same school are coded as friends if either
$i$ lists $j$ as a friend, or $j$ lists $i$ as a friend, or
both. While we analyze this data more thoroughly in
Section~\ref{sec:app}, we also use it here as basis for the simulation. 
For each simulation, we generate a network of size $n$ where we consider sample size $n \in \{500, 1000,
2000, 4000\}.$ For the small-world network, we generate a single
network of size $n$. For the Add Health network, we retain the original
network characteristics by randomly sampling schools with probability
proportional to its size until the total sample size reaches $n$. The
average degree of the Add Health network ranges from $3.82$ to $5.95$,
and its average number of the peers-of-peers ranges from $20.85$
to $33.24$, both of which are in the middle of the low-density
small-world network (average degree = 4) and the high-density small
world network (average degree = 8). The density of the Add Health
network ranges from $0.15$ to $0.77$ \%, which are close to the
density of the low-density small-world network. Thus, these three different types of networks jointly cover a wide
range of network density and size. See Table~\ref{tab:sim} for more details.

Given a network, we simulate data with the following data-generating mechanism: For units $i = 1, \ldots, n,$
\begin{itemize}
\item[(1)] Unobserved confounder with network dependence: 
  $U_i = \sum_{s \geq 0} \zeta^s \sum_{j \in \cN(i; s)} \widetilde{U}_{j}/|\cN(i; s)|$ where
  $\zeta = 0.8$ and $\widetilde{U}_{j} \iid \textsf{Normal} (0,
  1)$. This data generating process for a network-dependent variable
  follows a simulation setup of \cite{kojevnikov2020}. 
\item[(2)] Observed covariates with network dependence:
  $\bX_i = (X_{i1}, X_{i2}, X_{i3})$ where, for $k \in \{1, 2, 3\}$, $X_{ik} = \sum_{s \geq 0}
  \zeta^s \sum_{j \in \cN(i; s)} \widetilde{X}_{jk}/|\cN(i; s)|$,
  $\zeta = 0.8$, and $\widetilde{X}_{jk} \iid \textsf{Normal}(0,
  1)$.
\item[(3)] Observed auxiliary variable:  $C_i = U_i + \beta_c^\top \bX_i + \epsilon_{i0}$ where  $\epsilon_{i0} \iid
  \textsf{Normal} (0, 1)$ and $\beta_c = (0.05, 0.05, 0.05)$.
\item[(4)] Focal behavior at the baseline: $Y_{i1} = U_i + 0.05 C_i
  + \beta_1^\top \bX_i + \epsilon_{i1}$ where  $\epsilon_{i1} \iid
  \textsf{Normal} (0, 1)$ and $\beta_1 = (-1, -1, -1)$.
\item[(5)] Focal behavior at the follow-up: $Y_{i2} = \tau A_i +
  0.2 Y_{i1} + 3 U_i + 0.05 C_i + \beta_2^\top \bX_i + \epsilon_{i2}$ where  $\epsilon_{i2} \iid
  \textsf{Normal} (0, 1)$, and $\beta_2 = (-1, -1, -1)$. The treatment variable $A_i$ is defined as
  $A_i = \sum_{j \in \cN(i; 1)} Y_{j1} /|\cN(i; 1)|.$
\end{itemize}
The above models imply that the ACPE is $\tau$, which we set to be
$0.3$. We can use $W_i = C_i$ as the NCO, and $Z_i = \sum_{j \in \cN(i; 1)} C_{j}
/|\cN(i; 1)|$ as the NCE. Under this setup, the linear confounding bridge function, $h(W_{i}, A_i, \bX_i; \gamma) = \gamma_\alpha + \tau A_i
+ \gamma_W W_{i} + \gamma_X ^\top \bX_i$, satisfies Assumption~\ref{net-assum}. 

We evaluate the performance of the proposed DNC estimator and the
network HAC variance estimator. We evaluate two choices of
bandwidth for the network HAC variance estimator. First, we use
the bandwidth of $2$, which we analytically derive based on
Lemma~\ref{hac-choice}. The required Assumption~\ref{ana-var}
holds in this simulation design. Second, we also use the default bandwidth
$b_n$ (equation~\eqref{eq:band}) suggested in
\cite{kojevnikov2020}. We use the
Parzen kernel function.\footnote{$\omega(x) = 1 - 6x^2 + 6|x|^3$ if $0
  \leq |x| \leq 1/2$, $\omega(x) = 2(1 - |x|)^3$ if $1/2 <
  |x| \leq 1$, and $\omega(x) = 0$ if $1 < |x|$.}

For reference, we also report the ordinary
least squares estimator where we regress $Y_{i2}$ on the treatment
variable $A_i$ and a set of observed variables $(Y_{i1}, C_i, X_{i1},
X_{i2}, X_{i3}).$ This estimator is consistent only under conditional ignorability, which is violated due to unmeasured network
confounder $U_i$ under this simulation setup. Thus, this OLS estimator
quantifies the amount of network confounding that the DNC estimator
has to correct for.

\begin{table}[!t]
\centering
\scalebox{0.7}{
    \begin{tabular}{lcccc|ccccc|ccc}
      \hline
 \multicolumn{5}{c|}{\textbf{Simulation Design}} & \multicolumn{5}{c|}{\textbf{DNC}} &
                                                                          \multicolumn{3}{c}{\textbf{OLS}} \\
      \hline
      \multirow{2}{*}{Network} & Sample & Average & Average &
                                                              \multirow{2}{*}{Density} & 
                                                        \multirow{2}{*}{Bias}
                                                              &
                                                                Standard
                                                                     &
                                                                       \multirow{2}{*}{RMSE}
                       &
                         Coverage & Coverage
                                        & \multirow{2}{*}{Bias} & Standard & \multirow{2}{*}{RMSE}\\[-3pt]
                  & Size & $|\cN_n(i;1)|$ & $|\cN_n(i;2)|$ & & & Error &
                       & (Analytical) & (Default) & & Error  & \\[3pt]
  \hline
      SW-4 & 500 & 4.00 & 10.05 & 0.80 & 0.14 & 0.63 & 0.65 & 0.96 & 0.96 & 0.98 & 0.26 & 1.02 \\ 
                                                 & 1000 & 4.00 & 10.10 & 0.40 & 0.07 & 0.42 & 0.42 & 0.95 & 0.95 & 1.00 & 0.18 & 1.01 \\ 
                                                 & 2000 & 4.00 & 10.13 & 0.20 & 0.03 & 0.28 & 0.28 & 0.95 & 0.94 & 1.01 & 0.13 & 1.01 \\ 
                                                 & 4000 & 4.00 & 10.14 & 0.10 & 0.02 & 0.20 & 0.20 & 0.95 & 0.94 & 1.00 & 0.09 & 1.01 \\[10pt]   
      SW-8 & 500 & 8.00 & 35.93 & 1.60 & 0.25 & 1.23 & 1.26 & 0.96 & 0.96 & 0.88 & 0.35 & 0.95 \\ 
                                                 & 1000 & 8.00 & 36.64 & 0.80 & 0.11 & 0.58 & 0.59 & 0.96 & 0.96 & 0.88 & 0.25 & 0.91 \\ 
                                                 & 2000 & 8.00 & 37.01 & 0.40 & 0.05 & 0.38 & 0.39 & 0.94 & 0.94 & 0.90 & 0.17 & 0.92 \\ 
                                                 & 4000 & 8.00 & 37.17 & 0.20 & 0.02 & 0.26 & 0.26 & 0.95 & 0.95 & 0.89 & 0.12 & 0.90 \\[10pt]   
      Add Health & 500 & 3.82 & 20.85 & 0.77 & 0.15 & 0.72 & 0.74 & 0.96 & 0.87 & 0.98 & 0.29 & 1.02 \\ 
                                                 & 1000 & 4.80 & 26.72 & 0.48 & 0.06 & 0.46 & 0.46 & 0.95 & 0.94 & 0.95 & 0.20 & 0.97 \\ 
                                                 & 2000 & 5.69 & 31.88 & 0.28 & 0.03 & 0.31 & 0.31 & 0.95 & 0.95 & 0.93 & 0.15 & 0.94 \\ 
                                                 & 4000 & 5.95 & 33.24 & 0.15 & 0.02 & 0.22 & 0.22 & 0.94 & 0.94 & 0.92 & 0.10 & 0.92 \\ 
   \hline
    \end{tabular}}
  \caption{Operating Characteristics of Estimators under Different Networks.}\label{tab:sim}
  \vspace{-0.1in}
  \begin{flushleft}
    \spacingset{1}{{\footnotesize \noindent \textit{Note}: We consider
        three different networks; the small
        world network model with the average degree of four (SW-4) and
        eight (SW-8),
        and the Add Health network. We report the average degree, the average number of the
        peers-of-peers, and the density for each
        network with each sample size. For the DNC estimator, we
        report the absolute mean bias, the standard error, the RMSE, and
        coverage of the 95\% confidence intervals based on the
        analytical bandwidth and the default bandwidth. For
        reference, we also report the absolute mean bias, the standard error,
        and the RMSE of the OLS estimator. The absolute mean bias, the standard error,
        and the RMSE for both estimators are standardized by the true ACPE. 
      }}
  \end{flushleft}
\end{table}

\paragraph{Results.}
We generate 2000 simulations and evaluate estimators in terms of 
absolute mean bias, standard error (computed as standard
deviation of point estimates across simulations), root mean squared error (RMSE), and coverage of
95\% confidence intervals based on the network HAC variance
estimator. We standardize the first three quantities by the true
ACPE to ease interpretation. Table~\ref{tab:sim} summarizes the results of the
simulation study. The performance of the OLS estimator for coverage is not
shown because all are close to zero due to substantial unmeasured confounding. 

Our proposed DNC estimator remained stable with relatively small bias
across all scenarios, and the bias reduced as sample size
increased. As expected, standard errors of the proposed DNC estimators
were larger than the biased OLS estimators, but the RMSE of the DNC
estimator was smaller due to smaller bias. Compared to the
low-density small-world network (SW-4), bias, standard errors,
and RMSE were larger in the high-density small-world network
(SW-8). Results for the Add Health network fell somewhere in between. The
coverage of 95\% confidence intervals
was close to the nominal level when the analytical bandwidth was 
chosen. While coverage with default bandwidth tended to under-cover slightly at smaller sample
sizes in the Add Health network structure, it improved as sample size
increased. They indicated that our proposed standard error estimation
provided valid inference. These results confirmed our theoretical results in
finite sample and demonstrated the advantages of the proposed DNC estimator.  

\section{Empirical Application: Causal Peer Effects in Education}
\label{sec:app}
We apply our method to Add Health data to evaluate causal peer
effects of education outcomes in a friendship network. The study
of causal peer effects in education has a long history in the social
sciences, and many studies have shown moderate positive peer
effects \citep[e.g.,][]{epple2011peer, sacerdote2011peer}.
While some papers have used experimental or quasi-experimental methods where classmates or
roommates are randomly assigned by schools, the vast majority
of existing evidence comes from observational studies of causal peer
effects. In the absence of randomization, researchers have adjusted
for a variety of observed pre-treatment covariates and a host of fixed
effects (e.g., fixed effects for schools or network components). However,
such approaches rely on a conditional ignorability assumption, and assume away unmeasured latent homophily or contextual
confounding. For example, students who have higher education
performance might become friends with other high-performing students
due to unobserved characteristics. In this case, strong association
between one's education outcome and her friends' outcomes cannot be
interpreted as the causal peer effect. Potential bias due to such
unmeasured network confounding can undermine the validity of
causal conclusions. To explore the possibility of unmeasured network
confounding, we use the proposed double negative control approach to
identify and estimate the ACPE. 

\subsection{Data}
Add Health project collected
survey data from students in grades 7–12 from a nationally
representative sample of over 100 private and public schools in years
1994–1995 in the United States. There were two types of surveys conducted in years
1994-1995, both of which provide information about social
and demographic characteristics of respondents, education
level and occupation of their parents, and their friendship links
(i.e., their best friends, up to five females and up to five males). The first type was an in-school survey, which were administered
to students in schools from September 1994 until April 1995. It
includes over $90,000$ students across about 140 schools. Each school administration occurred on a single day within
one 45- to 60-minute class period. The second type was an in-home
survey, in which students answered a 90-minute in-home interview. It
includes over $20,000$ students. The in-home interview was conducted at least 90 days after
the in-school survey, except for 10 students who we exclude from
analysis. Our analysis focuses on $10,264$ students who
completed both in-school and in-home surveys and answered questions
related to variables we use below. The in-school survey serves as the
baseline time period and the in-home survey as the follow-up
period. 

We examine a network based on the friendship information collected in the in-school
survey. We define friendships as symmetric relationships: the pair of
students $i$ and $j$ in the same school are coded as friends if either $i$ lists $j$ as a
friend, or $j$ lists $i$ as a friend, or both. The average degree of the friendship network is $6.20$.

\subsection{Setup}
We use the grade-point average (GPA) at follow-up period $Y_{i2}$
as the outcome variable. GPA ranges between 1 and 4, and is
computed based on the average grade-point of four subjects; English,
Mathematics, History/Social Studies, and Science. The treatment
variable is the average GPA of the network peers
at baseline $A_i = \sum_{j \in \cN(i;1)} Y_{j1}/|\cN(i;
1)|$ where $|\cN(i; s)|$ denotes the number of the $s$-th order network peers of unit $i$. 

Following Section~\ref{subsubsec:select-nc}, we consider two sets of negative
controls. First, we consider focal behaviors of peers and those
measured at baseline as negative controls. In particular, we use the GPA at baseline
$Y_{i1}$ as the NCO, i.e., $W_i = Y_{i1}$. We then use the average GPA
of peers-of-peers at baseline as the NCE, i.e., $Z_{i} = \sum_{j \in \cN(i;
  2)} Y_{j1}/|\cN(i; 2)|$. These are valid negative controls
when the GPAs of students are recorded concurrently
within schools. Second, we also use an auxiliary variable for
negative controls. To make the assumption
about the confounding bridge function (Assumption~\ref{net-assum}.3)
more plausible, we again use GPA at baseline as the
NCO, i.e., $W_i = Y_{i1}$, while we use level of peers' headaches as the
NCE. Specifically, we use the average level of headaches of peers as NCE, i.e., $\widetilde{Z}_{i} = \sum_{j \in
  \cN(i; 1)} C_{j1}/|\cN(i; 1)|$ where $C_{j1}$ captures level
of headaches by student $j$ measured at baseline. According to
\cite{cohen2008}, whether a student has headaches is a plausible
negative control because it is unlikely (a) to causally affect whether
students are friends to each other and (b) to causally affect headaches of other units.  

In accordance with prior analyses of these data in the literature, we also include a series of
observed pre-treatment covariates $\bX_i$, including fixed effects for
each school, the number of friends, age, gender, race, born in the US, health status, physical fitness, school attendance, student
grade, motivation in education, school attachment, homework, self
esteem, household size, living with mother, living with father, parental care, mother's
education, father's education, whether both parents work, relationship
with teachers, and relationship with friends.

Following Section~\ref{subsec:l-dnc}, we use a linear DNC
estimator. In particular, we specify a linear confounding bridge as,
\begin{equation*}
  h(W_{i}, A_i, \bX_i; \gamma) = \gamma_\alpha + \gamma_A A_i +
  \gamma_W W_{i} + \gamma_X^\top \bX_i,
\end{equation*}
where $(A_i,  W_{i})$ are the treatment variable and the NCO, and
$\bX_i$ represent observed pre-treatment covariates
defined above, including fixed effects for schools. 

Our causal estimand is $\gamma_A,$ which captures the causal effect on
a given student's GPA induced by one point increase in the average GPA of her peers. We use the proposed network HAC variance
estimator to compute standard errors and 95\% confidence intervals. We
report standard errors based on the analytical choice of the bandwidth
described in Section~\ref{subsec:l-dnc}, while results for the default bandwidth selection are similar and therefore not reported. 

We compare our proposed DNC estimator to the OLS regression estimator, which relies on the
assumption of conditional ignorability where we adjust for
pre-treatment outcome $Y_{i1}$ and $\bX_i$. In absence of unmeasured network confounding, we expect
DNC estimates and OLS estimates to be comparable (i.e. within sampling variability). For the OLS estimator, we apply the network HAC variance estimator
\citep{kojevnikov2020} to residuals in order to make the comparison
clear. We note that the estimated standard errors for the OLS are valid only when
there is no unmeasured network confounding. 

\begin{table}[!t]
  \centering
  \begin{tabular}{lcccc}
    \Xhline{1.5pt}
    \textbf{Method} & \textbf{Estimate} & \textbf{Stand. Error} &
                                                                  \textbf{p-value} & \textbf{95\% CI}\\ 
    \hline\hline
    OLS under & \multirow{2}{*}{0.176} & \multirow{2}{*}{0.015}
                                                                & \multirow{2}{*}{0.000} &
                                                                                           \multirow{2}{*}{(0.147,
                                                                                           0.206)}\\[-3pt] 
    \ \ conditional ignorability & & & & \\[3pt]
    \hline
    DNC Estimator with NCE $Z_i$: & 0.033 & 0.049 & 0.497 & (-0.063, 0.129) \\[-2pt] 
    \ \  the average GPA of $\cN(i;2)$ & & & &\\[3pt]
    DNC Estimator with NCE $\widetilde{Z}_i$: & 0.078 & 0.183 & 0.668 & (-0.280, 0.437) \\[-2pt] 
    \ \  the average level of headaches of $\cN(i;1)$ & & & &\\[3pt]
    \Xhline{1.5pt}
  \end{tabular}
  \caption{Estimated Average Causal Peer Effects on
    GPA.}\label{tab:res}
  \vspace{-0.1in}
  \begin{flushleft}
    \spacingset{1}{{\footnotesize \noindent \textit{Note}: We report point estimates, standard errors,
        p-values, and 95\% confidence intervals for each method.}}
  \end{flushleft}
\end{table}

\subsection{Results}
Table~\ref{tab:res} reports estimates of the ACPE, standard
errors, p-values, and 95\% confidence intervals for each method. The
OLS estimate suggests that the estimated ACPE is as large as $0.176$
and statistically significant. However, our proposed DNC estimator
indicates that there may be a large amount of unmeasured network confounding operating in this network which cannot be accounted for by a standard regression analysis, despite having accounted for a large number of pre-treatment
covariates. The DNC estimate based on NCE $Z_i$
is $0.033$, is less than 20\% of the OLS estimate, and is not
statistically significant. The DNC estimate based on NCE
$\widetilde{Z}_i$ shows a similar pattern: a point estimate is $0.078$
and is not statistically significant. The standard error based on the
second NCE $\widetilde{Z}_i$ is larger than the one based on the first
NCE partly because the association between the NCO and NCE is weaker
for the second NCE \citep{miao2018confounding}. These results show that
the OLS estimator under conditional ignorability can suffer from more
than 100\% bias, which is consistent with previous validation studies on
peer effects in the literature \citep[e.g.,][]{eckles2017bias}.

\section{Extension: Higher-order Peer Effects}
\label{sec:ext}
Following standard causal peer effect literature, we have focused
on the causal effect from peers as the causal estimand of primary interest (the ACPE
defined in equation~\eqref{eq:acpe}). It is important to emphasize that all results in Section~\ref{sec:net} do
not rule out causal effects from higher-order
peers (e.g., peers-of-peers). If they exist, one can simply adjust for focal behaviors of higher-order peers as observed pre-treatment covariates $\bX_i$. We have considered such
higher-order peer effects as nuisance when studying identification and
estimation of the ACPE. In this section, we clarify that the proposed
double negative control approach can also be used for identification
and estimation of higher-order causal peer effects as well. 

The study of such higher-order peer effects can be important for several reasons. First, in some applications, focal
behaviors might be directly affected by higher-order peers even if peers might not change their behaviors. For example,
information can diffuse from higher-order peers even if there is no
behavioral change among peers. Second, estimation
of higher-order peer effects can account for some forms of
misspecification of underlying networks. It is possible that 
observed network and time might not perfectly match the underlying
process through which units causally affect peers. For example, it is possible that units affect
peers faster, and units can affect their peers-of-peers within one
observed time interval. Additionally, the observed network might miss
some ties between units, and thus, two units with the observed
shortest distance of two might in fact be connected directly in the underlying
true network. In such cases, we want to estimate causal effects
from peers and peers-of-peers jointly. 

One can explicitly include focal behaviors of higher-order peers into the potential
outcome. Suppose we are interested in causal effects from all units
within network distance $\sd.$ We define a vector of the 
treatment variable $\tilde{A}_i = (A_{i1}, \ldots, A_{i\sd})$ where  $A_{is}
= \phi(\{Y_{j1}: j \in \cN(i;s)\}) \in \mathbb{R}$, $s \in \{1,
\ldots, \sd\}$, and function $\phi$ is specified by a researcher based
on subject matter knowledge. When $\sd = 1$, this setup reduces to the one in
Section~\ref{sec:net}. The potential outcome $Y_{i2}(\tilde{a})$ is defined
as the outcome that would realize when the treatment vector is set to
$\tilde{A}_i = \tilde{a}$.  We can then define the higher-order ACPE as
\begin{equation}
  \tau (\tilde{a}, \tilde{a}^\prime) \coloneqq \frac{1}{n} \sum_{i =1}^n
  \E \left\{Y_{i2}(\tilde{a}) - Y_{i2}(\tilde{a}^\prime)\right\} \label{eq:gacpe} 
\end{equation}
where $\tilde{a}, \tilde{a}^\prime \in \widetilde{\mathcal{A}}$ where $\widetilde{\mathcal{A}}$ is the support
of $\tilde{A}$. For example, $\tau ((a_{1}, a_2), (a_1, a_2^\prime))$ captures the second-order peer effect by fixing the treatment value
of peers and changing the treatment value of
peers-of-peers. Importantly, while this setup considers up to the
$\sd$-th order peer effects as the
causal estimand, this does not assume the absence of causal effects from
peers at distance more than $\sd$. We only view them as nuisance.

We can straightforwardly generalize Assumption~\ref{net-assum} and
Theorem~\ref{iden-net} to this
setting of higher-order peer effects by replacing $A_i$ with $\tilde{A}_i$. The selection of
negative controls can also proceed in similar fashion. A
plausible candidate is again an auxiliary variable $C_i$ that (a) does not affect network
relationships and (b) does not affect variables of other units. For example, even if we add the second-order peer
effects to Figure~\ref{fig:dag-net-genNC}.(i) (i.e., a causal arrow from
$Y_{41}$ to $Y_{22}$), the original choice of negative controls ---
$C_2$ as the NCO and $\{C_1, C_3, C_4\}$ as the NCEs --- remains valid.

Another candidate for negative controls is the focal behavior itself. For example, if one were to
add the second-order peer effects
to Figure~\ref{fig:dag-net-genNC}.(ii) (i.e., an causal arrow from
$Y_{41}$ to $Y_{22}$), the original choice of NCO $Y_{21}$ would remain
valid, while the original choice of NCEs $\{Y_{41}, Y_{42}\}$ would no
longer be valid. If all third-order peer effects are absent, focal behaviors of third-order peers would be a plausible
candidate for the NCEs. In summary, while the specific choice of
negative controls need to be adjusted when examining higher-order ACPE, the
two primary ways of selecting negative controls we discussed in Section~\ref{sec:net} continue to be useful. 

Finally, estimation and inference can proceed as in Theorems~\ref{est-net} and~\ref{hac} can be extended by
replacing $A_i$ with $\tilde{A}_i$ in the definition of $\bL_{n,i}$.

\section{Concluding Remarks}
\label{sec:con}
In this article, we have developed the double negative control
approach to identification and estimation of causal peer
effects. In contrast to existing literature, we take into account both
unmeasured network confounding and network dependence of
observations. We discuss two general approaches for selecting negative
controls from network data in practice. One is based on an auxiliary variable that (a) does not affect network
relationships and (b) does not affect variables of other units. The other plausible
candidates for negative controls are focal behaviors of peers and
those measured at baseline. We then provide a GMM estimator for
the average causal peer effect and establish its consistency and
asymptotic normality under conditions of $\psi$-network dependence. We
also derive the network HAC variance estimator, with which researchers can
construct asymptotic confidence intervals.

Our findings have established a theoretical basis for future research
on nonparametric estimation of causal peer effects with double negative controls for unmeasured network confounding. This will be
able to extend previous studies that consider double negative control
adjustment of unmeasured
confounding in i.i.d. or panel data settings \citep{deaner2018proxy, shi2020dnc, cui2020,tchetgen2020proximal,
  ghassami2021minimax} and methods that examine network effects without
unmeasured network confounding \citep{van2014causal, ogburn2017causal,
  fora2020, ogburn2020causal, tchetgen2020auto}. Another interesting open question is identification
and estimation of causal peer effects in complex longitudinal studies
with time-varying treatments \citep[e.g.,][]{robins2000marginal,
  tchetgen2020proximal}.

\section*{Supplementary Materials}
The supplemental materials contain proofs of all the results described
in the main text, as well as auxiliary results and proofs used to
demonstrate the desired theoretical properties. We also provide
additional simulation results.

\section*{Acknowledgments}
Research reported in this publication was supported by the National Institutes of Health (award R01AI27271, R01CA222147, R01GM139926,
R01AG065276, to Eric Tchetgen Tchetgen). This research uses data from Add Health, a program project designed by
J. Richard Udry, Peter S. Bearman, and Kathleen Mullan Harris, and
funded by a grant P01-HD31921 from the Eunice Kennedy Shriver National
Institute of Child Health and Human Development, with cooperative
funding from 17 other agencies. Special acknowledgement is due to
Ronald R. Rindfuss and Barbara Entwisle for assistance in the original
design. Persons interested in obtaining Data Files from Add Health
should contact Add Health, The University of North Carolina at Chapel
Hill, Carolina Population Center, Carolina Square, Suite 210, 123
W. Franklin Street, Chapel Hill, NC 27516 (addhealth\_contracts@unc.edu). No direct support was received from
grant P01-HD31921 for this analysis.

\newpage
\pdfbookmark[1]{References}{References}
\bibliography{egami}

\clearpage
\appendix

\spacingset{1.6}
\setcounter{table}{0}
\setcounter{equation}{0}
\setcounter{figure}{0}
\renewcommand {\thetable} {A\arabic{table}}
\renewcommand {\thefigure} {A\arabic{figure}}
\renewcommand {\theequation} {A.\arabic{equation}}

\vspace{-0.5in}

\begin{center}
  {\fontsize{15}{15}\selectfont Supplementary Materials for \\
  ``Identification and Estimation of Causal Peer Effects
  Using \\ Double Negative Controls for Unmeasured Network Confounding''}
\end{center}

\section{Identification}
\label{appsec:iden}
\subsection{Proof of Lemma~\ref{existence}}
\label{subsec:proof-existence}
In this proof, to make a discussion general, we use $Y$ to denote the outcome and use $A$
to denote the treatment instead of using $(Y_{12}, Y_{21})$ and
$(Y_{i2}, A_i)$, which we use in Section~\ref{sec:dyad} and
Section~\ref{sec:net}, respectively.
To provide rigorous discussions on the existence of a solution to a
Fredholm integral equation of the first kind, we rely on Picard’s
theorem \citep[Theorem 15.18]{kress1989}.

\begin{lemma}[Picard’s theorem (Kress, 1989, Theorem 15.18)]
  \label{picard}
  Given Hilbert spaces $\cS_1$ and $\cS_2$, let $K: \cS_1 \rightarrow
  \cS_2$ be a compact operator with singular system $(\nu_{p},
  \upsilon_{p}, \kappa_{p})_{p=1}^{+\infty}.$ Define its adjoint
  to be $K^\ast: \cS_2 \rightarrow \cS_1.$ Then, for $h \in \cS_1$ and
  $\tilde{h} \in
  \cS_2$, there exists a solution to a Fredholm integral equation of the first kind $K h =
  \tilde{h}$ if and only if (1) $\tilde{h} \in
  \mbox{\normalfont Null}(K^\ast)^{\perp}$ and (2) $\sum_{p=1}^{+\infty}
  \frac{1}{\nu_p^2} |\langle \tilde{h}, \kappa_{p}\rangle|^2 \ < \ +
  \infty,$ where the inner product is defined for a Hilbert space
  $\cS_2$, $\mbox{\normalfont Null}(K^\ast) = \{\tilde{
    h}: K^\ast \tilde{h} = 0\}$ is the null space of $K^\ast$, and $\perp$
  represents the orthogonal complement to a subset. 
\end{lemma}

To apply this Picard’s theorem, we need to provide some additional
notations. We use $F$ and $dF$ to denote the cumulative
distribution function and the Radon-Nikodym derivative of $F$. We
define $L^2\{F (t)\}$ to be the space of all square integrable
functions of $t$ with respect to a cumulative distribution function
$F(t)$, which is a Hilbert space with the inner product
\begin{equation*}
  \langle h_1, h_2 \rangle \coloneqq \int_{-\infty}^{+\infty}
  h_1(t) h_2(t) dF(t) \hspace{0.2in} \mbox{for all } h_1, h_2 \in L^2\{F (t)\}.
\end{equation*}
We define a kernel
\begin{equation*}
  K(w, u, a, \bx) = \cfrac{dF(w, u \mid a, \bx)}{dF(w \mid a, \bx) dF(u \mid a, \bx)}.
\end{equation*}
We then define the linear operators $K_{a,\bx}: L^2\{F(w \mid a, \bx)\} \rightarrow L^2\{F(u \mid a, \bx)\}$ by
\begin{equation*}
  K_{a,\bx} h = \int^{+ \infty}_{-\infty} K(w, u, a, \bx) h(w) d F(w
  \mid a, \bx) = \E\{h(w) \mid u, a, \bx\}
\end{equation*}
for $h \in L^2\{F(w \mid a, \bx)\}.$ The adjoint of this linear
operator $K^{\ast}_{a,\bx}: L^2\{F(u \mid a,
\bx)\} \rightarrow L^2\{F(w \mid a, \bx)\}$ is given by
\begin{equation*}
  K^{\ast}_{a,\bx} \tilde{h} = \int^{+ \infty}_{-\infty} K(w, u, a, \bx) \tilde{h}(u) d F(u
  \mid a, \bx) = \E\{\tilde{h}(u) \mid w, a, \bx\}
\end{equation*}
for $\tilde{h} \in L^2\{F(u \mid a, \bx)\}.$

We first assume that $W$ is relevant for $U$.
\begin{assumption}[Relevance of $W$ for $U$]
  \label{com-wu}
  For any square integrable function $f$ and any $a$ and $\bx$, if $\E\{f(U) \mid W = w, A = a, \bX = \bx\}
  = 0$ for almost all $w$, then $f(U) = 0$ almost surely.
\end{assumption}
This is formally known as the completeness condition, and can be
interpreted similarly to Assumption~\ref{com}. We also introduce
regularity conditions related to the singular value decomposition.
\begin{assumption}[Regularity Conditions]
  \label{assum-exis}
    \begin{eqnarray}
      && \int^{+\infty}_{-\infty} dF(u \mid w, a, \bx) dF(w
         \mid u, a, \bx) dw du \ < \ + \infty \label{eq:r-1}\\
      && \int^{+\infty}_{-\infty} \E(Y \mid a, u, \bx)^2
         dF(u \mid a, \bx) du \ < \ + \infty \label{eq:r-2}\\
      && \sum_{p=1}^{+\infty} \frac{1}{\nu_{a, \bx, p}^2} |\langle\E(Y \mid a, u, \bx),
         \kappa_{a,\bx,p}\rangle|^2 \ < \ + \infty \label{eq:r-3}
    \end{eqnarray}
    where $\nu_{a,\bx, p}$ is the $p$-th singular value of $K_{a,
      \bx}$, and $\kappa_{a, \bx, p} \in L^2\{F(u \mid a, \bx)\}$ is
    an orthogonal sequence.
\end{assumption}
Under Assumptions~\ref{com-wu} and~\ref{assum-exis}, we prove the existence of a
solution to the following Fredholm integral equation of the first kind.
\begin{equation}
  \E(Y \mid A = a, U = u, \bX = \bx)  = \E\{h(W, a, \bx) \mid  A = a,
  U = u, \bX = \bx\}.  \label{eq:ex-1}
\end{equation}

First, we can re-write equation~\eqref{eq:ex-1} as follows using the
notations introduced above.
\begin{equation}
  K_{a, \bx} h  = \E(Y \mid A = a, U = u, \bX = \bx).
\end{equation}
Therefore, to use Picard's theorem, we need to prove (i) $K_{a, \bx}$
is a compact operator, (ii) $\E(Y \mid A = a, U = u, \bX = \bx) \in
L^2\{F(u \mid a, \bx)\}$, (iii) $\E(Y \mid A = a, U = u, \bX = \bx) \in
\mbox{\normalfont Null}(K_{a,\bx}^\ast)^{\perp}$, and (iv) $\sum_{p=1}^{+\infty}
\frac{1}{\nu_{a, \bx, p}^2} |\langle \E(Y \mid A = a, U = u, \bX = \bx), \kappa_{a, \bx, p}\rangle|^2 \ < \ +
\infty,$ where $\nu_{a,\bx, p}$ is the $p$-th singular value of $K_{a,
  \bx}$, and $\kappa_{a, \bx, p} \in L^2\{F(u \mid a, \bx)\}$ is an
orthogonal sequence. 

Proof of (i): We note that $K_{a,\bx}$ and $K^\ast_{a, \bx}$ are compact
operators under equation~\eqref{eq:r-1} \citep[Example 2.3 on page
5659]{carrasco2007linear}. Therefore, there exists a singular system $(\nu_{a,\bx, p}, \upsilon_{a, \bx, p}, \kappa_{a, \bx,
  p})$ of $K_{a, \bx}$ according to \citet[Theorem
15.16]{kress1989} where $\nu_{a,\bx, p}$ is the $p$-th singular value
of  $K_{a, \bx}$, and $\upsilon_{a, \bx, p} \in L^2\{F(w \mid a, \bx)\}$
and $\kappa_{a, \bx, p} \in L^2\{F(u \mid a, \bx)\}$ are orthogonal
sequences.

Proof of (ii): Under equation~\eqref{eq:r-2}, we have $\E(Y \mid a, u, \bx)
\in L^2\{F(u \mid a, \bx)\}$.

Proof of (iii): We show that $\mbox{Null}(K^\ast_{a, \bx})^{\perp} = L^2\{F(u \mid a, \bx)\}.$ For any $\tilde{h} \in
\mbox{Null}(K^\ast_{a, \bx}),$ we have $K^\ast_{a, \bx} \tilde{h}  = \E\{\tilde{h}(u) \mid w, a, \bx\} = 0$ almost surely by
the definition of the null space. Under Assumption~\ref{com-wu} (Relevance of $W$ for $U$), we have $\tilde{h}(u) = 0$ almost surely. Therefore, $\mbox{Null}(K^\ast_{a,
  \bx})^{\perp} = L^2\{F(u \mid a, \bx)\}.$ Based on (ii), we have $\E(Y \mid a, u, \bx) \in L^2\{F(u
\mid a, \bx)\}$ under equation~\eqref{eq:r-2}, and therefore, $\E(Y \mid a, u, \bx) \in \mbox{Null}(K^\ast_{a, \bx})^{\perp}$.

Proof of (iv): Finally, this key condition for Picard's theorem is
directly implied by equation~\eqref{eq:r-3}, which completes the
proof. \qed

\subsection{Details on Completeness Conditions}
\label{subsec:more-com}
In this section, to make discussions simpler, we only focus on two
random variables $W$ and $Z$. We say that $Z$ is complete with respect
to $W$ if $\forall f(W) \in L^2\{F(W)\},$ 
\begin{eqnarray}
  && \E\{f(W) \mid Z \} = 0 \hspace{0.1in} \mbox{almost surely} \
     \Longrightarrow  \ f(W) = 0 \hspace{0.1in} \mbox{almost surely.} \label{eq:com-def}
\end{eqnarray}
This completeness condition, also known as $L^2-$completeness,
requires that the conditional expectation projection operator $K:
L^2\{F(W)\} \rightarrow L^2\{F(Z)\}$ be injective (i.e., Null($K$) =
$\{0\}$). Intuitively, this means that no information has been lost
through projection of $W$ on $Z$. A necessary and sufficient condition
of completeness is given by the following lemma. 
\begin{lemma}[\cite{severini2006, andrews2017examples}]
  $Z$ is complete with respect to $W$ if and only if every
  non-constant random variable $\lambda(W) \in L^2\{F(W)\}$ is
  correlated with some random variable $\tilde{\lambda}(Z) \in L^2\{F(Z)\}.$
\end{lemma}
This formally captures the notion that completeness ensures that
there is no loss of information through projection of $W$ on $Z$.

As explained in Section~\ref{subsec:iden}, the completeness condition
has been long used in statistics and econometrics. Originally in statistics, \cite{lehmann2012comp1, lehmann2012comp2} introduced the
concept of completeness and used it to define estimators with
minimal risk within unbiased estimators. They defined completeness as
$\E_{\theta}(f(V)) = 0$ for any $\theta \in \Theta$ implying $f(V) =
0$ a.s. with respect to some parameter space $\Theta$ parameterizing
the distribution space. \cite{shao2003} defined completeness with
respect to a family of distributions, i.e., $\E_{P} (f(V)) = 0$ for any
$P \in \cP$ implying $f(V) = 0$ a.s. with respect to some family of
$\cP$.  In our definition of the completeness (Assumption~\ref{com} and
Assumption~\ref{net-assum}.4), we set $\cP$ to be the conditional
distribution. If we define a family of distributions to be
$\cP = \{F(W \mid Z): Z \in \cZ\}$ of random variable $W$, the
connection between our definition of completeness and the traditional
completeness condition given in
\cite{lehmann2012comp1, lehmann2012comp2} becomes clear. In
particular, we say that a family of distributions
$\cP = \{F(W \mid Z): Z \in \cZ\}$ of random variable $W$ is
complete with respect to $Z$ if $\forall f(W) \in L^2\{F(W)\},$ 
$\E_{F(W \mid Z)}\{f(W)\} = \E\{f(W) \mid Z\} = 0$ for almost all $Z$
implies that $f(W) = 0$ almost surely. This is equivalent to our definition given in
equation~\eqref{eq:com-def}.

Recently, completeness conditions have been extensively applied in the
econometrics literature to obtain identification for a variety of
nonparametric and semi-parametric models, most famously, in nonparametric models with instrumental variables
\citep[e.g.,][]{ai2003, newey2003, chernozhukov2007instrumental,
  darolles2011}. Other examples include measurement error models
\citep[e.g.,][]{hu2008instrumental} and panel or dynamic models
\citep[e.g.,][]{hu2012nonparametric, freyberger2018}.

Finally, as in our paper, completeness conditions have been essential
in the literature of negative controls and proximal
causal learning \citep{tchetgen2020proximal}. \cite{miao2018identifying} make two completeness conditions
(a) the completeness of $W$ with respect to $Z$, (b) the completeness
of $Z$ with respect to $U$ (see Conditions 2 and 3 in their
paper). \cite{deaner2018proxy, shi2020dnc, kallus2021causal} make alternative two completeness conditions
(a) the completeness of $W$ with respect to $U$, (b) the completeness
of $Z$ with respect to $U$ (see Assumption 3 in
\cite{deaner2018proxy}, Assumption 4 in \cite{shi2020dnc}, and Example
6 in \cite{kallus2021causal}). \cite{miao2018confounding} make one completeness condition
(the completeness of $Z$ with respect to $W$; see Assumption 5 in
their paper) along with the assumption of the existence of an outcome
confounding bridge function, which can be justified by another
completeness condition (the completeness of $W$ with respect to $U$).

In Sections~\ref{sec:dyad} and~\ref{sec:net}, we followed \cite{miao2018confounding} and
made Assumptions~\ref{cb} and~\ref{com} and
Assumptions~\ref{net-assum}.3 and~\ref{net-assum}.4, respectively. We
prove nonparametric identification of the ACPE under those assumptions
in Section~\ref{subsec:prove-iden} below. We also prove
nonparametric identification of the ACPE under an alternative set of
completeness conditions in Section~\ref{subsec:prove-iden-2} as well.

\subsection{Proof of Lemma~\ref{nc-aux}}
First, equation~\eqref{eq:a-1} implies that
\begin{eqnarray}
  && C_{i} \ \indep \ (\{C_j: j \neq i\}, A_i) \ \mid \ U_i, \bX_i, \label{eq:aux-0}\\
  \Longrightarrow && W_{i} \ \indep \ A_i \ \mid \ U_i, \bX_i, \label{eq:aux-1}
\end{eqnarray}
as we define $W_i = C_i$ and $A_i = \phi(\{Y_{j1}: j \in \cN(i;1)\}) \in \mathbb{R}$.
Then, equation~\eqref{eq:aux-0} also implies that
\begin{eqnarray}
  && C_{i} \ \indep \ \{C_j: j \neq i\} \ \mid \ A_i, U_i, \bX_i, \nonumber\\
  \Longrightarrow && W_{i} \ \indep \ \bZ_i \ \mid \ A_i, U_i, \bX_i, \label{eq:aux-2}
\end{eqnarray}
as we define $W_i = C_i$ and $\bZ_i = \{C_{j}: j \neq i\}.$ Finally, equation~\eqref{eq:a-2} implies that
\begin{eqnarray}
  && Y_{i2} \ \indep \ \{C_j: j \neq i\} \ \mid \ A_i, U_i, \bX_i, \nonumber\\
  \Longrightarrow && Y_{i2} \ \indep \ \bZ_i \ \mid \ A_i, U_i, \bX_i, \label{eq:aux-3}
\end{eqnarray}
where $\bZ_i = \{C_{j}: j \neq i\}.$ Therefore,
equations~\eqref{eq:aux-1}--\eqref{eq:aux-3} are equivalent to
Assumption~\ref{net-assum}.2, which completes the proof. \qed

\subsection{Proof of Lemma~\ref{nc-focal}}

First, equation~\eqref{eq:f-1} implies that
\begin{eqnarray}
  && Y_{i1} \ \indep \ (A_i, \{Y_{j1}: j \in \cN(i;s), s \geq 2\}) \ \mid \ U_i, \bX_i, \label{eq:focal-0}\\
  \Longrightarrow && W_{i} \ \indep \ A_i \ \mid \ U_i, \bX_i, \label{eq:focal-1}
\end{eqnarray}
as we define $W_i = Y_{i1}.$ Then, equation~\eqref{eq:focal-0} also implies that
\begin{eqnarray}
  && Y_{i1} \ \indep \ \{Y_{j1}: j \in \cN(i;s), s \geq 2\} \ \mid \ A_i, U_i, \bX_i, \nonumber\\
  \Longrightarrow && W_{i} \ \indep \ \bZ_i \ \mid \ A_i, U_i, \bX_i, \label{eq:focal-2}
\end{eqnarray}
as we define $W_i = Y_{i1}$ and $\bZ_i = \{Y_{j1}: j \in
\cN(i;s), s \geq \tilde{s} \}$ where $\tilde{s} \geq 2.$

Finally, equation~\eqref{eq:f-2} states that
\begin{eqnarray}
  && Y_{i2} \ \indep \ \{Y_{j1}: j \in \cN(i;s), s \geq \tilde{s} \} \ \mid \ A_i, U_i, \bX_i, \nonumber\\
  \Longrightarrow && Y_{i2} \ \indep \ \bZ_i \ \mid \ A_i, U_i, \bX_i, \label{eq:focal-3}
\end{eqnarray}
where $\bZ_i = \{Y_{j1}: j \in \cN(i;s), s \geq \tilde{s}\}.$ Therefore,
equations~\eqref{eq:focal-1}--\eqref{eq:focal-3} are equivalent to
Assumption~\ref{net-assum}.2, which completes the proof. \qed

\subsection{Proof of Theorem~\ref{iden-net}}
\label{subsec:prove-iden}
Here, we prove identification of $\E\{Y_{i2}(a)\}$ for $a
\in \cA$ and a given unit $i \in N_n$, which is sufficient for proving
identification of the ACPE. The proof of Theorem~\ref{iden} is a
special case of the proof we provide below.

This proof adopts the proof by \cite{miao2018confounding} to our
network setting. First, we prove that the mean potential outcomes can be
identified as the mean of the outcome confounding bridge function.  
\begin{equation*}
  \E\{Y_{i2}(a)\} = \E\{h(W_{i}, a, \bX_i)\}.
\end{equation*}
\paragraph{Proof:}
Under Assumption~\ref{net-assum}.1,
\begin{eqnarray*}
  \int \E(Y_{i2} \mid A_i = a, U_i = u, \bX_i = \bx) dF(U_i = u, \bX_i = \bx) 
  & = & \int \E\{Y_{i2} (a) \mid U_i = u, \bX_i = \bx\}  dF(U_i = u, \bX_i = \bx) \\
  & = & \E\{Y_{i2} (a)\}.
\end{eqnarray*}
Under Assumption~\ref{net-assum}.2,
\begin{eqnarray*}
  && \int \E\{h(W_{i}, a, \bX_i) \mid A_i = a, U_i = u, \bX_i = \bx) dF(U_i = u, \bX_i = \bx) \\
  & = & \int \E\{h(W_{i}, a, \bX_i) \mid U_i = u, \bX_i = \bx) dF(U_i = u, \bX_i = \bx)  \\
  & = & \E\{h(W_{i}, a, \bX_i)\}.
\end{eqnarray*}
Under Assumption~\ref{net-assum}.3, $\E(Y_{i2} \mid A_{i} = a, U_i =
u, \bX_i = \bx)  = \E\{h(W_{i}, a, \bX_i) \mid  A_{i} = a, U_i = u, \bX_i = \bx\}$, and therefore,
\begin{equation*}
  \E\{Y_{i2}(a)\} = \E\{h(W_{i}, a, \bX_i)\},
\end{equation*}
which completes the proof.  \qed

Next, we prove that the confounding bridge function is identified as follows. 
\begin{equation}
    \E(Y_{i2} \mid \bZ_{i}, A_{i}, \bX_i) = \E\{h(W_{i}, A_{i},
    \bX_i) \mid \bZ_{i}, A_{i}, \bX_i\}. \label{eq:nce-iden}
  \end{equation}
\paragraph{Proof:}
Under Assumption~\ref{net-assum}.2,
\begin{eqnarray*}
  && \int \E(Y_{i2} \mid A_i, U_i = u, \bX_i) dF(U_i = u \mid \bZ_i, A_i, \bX_i) \\
  & = & \int \E(Y_{i2} \mid \bZ_i, A_i, U_i = u, \bX_i) dF(U_i = u \mid \bZ_i, A_i, \bX_i)  \\
  & = & \E(Y_{i2} \mid \bZ_{i}, A_{i}, \bX_i). 
\end{eqnarray*}
Under Assumption~\ref{net-assum}.2,
\begin{eqnarray*}
  && \int \E\{h(W_{i}, A_i, \bX_i) \mid A_i, U_i = u, \bX_i\} dF(U_i = u \mid \bZ_i, A_i, \bX_i)\\
  & = & \int \E\{h(W_{i}, A_i, \bX_i) \mid \bZ_i, A_i, U_i = u,
        \bX_i\} dF(U_i = u \mid \bZ_i, A_i, \bX_i) \\
  & = &  \E\{h(W_{i}, A_{i}, \bX_i) \mid \bZ_{i}, A_{i}, \bX_i\}.
\end{eqnarray*}
Under Assumption~\ref{net-assum}.3, $\E(Y_{i2} \mid A_{i}, U_i, \bX_i)  = \E\{h(W_{i},
A_i, \bX_i) \mid  A_{i}, U_i, \bX_i\}$, and therefore,
\begin{equation*}
  \E(Y_{i2} \mid \bZ_{i}, A_{i}, \bX_i) = \E\{h(W_{i}, A_{i}, \bX_i) \mid \bZ_{i}, A_{i}, \bX_i\}.
\end{equation*}
We finally demonstrate that the solution to
equation~\eqref{eq:nce-iden} is unique and identifies the outcome
confounding bridge function $h$ under Assumption~\ref{net-assum}.4. Suppose there are two
functions $h(W_{i}, A_{i}, \bX_i)$ and $h^\prime(W_{i}, A_{i},
\bX_i)$ that satisfy equation~\eqref{eq:nce-iden}. Then,
\begin{equation*}
  \E\{h(W_{i}, A_{i}, \bX_i) - h^\prime(W_{i}, A_{i}, \bX_i) \mid
  \bZ_{i} = \bz, A_{i} = a, \bX_i = \bx\} = 0
\end{equation*}
for all $a$, $\bx$, and almost all $\bz$. Then, under Assumption~\ref{net-assum}.4, $h(W_{i}, A_{i}, \bX_i) =  h^\prime(W_{i},
A_{i}, \bX_i)$ almost surely. Thus, the solution to
equation~\eqref{eq:nce-iden} identifies the outcome confounding bridge
function. 
\qed

\subsection{Identification of the ACPE under Alternative Assumptions}
\label{subsec:prove-iden-2}
Here, we show that the same identification formula for the ACPE can be
proven based on an alternative set of assumptions. The main difference
is that we first define an outcome bridge function as a solution to 
the following Fredholm integral equation of the first kind. 
\begin{equation}
  \E(Y_{i2} \mid \bZ_{i}, A_{i}, \bX_i) = \E\{h^\dagger(W_{i}, A_{i}, \bX_i)
  \mid \bZ_{i}, A_{i}, \bX_i\}. \label{eq:alt-1}
\end{equation}
Then, we show, under some assumptions, this outcome bridge
function satisfies
\begin{equation}
  \E(Y_{i2} \mid A_{i} = a, U_i, \bX_i)  = \E\{h^\dagger (W_{i}, a, \bX_i) \mid  A_{i} = a, U_i, \bX_i\}. \label{eq:alt-2}
\end{equation}
This approach is opposite to the approach we used in the main paper
and proved in Section~\ref{subsec:prove-iden} where we defined an
outcome bridge function as a solution to equation~\eqref{eq:alt-2} and
then showed that it satisfies equation~\eqref{eq:alt-1}. The approach
used in this section is similar to the one used in \cite{deaner2018proxy, miao2018identifying, shi2020dnc}.

We see below that this difference in the proof approaches lead to a
different set of assumptions, while they both result in the same
identification formula for the ACPE. 

In particular, while we maintain Assumption~\ref{net-assum}.1 and Assumption~\ref{net-assum}.2, we replace Assumption~\ref{net-assum}.3 and
Assumption~\ref{net-assum}.4 with two different assumptions below
(Assumptions~\ref{exist-cb-2} and~\ref{com-zu}). 

\begin{assumption}[Outcome Confounding Bridge $h^\dagger$]
  \label{exist-cb-2}
  There exists some function $h^\dagger (W_{i}, A_{i}, \bX_i)$ such
  that for all $a \in \cA$, and all $i \in N_n,$
  \begin{equation}
      \E(Y_{i2} \mid A_{i} = a, \bZ_i = \bz, \bX_i)  = \E\{h^\dagger(W_{i}, a,
      \bX_i) \mid  A_{i} = a, \bZ_i = \bz, \bX_i\}. \label{eq:cb-2}
    \end{equation}
  \end{assumption}
  \begin{assumption}[Relevance of $\bZ$ for $U$]
  \label{com-zu}
  For any square integrable function $f$ and for any $a$ and $\bx$, if $\E\{f(U_i) \mid \bZ_i = \bz, A_i = a, \bX_i = \bx\}
  = 0$ for almost all $\bz$, then $f(U_i) = 0$ almost surely.
\end{assumption}

\begin{theorem}
  \label{iden-net-2}
  Under
  Assumptions~\ref{net-assum}.1,~\ref{net-assum}.2,~\ref{exist-cb-2}
  and~\ref{com-zu}, an outcome confounding bridge 
  function $h^\dagger$ $($defined in equation~\eqref{eq:cb-2}$)$ satisfies the
  following equality for all $a \in \cA$, and all $i \in N_n,$
  \begin{equation*}
    \E(Y_{i2} \mid U_{i}, A_{i} = a, \bX_i) = \E\{h^\dagger(W_{i}, a, \bX_i) \mid U_{i}, A_{i} = a, \bX_i\},
  \end{equation*}
  and, the ACPE is identified by 
  \begin{equation*}
    \tau(a, a^\prime) = \frac{1}{n} \sum_{i=1}^n \E\{h^\dagger (W_{i}, a,
    \bX_i)- h^\dagger(W_{i}, a^\prime, \bX_i)\}.
  \end{equation*}
\end{theorem}

Finally, like Lemma~\ref{existence}, we can also prove Assumption~\ref{exist-cb-2} under a completeness
condition and associated regularity conditions
(Assumption~\ref{assum-reg2} defined below in Section~\ref{subsub:exist-2}).
\begin{lemma}
  \label{existence-2}
  Under Assumptions~\ref{com-wu} and~\ref{assum-reg2}, there exists a
  function $h^\dagger(W_{i}, a, \bX_i)$ such that for all $a \in \cA$ and all $i \in N_n,$, equation~\eqref{eq:cb-2} holds. 
\end{lemma}

\subsubsection{Proof of Theorem~\ref{iden-net-2}}
First, we show that an outcome confounding bridge function $h^\dagger$ defined in
equation~\eqref{eq:cb-2} satisfies the following equality.
\begin{equation}
  \E\{h^\dagger(W_{i}, a, \bX_i) \mid A_i= a, U_i = u, \bX_i = \bx)  =
  \E\{Y_{i2}(a) \mid A_i = a, U_i = u, \bX_i = \bx\} \label{eq:iden-20}
\end{equation}
We have 
\begin{eqnarray}
  && \E\{h^\dagger(W_i, a, \bX_i) \mid A_i= a, \bZ_i = \bz, \bX_i = \bx) \nonumber\\
  & = & \int \E\{h^\dagger(W_i, a, \bX_i) \mid A = a, U_i = u, \bZ_i = \bz, \bX_i = \bx) dF(U_i = u \mid A= a, \bZ_i = \bz, \bX_i = \bx) \nonumber\\
  & = & \int \E\{h^\dagger(W_i, a, \bX_i) \mid A = a, U_i = u, \bX_i =
        \bx) dF(U_i = u \mid A= a, \bZ_i = \bz, \bX_i = \bx) \label{eq:iden-21}
\end{eqnarray}
where the first equality follows from iterated expectations, and the
second from Assumption~\ref{net-assum}.2. We also have
\begin{eqnarray}
  && \E(Y_{i2} \mid A_i = a, \bZ_i = \bz, \bX_i = \bx) \nonumber \\
  & = & \int \E(Y_{i2} \mid A_i = a, U_i =u, \bZ_i = \bz, \bX_i = \bx)
        dF(U_i = u \mid A_i= a, \bZ_i = \bz, \bX_i = \bx) \nonumber \\
  & = & \int \E(Y_{i2} \mid A_i = a, U_i =u, \bX_i = \bx)  dF(U_i = u
        \mid A_i= a, \bZ_i = \bz, \bX_i = \bx) \label{eq:iden-22}
\end{eqnarray}
where the first equality follows from iterated expectations, and the
second from Assumption~\ref{net-assum}.2.

Under Assumption~\ref{exist-cb-2}, we have
\begin{eqnarray*}
  && \E(Y_{i2} \mid A_{i} = a, \bZ_i = \bz, \bX_i)  = \E\{h^\dagger(W_{i}, a, \bX_i) \mid  A_{i} = a, \bZ_i = \bz, \bX_i\}\\
  &\Longleftrightarrow &  \int \{\E(Y_{i2} \mid A_i = a, U_i =u, \bX_i = \bx) - \E\{h^\dagger(W_i, a, \bX_i)
                         \mid A_i = a, U_i = u, \bX_i = \bx)\}
                         \nonumber \\[-8pt]
  && \hspace{0.6in} \times dF(U_i = u \mid A_i= a, \bZ_i = \bz,
     \bX_i = \bx) = 0 \nonumber\\
&\Longrightarrow& \E(Y_{i2}(a) \mid A_i = a, U_i =u, \bX_i = \bx) = \E\{h^\dagger(W_i,
                      a, \bX_i) \mid A_i = a, U_i = u, \bX_i = \bx\}
\end{eqnarray*}
where the first equivalence comes from equations~\eqref{eq:iden-21}
and~\eqref{eq:iden-22}, and the final line follows from
Assumption~\ref{com-zu} and the consistency of the potential
outcomes.

Next, by using equation~\eqref{eq:iden-20}, we prove that
\begin{equation*}
  \E\{Y_{i2}(a)\} = \E\{h^\dagger(W_{i}, a, \bX_i)\}.
\end{equation*}
Under Assumption~\ref{net-assum}.1, we have 
\begin{eqnarray*}
  && \int \E(Y_{i2}(a) \mid A_i = a, U_i = u, \bX_i = \bx) dF(U_i = u, \bX_i = \bx) \\
  & = & \int \E\{Y_{i2} (a) \mid U_i = u, \bX_i = \bx\}  dF(U_i = u, \bX_i = \bx) \\
  & = & \E\{Y_{i2} (a)\}.
\end{eqnarray*}
Under Assumption~\ref{net-assum}.2, we have 
\begin{eqnarray*}
  && \int \E\{h^\dagger(W_{i}, a, \bX_i) \mid A_i = a, U_i = u, \bX_i = \bx) dF(U_i = u, \bX_i = \bx) \\
  & = & \int \E\{h^\dagger(W_{i}, a, \bX_i) \mid U_i = u, \bX_i = \bx) dF(U_i = u, \bX_i = \bx)  \\
  & = & \E\{h^\dagger(W_{i}, a, \bX_i)\}.
\end{eqnarray*}
Equation~\eqref{eq:iden-20} states that $\E(Y_{i2}(a) \mid A_{i} = a, U_i =
u, \bX_i = \bx)  = \E\{h^\dagger(W_{i}, a, \bX_i) \mid  A_{i} = a, U_i = u, \bX_i = \bx\}$, and therefore,
\begin{equation*}
  \E\{Y_{i2}(a)\} = \E\{h^\dagger(W_{i}, a, \bX_i)\},
\end{equation*}
which completes the proof.  \qed

\subsubsection{Proof of Lemma~\ref{existence-2}}
\label{subsub:exist-2}
In this proof, to make a discussion general, we use $Y$ to denote the outcome and use $A$
to denote the treatment instead of using $(Y_{i2}, A_i)$, which we use
in Section~\ref{sec:net}. To provide rigorous discussions on the existence of a solution to a
Fredholm integral equation of the first kind, we keep using some
notations introduced in Section~\ref{subsec:proof-existence}.

Using general notations, we re-state Lemma~\ref{existence-2} as follows. 
Under Assumptions~\ref{com-wu} and~\ref{assum-reg2}, there exists
a function $h(W, a, \bX)$ such that for all $a \in \cA$, a solution to the following Fredholm integral equation of the first
kind exists. 
\begin{equation}
  \E(Y \mid \bZ = \bz, A = a, \bX = \bx) = \E\{h^\dagger(W, a, \bx)
  \mid \bZ = \bz, A = a, \bX = \bx\}. \label{eq:ex-2}
\end{equation}
We also introduce regularity conditions related to the singular value decomposition.
\begin{assumption}[Regularity Conditions II]
  \label{assum-reg2}
  \begin{eqnarray}
    && \int^{+\infty}_{-\infty} dF(\bz \mid w, a, \bx) dF(w
       \mid \bz, a, \bx) dw d\bz \ < \ + \infty \label{eq:r-21}\\
    && \int^{+\infty}_{-\infty} \E(Y \mid a, \bz, \bx)^2
       dF(\bz \mid a, \bx) d\bz \ < \ + \infty \label{eq:r-22}\\
    && \sum_{p=1}^{+\infty} \frac{1}{\tilde{\nu}_{a, \bx, p}^2} |\langle\E(Y \mid a, \bz, \bx),
       \tilde{\kappa}_{a,\bx,p}\rangle|^2 \ < \ + \infty \label{eq:r-23}
  \end{eqnarray}
  where $\tilde{\nu}_{a,\bx, p}$ is the $p$-th singular value of $\widetilde{K}_{a,
    \bx}$, and $\tilde{\kappa}_{a, \bx, p} \in L^2\{F(\bz \mid a, \bx)\}$ is
  an orthogonal sequence.
\end{assumption}
\paragraph{Proof:}
We start by defining a kernel
  \begin{equation*}
    \widetilde{K}(w, \bz, a, \bx) = \cfrac{dF(w, \bz \mid a, \bx)}{dF(w \mid a, \bx) dF(\bz \mid a, \bx)}.
  \end{equation*}
  We then define the linear operators $\widetilde{K}_{a,\bx}: L^2\{F(w \mid a, \bx)\} \rightarrow L^2\{F(\bz \mid a, \bx)\}$ by
  \begin{equation*}
    \widetilde{K}_{a,\bx} h = \int^{+ \infty}_{-\infty} \widetilde{K}(w, \bz, a, \bx) h(w) d F(w
    \mid a, \bx) = \E\{h(w) \mid \bz, a, \bx\}
  \end{equation*}
  for $h \in L^2\{F(w \mid a, \bx)\}.$

  The adjoint of this linear operator $\widetilde{K}^{\ast}_{a,\bx}: L^2\{F(\bz \mid a,
  \bx)\} \rightarrow L^2\{F(w \mid a, \bx)\}$ is given by
  \begin{equation*}
    \widetilde{K}^{\ast}_{a,\bx} \tilde{h} = \int^{+ \infty}_{-\infty} \widetilde{K}(w, \bz, a, \bx) \tilde{h}(\bz) d F(\bz
    \mid a, \bx) = \E\{\tilde{h}(\bz) \mid w, a, \bx\}
  \end{equation*}
  for $\tilde{h} \in L^2\{F(\bz \mid a, \bx)\}.$

Using the introduced notations, we can re-write equation~\eqref{eq:ex-2} as follows using the notations introduced above.
\begin{equation}
  \widetilde{K}_{a, \bx} h  = \E(Y \mid A = a, \bZ = \bz, \bX = \bx).
\end{equation}
Therefore, to use Picard's theorem (Lemma~\ref{picard}), we need to prove (i) $\widetilde{K}_{a, \bx}$
is a compact operator, (ii) $\E(Y \mid A = a, \bZ = \bz, \bX = \bx) \in
L^2\{F(\bz \mid a, \bx)\}$, (iii) $\E(Y \mid A = a, \bZ = \bz, \bX = \bx) \in
\mbox{\normalfont Null}(\widetilde{K}_{a,\bx}^\ast)^{\perp}$, and (iv) $\sum_{p=1}^{+\infty}
\frac{1}{\widetilde{\nu}_{a, \bx, p}^2} |\langle \E(Y \mid A = a, \bZ = \bz, \bX = \bx), \widetilde{\kappa}_{a, \bx, p}\rangle|^2 \ < \ +
\infty,$ where $\widetilde{\nu}_{a,\bx, p}$ is the $p$-th singular value of $\widetilde{K}_{a,
  \bx}$, and $\widetilde{\kappa}_{a, \bx, p} \in L^2\{F(\bz \mid a,
\bx)\}$ is an orthogonal sequence. 

Proof of (i): We note that $\widetilde{K}_{a,\bx}$ and $\widetilde{K}^\ast_{a, \bx}$ are compact
operators under equation~\eqref{eq:r-21} \citep[Example 2.3 on page
5659]{carrasco2007linear}. Therefore, there exists a singular system $(\widetilde{\nu}_{a,\bx, p}, \widetilde{\upsilon}_{a, \bx, p}, \widetilde{\kappa}_{a, \bx,
  p})$ of $\widetilde{K}_{a, \bx}$ according to \citet[Theorem
15.16]{kress1989} where $\widetilde{\nu}_{a,\bx, p}$ is the $p$-th singular value
of  $\widetilde{K}_{a, \bx}$, and $\widetilde{\upsilon}_{a, \bx, p} \in L^2\{F(w \mid a, \bx)\}$
and $\widetilde{\kappa}_{a, \bx, p} \in L^2\{F(\bz \mid a, \bx)\}$ are orthogonal
sequences.

Proof of (ii): Under equation~\eqref{eq:r-22}, we have $\E(Y \mid a, \bz, \bx)
\in L^2\{F(\bz \mid a, \bx)\}$.

Proof of (iii): To show that $\E(Y \mid A = a, \bZ = \bz, \bX = \bx) \in
\mbox{\normalfont Null}(\widetilde{K}_{a,\bx}^\ast)^{\perp}$, we first define $\tilde{h} \in
\mbox{Null}(\widetilde{K}^\ast_{a, \bx}).$ Then, we show below that,
for any $\tilde{h} \in \mbox{Null}(\widetilde{K}^\ast_{a, \bx}),$
\begin{equation*}
\langle \E(Y \mid A = a, \bZ = \bz, \bX = \bx), \tilde{h}(\bz) \rangle  = 0. 
\end{equation*}

We begin by showing $\E\{\tilde{h}(\bz) \mid U, a, \bx\} = 0$. First, we have 
\begin{eqnarray}
  \E\{\tilde{h}(\bz) \mid w, a, \bx\}
  & = & \E\{\E\{\tilde{h}(\bz) \mid U, w, a, \bx\} \mid w, a, \bx\} \nonumber\\
  & = & \E\{\E\{\tilde{h}(\bz) \mid U, a, \bx\} \mid w, a, \bx\} \label{eq:ex-new-0}
\end{eqnarray}
where the first equality follows from iterated expectations, and
the second from Assumption~\ref{net-assum}.2. By
definition of the null space, we have $\widetilde{K}^\ast_{a, \bx}
\tilde{h}  = \E\{\tilde{h}(\bz) \mid w, a, \bx\} = 0$ almost
surely. Therefore, 
\begin{eqnarray}
  \E\{\tilde{h}(\bz) \mid w, a, \bx\} = 0 
  & \Longleftrightarrow & \E\{\E\{\tilde{h}(\bz) \mid U, a, \bx\} \mid w, a,
                          \bx\} = 0 \nonumber\\
  & \Longrightarrow & \E\{\tilde{h}(\bz) \mid U, a, \bx\} = 0
                      \hspace{0.2in} \mbox{almost surely.} \label{eq:ex-new}
\end{eqnarray}
where the first equivalence follows from equation~\eqref{eq:ex-new-0}, and
the second line follows from the relevance of $W$ for $U$
(Assumption~\ref{com-wu}). Finally, we now show $\langle \E(Y
\mid A = a, \bZ = \bz, \bX = \bx), \tilde{h}(\bz)
     \rangle = 0.$
\begin{eqnarray*}
  && \langle \E(Y \mid A = a, \bZ = \bz, \bX = \bx), \tilde{h}(\bz)
     \rangle \\
  & \coloneqq & \E\{\E(Y \mid A = a, \bZ = \bz, \bX = \bx) \tilde{h}(\bz)\mid A = a, \bX = \bx\} \\
  & = & \E\{\E\{\E(Y \mid U, A = a, \bZ = \bz, \bX = \bx) \mid A = a,
        \bZ = \bz, \bX = \bx\} \tilde{h}(\bz) \mid A = a, \bX = \bx\} \\
  & = & \E\{\E\{\E(Y \mid U, A = a, \bX = \bx) \mid A = a,
        \bZ = \bz, \bX = \bx\} \tilde{h}(\bz)\mid A = a, \bX = \bx\} \\
  & = & \E\{ \E\{\E\{\E(Y \mid U, A = a, \bX = \bx) \mid A = a,
        \bZ = \bz, \bX = \bx\} \tilde{h}(\bz) \mid U, A=a, \bX=\bx\} \mid A = a, \bX = \bx\} \\
  & = & \E\{\E(Y \mid U, A = a, \bX = \bx) \E\{\tilde{h}(\bz) \mid U, A=a, \bX=\bx\} \mid A = a, \bX = \bx\} \\
  & = & 0,
\end{eqnarray*}
where the first line follows from the definition of the inner product
in a Hilbert space, the second from iterated expectations
applied to $\E(Y \mid A = a, \bZ = \bz, \bX = \bx)$, the third from
Assumption~\ref{net-assum}.2, the fourth from iterated
expectations, the fifth from conditioning on $(U, A=a, \bX=\bx)$, and
finally, the sixth follows from equation~\eqref{eq:ex-new}. Therefore,
this shows that $\E(Y \mid A = a, \bZ = \bz, \bX = \bx) \in \mbox{Null}(\widetilde{K}^\ast_{a, \bx})^{\perp}.$

Proof of (iv): Finally, this key condition for Picard's theorem is
directly implied by equation~\eqref{eq:r-23}, which completes the
proof. \qed

\section{Asymptotic Properties of the DNC Estimator}
\label{appsec:asym}
\subsection{Setup and Regularity Conditions}
To derive asymptotic properties of our estimator, we assume the standard
GMM regularity conditions \citep{hansen1982gmm, newey1994}.
\vspace{0.1in}\\
\spacingset{1.2}{
  \noindent The GMM regularity conditions:
\begin{itemize}
\item Parameter space $\Theta$ is compact.
\item $m(\bL_{n,i}; \theta)$ is differentiable in
  $\theta \in \Theta$ with probability one. 
\item $m(\bL_{n,i}; \theta)$ and $\frac{\partial}{\partial \theta}
  m(\bL_{n,i}; \theta)$ are continuous at each $\theta \in \Theta$ with probability one.
\item $\E \left\{m(\bL_{n,i}; \theta) \right\} = 0$ only when $\theta = \theta_{0},$ and $\theta_{0}$ is in the interior of $\Theta$.
\item $\E\{m(\bL_{n,i}; \theta)\}$ and $\E\{\frac{\partial}{\partial
    \theta} m(\bL_{n,i}; \theta)\}$ are continuous in $\theta$.
\item $\frac{1}{n} \sum_{i=1}^n m(\bL_{n,i}; \theta)$ is stochastically equicontinuous on $\Theta$.
\item $\frac{1}{n} \sum_{i=1}^n \frac{\partial}{\partial \theta}
  m(\bL_{n,i}; \theta)$ is stochastically equicontinuous on $\Theta$.
\item $M_{0} = \frac{1}{n} \sum_{i=1}^n
  \E\left\{\frac{\partial}{\partial \theta} m(\bL_{n,i};
    \theta_0)\right\}$ is full rank.
\item For $p$ that satisfies Assumption~\ref{net1}, $\sup_{n
    \geq 1} \max_{i \in N_n} \E\{|c^\top m(\bL_{n,i}; \theta)|^p\} <
  \infty$ for any $c$ with $||c||_2 = \sqrt{c^\top c} = 1$ for all $\theta \in \Theta$. 
\item For $p$ that satisfies Assumption~\ref{net1}, $\sup_{n
    \geq 1} \max_{i \in N_n} \E\{|\tilde{c}^\top \frac{\partial}{\partial \theta} m(\bL_{n,i}; \theta)|^p\} <
  \infty$ for any $\tilde{c}$ with $||\tilde{c}||_2 =
  \sqrt{\tilde{c}^\top \tilde{c}} = 1$ for all $\theta \in \Theta$. 
\end{itemize}}
\vspace{0.1in}

\noindent We first define the GMM objective function: 
\begin{eqnarray*}
  Q_n(\theta) & = &\left\{\frac{1}{n} \sum_{i=1}^n \E\{m(\bL_{n,i};
    \theta)\} \right\}^\top \Omega  \left\{\frac{1}{n} \sum_{i=1}^n \E\{m(\bL_{n,i};
    \theta)\} \right\},\\
  \widehat{Q}_n(\theta) & = & \left\{\frac{1}{n} \sum_{i=1}^n m(\bL_{n,i};
    \theta)\right\}^\top \Omega  \left\{\frac{1}{n} \sum_{i=1}^n m(\bL_{n,i};
    \theta)\right\}.
\end{eqnarray*}
Then, the GMM estimator of $\theta$ can be written as:  
\begin{equation}
  \widehat{\theta} = \mbox{argmin}_{\theta \in \Theta} \widehat{Q}_n(\theta). 
\end{equation}

\subsection{Proof of Theorem~\ref{est-net}}
Given that our DNC estimator $\widehat{\tau}(a, a^\prime)$ for $\tau(a,
a^\prime)$ corresponds to the first element of $\widehat{\theta},$
we state theoretical properties in terms of $\widehat{\theta},$
which imply Theorem~\ref{est-net}. 
\paragraph{Consistency.} We first want to show consistency of the GMM estimator:
$$\widehat{\theta} \xrightarrow{p} \theta_0.$$
\paragraph{Proof:}
Under Assumption~\ref{net1}, Proposition 3.1 by
\cite{kojevnikov2020} implies point-wise convergence of
$\frac{1}{n} \sum_{i=1}^n m(\bL_{n,i}; \theta).$ That is, for all
$\theta \in \Theta,$
\begin{equation}
    \frac{1}{n} \sum_{i=1}^n \left\{m(\bL_{n,i}; \theta) -
      \E\{m(\bL_{n,i}; \theta)\} \right\} \xrightarrow{p} 0.
  \end{equation}

Under the stochastic equicontinuity, the compactness of the
parameter space, and the continuity of moment, we establish the
uniform convergence \citep{newey1994}.
\begin{equation}
  \sup_{\theta \in \Theta} \left\vert\frac{1}{n} \sum_{i=1}^n
    \left\{m(\bL_{n,i}; \theta) -
      \E\{m(\bL_{n,i}; \theta)\} \right\} \right\vert
  \xrightarrow{p} 0.
\end{equation}
Therefore, under the GMM regularity conditions described above,
\begin{equation}
  \sup_{\theta \in \Theta} \left\vert \widehat{Q}_n(\theta)  -
    Q_n(\theta) \right\vert \xrightarrow{p} 0. \label{eq:unif-q}
\end{equation}
Finally, under the GMM regularity conditions described above, we have
(i) $Q_n(\theta)$ is uniquely minimized at $\theta_{0}$, (ii)
parameter space $\Theta$ is compact, (iii) $Q_n(\theta)$ is
continuous, and (iv) the uniform convergence
(equation~\eqref{eq:unif-q}). Therefore, Theorem 2.1 of
\cite{newey1994} implies $$\widehat{\theta} \xrightarrow{p}
\theta_{0},$$ which completes the proof of consistency.\qed

\paragraph{Asymptotic Normality.} Next, we show asymptotic normality.
\begin{equation*}
  \sqrt{n} (\widehat{\theta} - \theta_0)
  \xrightarrow{d} \normalfont{\textsf{Normal}}(0, \Sigma)
  \end{equation*}
  where 
  \begin{eqnarray*}
    && \Sigma  = \Gamma_{0}
    \Var\left(\frac{1}{\sqrt{n}} \sum_{i=1}^n m(\bL_{n,i};
        \theta_{0}) \right) \Gamma_{0}^\top,\\
&& \Gamma_{0} = (M_{0}^\top \Omega M_{0})^{-1} M_{0}^\top
\Omega, \hspace{0.1in} M_{0} = \frac{1}{n} \sum_{i=1}^n \E \left\{\frac{\partial}{\partial \theta}  m (\bL_{n,i};
   \theta_{0}) \right\}.
  \end{eqnarray*}
  \paragraph{Proof:} 
  By definition, we have
  \begin{equation*}
    \widehat{\theta} = \mbox{argmin}_{\theta \in \Theta} \widehat{Q}_n(\theta) 
  \end{equation*}
  We take the first order condition. 
  \begin{equation*}
    \cfrac{\partial \widehat{Q}_n(\widehat{\theta})}{\partial
      \theta} =  0 
  \end{equation*}  
  Using the mean-value expansion, we have 
  \begin{eqnarray*}
    \sqrt{n} (\widehat{\theta} - \theta_{0}) & = & - 
                                                       \left\{\cfrac{\partial^2 \widehat{Q}_n(\widetilde{\theta})}{\partial
                                                       \theta \theta^\top} \right\}^{-1}  \times
                                                       \sqrt{n} \cfrac{\partial \widehat{Q}_n(\theta_{0})}{\partial \theta}\\
                                                 & = & - 
                                                       \left\{\cfrac{\partial^2 \widehat{Q}_n(\widetilde{\theta})}{\partial
                                                       \theta \theta^\top} \right\}^{-1}  \times \l\{\frac{1}{n} \sum_{i=1}^n \frac{\partial}{\partial\theta^\top}
                                                       m(\bL_i; \theta_{0})\r\}^\top \Omega \frac{1}{\sqrt{n}} \sum_{i=1}^n m(\bL_i;\theta_{0})
  \end{eqnarray*}
  where $\widetilde{\theta}$ is a mean value, located between
  $\widehat{\theta}$ and $\theta_{0},$ and 
  \begin{eqnarray*}
    && \left[\cfrac{\partial^2 \widehat{Q}_n(\widetilde{\theta})}{\partial \theta
       \partial \theta^\top} \right]_{jk} = \ \l\{\frac{1}{n} \sum_{i=1}^n \frac{\partial}{\partial\theta_j}
       m(\bL_i; \widetilde{\theta})\r\}^\top \Omega \l\{\frac{1}{n} \sum_{i=1}^n \frac{\partial}{\partial\theta_k}
       m(\bL_i; \widetilde{\theta})\r\} \\ 
    && \hspace{1.7in}  + \l\{\frac{1}{n} \sum_{i=1}^n
       \frac{\partial^2}{\partial\theta_j \partial \theta_k}
       m(\bL_{n,i}; \widetilde{\theta})\r\}^\top \Omega
       \l\{\frac{1}{n} \sum_{i=1}^n m(\bL_{n,i}; \widetilde{\theta})\r\}.
  \end{eqnarray*}
  Therefore, under the GMM regularity conditions, 
  Assumption~\ref{net1}, and consistency of $\widehat{\theta}$, 
  \begin{eqnarray*}
    && \left\{\cfrac{\partial^2
        \widehat{Q}_n(\widetilde{\theta})}{\partial \theta
        \theta^\top} \right\}^{-1} \xrightarrow{p} (M_{0}^\top \Omega M_{0})^{-1}, \\
&& \l\{\frac{1}{n} \sum_{i=1}^n \frac{\partial}{\partial\theta^\top}
m(\bL_i; \theta_{0})\r\}^\top \Omega \xrightarrow{p} M_{0}^\top \Omega.
  \end{eqnarray*}
  Thus,
  \begin{equation*}
    \sqrt{n} (\widehat{\theta} - \theta_{0}) =  - (M_{0}^\top
    \Omega M_{0})^{-1} M_{0}^\top \Omega \times \frac{1}{\sqrt{n}}
    \sum_{i=1}^n m(\bL_{n,i}; \theta_{0}) + o_p(1).
  \end{equation*}
  Finally, under Assumption~\ref{net1}, the Cram\'{e}r–Wold device and
  the network CLT (Theorem 3.2)
  by \cite{kojevnikov2020} imply
  \begin{equation*}
    \frac{1}{\sqrt{n}} \sum_{i=1}^n m(\bL_{n,i}; \theta_{0,n})
    \xrightarrow{d} \cN\left(0, \Var\left(\frac{1}{\sqrt{n}}
        \sum_{i=1}^n m(\bL_{n,i}; \theta_{0}) \right)\right).
\end{equation*}
By combining the results using the Slutsky's theorem, we obtain the
desired result.
\begin{equation*}
  \sqrt{n} (\widehat{\theta} - \theta_0)
  \xrightarrow{d} \normalfont{\textsf{Normal}}(0, \Sigma),
\end{equation*}
which completes the proof. \qed
  
\subsection{Proof of Theorem~\ref{hac}}
We consider asymptotic properties of the network HAC variance estimator. In
addition to the regularity conditions required to prove
Theorem~\ref{est-net}, we also require the following regularity
conditions for the choice of kernel and bandwidth. With $p$ that satisfies Assumption~\ref{net1}, 
\begin{equation*}
  \lim_{n \rightarrow \infty} \sum_{s \geq 0}
  |\omega(s/b_n) - 1 | \rho_n(s) \beta_{n,s}^{1 -
    2/p} = 0 \ \ \mbox{a.s.,}  
\end{equation*}
where $\rho_n(s)$ measures the average number of network peers at
  the distance $s$, $\rho_n(s)
= \frac{1}{n} \sum_{i =1}^n \cN_n(i; s)$.

\paragraph{Proof:}
Given that $\widehat{\theta}$ is a consistent estimator of
$\theta_{0}$, using the continuous mapping theorem under the GMM regularity condition, we need to prove that
\begin{equation*}
\widetilde{\Lambda}_n = \sum_{s \geq 0} \omega(s/b_n) \left\{\frac{1}{n}
  \sum_{i \in N_n} \sum_{j \in \cN_n(i;s)} m (\bL_{n,i}; \theta_{0})
  m (\bL_{n,j}; \theta_{0})^\top \right\}.
\end{equation*}
is a consistent estimator of $\Lambda_0$.
Because we assume that $m (\bL_{n,i}; \theta_{0}) $ is
$\psi$-weakly dependent (Assumption~\ref{net1}), under the regularity
condition on the choice of kernel and bandwidth
(equation~\eqref{eq:reg-band}), Proposition 4.1 of
\cite{kojevnikov2020} implies that $\widetilde{\Lambda}_n$ is a consistent estimator for $\Lambda_0.$ 

Moreover, under Assumption~\ref{net1} and the GMM regularity
conditions, we obtain consistency of $\widehat{M}$: $\widehat{M} - M_{0} \xrightarrow{p} 0$, 
where $\widehat{M} = \frac{1}{n} \sum_{i=1}^n
\frac{\partial}{\partial \theta}  m (\bL_{n,i}; \widehat{\theta})$
and $M_{0} = \frac{1}{n} \sum_{i=1}^n \E
\left\{\frac{\partial}{\partial \theta}  m (\bL_{n,i}; \theta_{0})
\right\}.$ Finally, we can combine the results to obtain the desired result.
\begin{equation*}
  \widehat{\Sigma}  - \Sigma \xrightarrow{p} 0
\end{equation*}
where
\begin{eqnarray*}
  && \Sigma  = \Gamma_{0}
     \Lambda_0 \Gamma_{0}^\top, \ \ \ \widehat{\Sigma} = \widehat{\Gamma}
     \widehat{\Lambda}\widehat{\Gamma}^\top \\ 
  && \Gamma_{0} = (M_{0}^\top \Omega M_{0})^{-1}
     M_{0}^\top \Omega,  \ \ \ \widehat{\Gamma} = (\widehat{M}^\top
     \Omega \widehat{M})^{-1} \widehat{M}^\top \Omega,
\end{eqnarray*}
which completes the proof. \qed

\subsection{Proof of Lemma~\ref{hac-choice}}
Under Assumption~\ref{ana-var}.1, the ACPE can be represented as a
linear function of parameters $\gamma$ in the outcome confounding bridge
function. Under this setting, it is sufficient to obtain multivariate asymptotic normality and consistent variance estimator for
$\gamma$. As a result, we can simplify the moment function to be
\begin{equation*}
  \tm(\bL_{n,i}; \gamma) = 
  \{Y_{i2} - h(W_{i}, A_i, \bX_i; \gamma)\} \times \eta(A_i, \bZ_i,
  \bX_i).
\end{equation*}
Under Assumption~\ref{ana-var}.2, there exists integer $s^\ast$ such that for units $i,
j$ with the distance $d_n(i,j) \geq s^\ast$,
\begin{equation*}
  \bL_{n,j} \ \indep \ \bL_{n,i} \mid A_i, \bZ_i, \bX_i, U_i.
\end{equation*}
For such $s^\ast$ and units $i,j$, we have 
\begin{equation}
  \tm(\bL_{n,j}; \gamma_{0}) \ \indep \ \tm(\bL_{n,i}; \gamma_{0}) \mid A_i, \bZ_i, \bX_i,
  U_i. \label{eq:hac-ana-1}
\end{equation}
In addition, under Assumptions~\ref{net-assum}.2 and~\ref{net-assum}.3, we have 
\begin{equation}
  \E\{\tm(\bL_{n,i}; \gamma_{0}) \mid A_i, \bZ_i, \bX_i, U_i\} = 0. \label{eq:hac-ana-2}
\end{equation}
Combining equations~\eqref{eq:hac-ana-1} and~\eqref{eq:hac-ana-2}, we obtain
\begin{eqnarray*}
  && \E\{\tm(\bL_{n,i}; \gamma_{0}) \tm(\bL_{n,j}; \gamma_{0})^\top \mid A_i, \bZ_i, \bX_i, U_i\} = 0\\
  &\Longrightarrow& \E\{\tm(\bL_{n,i}; \gamma_{0}) \tm(\bL_{n,j}; \gamma_{0})^\top\} = 0.
\end{eqnarray*}
for integer $s^\ast$ and units $i,j$ with the distance $d_n(i,j) \geq s^\ast$.
Therefore, 
\begin{equation*}
  \Lambda_0 = \sum_{s = 0}^{s^\ast - 1} \Lambda_0(s)
\end{equation*}
where
\begin{equation*}
  \Lambda_0(s) = \left\{\frac{1}{n} \sum_{i \in N_n} \sum_{j \in
      \cN_n(i;s)} \E\{ \tm (\bL_{n,i}; \gamma_{0}) \tm (\bL_{n,j};
    \gamma_0)^\top \} \right\}. 
\end{equation*}
We can obtain its estimator as follows.
\begin{equation*}
  \widehat{\Lambda}_{s^\ast} = \sum_{s = 0}^{s^\ast - 1} \omega(s/b_n) \left\{\frac{1}{n}
    \sum_{i \in N_n} \sum_{j \in \cN_n(i;s)} \tm (\bL_{n,i};
    \widehat{\gamma}) \tm (\bL_{n,j}; \widehat{\gamma})^\top \right\}.
\end{equation*}
Finally, we obtain the variance estimator for $\widehat{\gamma}.$
\begin{equation}
  \widehat{\Var}(\widehat{\gamma}) = \frac{1}{n} \widehat{\Gamma}_{\gamma} \widehat{\Lambda}_{s^\ast}
  \widehat{\Gamma}_{\gamma}^\top. 
\end{equation}
where  $\widehat{\Gamma}_{\gamma} = (\widehat{M}_{\gamma}^\top
\Omega \widehat{M}_{\gamma})^{-1} \widehat{M}_{\gamma}^\top \Omega$, and
$\widehat{M}_{\gamma} = \frac{1}{n} \sum_{i=1}^n
\frac{\partial}{\partial \gamma} \tm (\bL_{n,i}; \widehat{\gamma})$,
which completes the proof. 
\qed

\subsection{Heterogeneous Expectation}
In Section~\ref{subsec:est-net}, we assume that the expectation of
the causal peer effect, $\E\{Y_{i2}(a) -  Y_{i2}(a^\prime) \mid
\cG_n\}$, is constant across units, while we allow for network-dependent
(non-independent) errors. Here, to examine the heterogeneous expectation, we
explicitly write out the conditioning on $\cG_n.$ In this section, we
allow for heterogeneous expectation across units. As we observe only one sample of
interconnected units in a single network, we have to make some assumptions to make progress. In
this vein, we assume that $\E\{Y_{i2}(a) - Y_{i2}(a^\prime) \mid
\cG_n\}$ depends only on a summary statistic of network $\cG_n$, which
we denote by vector $\bg_i$. For example, $\bg_i$ could be the
network-degree of unit $i$, centrality of unit $i$, or 
other network summary statistics. This is a common assumption scholars
make in practice, and is similar to the idea of the exposure mapping \citep{aronow2012interference}, which is used to reduce dimensionality of the potential outcomes.

Formally, we assume $\E\{Y_{i2}(a) -
Y_{i2}(a^\prime) \mid \cG_n\} = \E\{Y_{i2}(a) -
Y_{i2}(a^\prime) \mid \bg_i\}.$ We then posit a model for the 
conditional expectation $\E\{Y_{i2}(a) \mid \bg_i\}$ with shared
coefficients. This allows us to accommodate heterogeneous expectation across units in the network, while we can
still make statistical inference about the target estimand with network-dependent errors. 

As a concrete example, consider the following linear model with
coefficients $\varphi$. 
\begin{eqnarray*}
  \E\{Y_{i2}(a) \mid \bg_i\}  & = & \varphi_0 + \varphi_1 \cdot a
                                    + \{\ell(\bg_i)^\top \varphi_2\} \cdot a
\end{eqnarray*}
where $\ell(\bg_i)$ is a user-specified function of $\bg_i$. Under this model, we can re-write the ACPE as follows.  
\begin{equation*}
  \tau(a, a^\prime) \coloneqq \frac{1}{n} \sum_{i=1}^n \E\{Y_{i2}(a) - Y_{i2}(a^\prime)
                               \mid \cG_n \}  = \varphi_1 \cdot (a - a^\prime) + ( \overline{\ell}^\top \varphi_2) \cdot (a -a^\prime)
\end{equation*}
where $\overline{\ell} = \frac{1}{n} \sum_{i=1}^n \ell(\bg_i).$ To
estimate the ACPE, we first modify the moment function as follows. 
\begin{eqnarray*}
  m^\dagger(\bL_{n,i}; \theta) & = & \left\{\tau + \left(\ell(\bg_i) -
                                     \overline{\ell} \right)^\top \varphi_2 \cdot (a - a^\prime)\right\} - \{h(W_{i}, a, \bX_i; \gamma) - h(W_{i}, a^\prime, \bX_i; \gamma)\},
\end{eqnarray*}
where $\theta = (\tau, \varphi_2, \gamma).$
We then show that $\E\{m^\dagger(\bL_{n,i}; \theta) \mid \cG_n\} = 0$
for all $i \in N_n$. We start with the first term. 
\begin{eqnarray*}
\E\left\{ \tau + \left(\ell(\bg_i) -
                                     \overline{\ell} \right)^\top
  \varphi_2 \cdot (a - a^\prime) \mid \cG_n \right\} 
  & = & \varphi_1 \cdot (a - a^\prime) + \ell(\bg_i)^\top \varphi_2 \cdot (a -
        a^\prime) \\
  & = & \E\{Y_{i2}(a) - Y_{i2}(a^\prime) \mid \cG_n \}.
\end{eqnarray*}
We next consider the second term. Under Assumption~\ref{net-assum},
\begin{eqnarray*}
  \E\left(\{h(W_{i}, a, \bX_i; \gamma) - h(W_{i}, a^\prime, \bX_i;
  \gamma)\} \mid \cG_n \right)   & = & \E\{Y_{i2}(a) - Y_{i2}(a^\prime) \mid \cG_n\},
\end{eqnarray*}
which shows that $\E\{m^\dagger(\bL_{n,i}; \theta) \mid \cG_n\} = 0$ for all $i \in N_n$.
Therefore, we can use the following moment functions to estimate the
ACPE $\tau(a, a^\prime)$.
\begin{equation*}
  m^\ast(\bL_{n,i}; \theta) = \left\{
    \begin{array}{l}
      m^\dagger(\bL_{n,i}; \theta) \times \eta^\ast(\bg_i)     \\
      \{Y_{i2} - h(W_{i}, A_i, \bX_i; \gamma)\} \times \eta(A_i, \bZ_i, \bX_i)     
    \end{array}
  \right\},
\end{equation*}
where $\eta^\ast(\bg_i) = (1, \ell(\bg_i)^\top)^\top$. Under the same
assumption used in Section~\ref{subsec:est-net}, we can consistently
estimate the ACPE and construct an asymptotic confidence
interval. \qed

\clearpage
\section{Simulation Study under Violation of Assumptions}
Here, we provide additional simulation studies to investigate the
performance of the proposed DNC estimator in settings where some key
identification assumptions are violated. In
Section~\ref{subsec:vio-nc}, we consider violation of the negative
control assumption (Assumption~\ref{net-assum}.2). In
Section~\ref{subsec:vio-comp}, we consider violation of the
outcome confounding bridge assumption (Assumption~\ref{net-assum}.3)
due to violation of the completeness condition. 

\subsection{Violation of Negative Control Assumptions}
\label{subsec:vio-nc}

\paragraph{Setup.}
In this section, we consider violations of the negative control
assumption (Assumption~\ref{net-assum}.2). In particular, we modify the data generating mechanism of
Section~\ref{sec:sim} as follows. For units $i = 1, \ldots, n,$
\begin{itemize}
\item[(1)] Unobserved confounder with network dependence: 
  $U_i = \sum_{s \geq 0} \zeta^s \sum_{j \in \cN(i; s)} \widetilde{U}_{j}/|\cN(i; s)|$ where
  $\zeta = 0.8$ and $\widetilde{U}_{j} \iid \textsf{Normal} (0,
  1)$. This part is the same as the one used in Section~\ref{sec:sim}.
\item[(2)] Observed covariates with network dependence:
  $\bX_i = (X_{i1}, X_{i2}, X_{i3})$ where, for $k \in \{1, 2, 3\}$, $X_{ik} = \sum_{s \geq 0}
  \zeta^s \sum_{j \in \cN(i; s)} \widetilde{X}_{jk}/|\cN(i; s)|$,
  $\zeta = 0.8$, and $\widetilde{X}_{jk} \iid \textsf{Normal}(0,
  1)$. This part is the same as the one used in Section~\ref{sec:sim}.
\item[(3)] Observed auxiliary variable: $C_i = \sum_{s \geq 0} (\zeta_C)^s
  \sum_{j \in \cN(i; s)} \widetilde{C}_{j}/|\cN(i; s)|$ where 
  $\widetilde{C}_i = U_i + \beta_c^\top \bX_i + \epsilon_{i0}$ where  $\epsilon_{i0} \iid
  \textsf{Normal} (0, 1)$ and $\beta_c = (0.05, 0.05, 0.05)$. This
  part is the difference from the one used in Section~\ref{sec:sim}.
\item[(4)] Focal behavior at the baseline: $Y_{i1} = U_i + 0.05 C_i
  + \beta_1^\top \bX_i + \epsilon_{i1}$ where  $\epsilon_{i1} \iid
  \textsf{Normal} (0, 1)$ and $\beta_1 = (-1, -1, -1)$. This part is the same as the one used in Section~\ref{sec:sim}.
\item[(5)] Focal behavior at the follow-up: $Y_{i2} = \tau A_i +
  0.2 Y_{i1} + 3 U_i + 0.05 C_i + \beta_2^\top \bX_i + \epsilon_{i2}$ where  $\epsilon_{i2} \iid
  \textsf{Normal} (0, 1)$, and $\beta_2 = (-1, -1, -1)$. The treatment variable $A_i$ is defined as
  $A_i = \sum_{j \in \cN(i; 1)} Y_{j1} /|\cN(i; 1)|.$ This part is the same as the one used in Section~\ref{sec:sim}.
\end{itemize}
The main and only difference is in (3) where we allow for network
association between auxiliary variable $C$ across units. Because we use $W_i = C_i$ as NCO, and $Z_i = \sum_{j \in \cN(i; 1)} C_{j}
/|\cN(i; 1)|$ as NCE, this network association violates
assumptions for NCO and NCE (Assumption~\ref{net-assum}.2).

We consider three different levels of the violation using parameter
$\zeta_C \in \{0.02, 0.10, 0.50\}.$ We call them ``Small'',
``Moderate'', and ``Large'' violations in Table~\ref{tab:sim-nc}. We fix sample size to be 1000, and we generate 2000 simulations to evaluate estimators in terms of the
absolute mean bias, the standard error (computed as the standard
deviation of point estimates across simulations), the root mean squared error (RMSE), and coverage of
95\% confidence intervals based on the network HAC variance
estimator. We standardize the first three quantities by the true
ACPE to ease interpretation.

\paragraph{Results.}
Table~\ref{tab:sim-nc} summarizes the results of the simulation
study. Our proposed DNC estimator has small bias and has reasonable
coverage when the violation is ``small.'' However, as we expect, the
larger is the violation, the bias is larger and coverage performance
becomes poorer. 

\begin{table}[!p]
  \centering
  \scalebox{1}{
    \begin{tabular}{lc|ccccc}
      \hline
      \multicolumn{2}{c|}{\textbf{Simulation Design}} &
                                                        \multicolumn{5}{c}{\textbf{DNC}} \\
      \multirow{2}{*}{Network} & \multirow{2}{*}{Violation} &
                                             \multirow{2}{*}{Bias}
                                           &
                                             Standard
                                           &
                                             \multirow{2}{*}{RMSE}
                                           &
                                             Coverage & Coverage\\[-3pt]
                                                      & & & Error &
                                           & (Analytical) & (Default) \\[3pt]
      \hline
      SW-4 & Small & 0.04 & 0.38 & 0.39 & 0.94 & 0.93 \\ 
                                                      & Moderate & 0.32 & 0.30 & 0.45 & 0.78 & 0.78 \\ 
                                                      & Large & 0.74 & 0.25 & 0.78 & 0.12 & 0.12 \\[10pt]   

      SW-8 & Small & 0.02 & 0.53 & 0.53 & 0.95 & 0.94 \\ 
                                                      & Moderate & 0.26 & 0.43 & 0.50 & 0.87 & 0.86 \\ 
                                                      & Large & 0.66 & 0.35 & 0.74 & 0.46 & 0.46 \\[10pt]   

      Add Health & Small & 0.03 & 0.41 & 0.41 & 0.94 & 0.93 \\ 
                                                      & Moderate & 0.34 & 0.34 & 0.48 & 0.80 & 0.78 \\ 
                                                      & Large & 0.81 & 0.28 & 0.86 & 0.13 & 0.13 \\ 
      \hline
    \end{tabular}}
  \caption{Operating Characteristics when the Negative
    Control Assumptions are Violated.}\label{tab:sim-nc}
  \vspace{-0.1in}
  \begin{flushleft}
    \spacingset{1}{{\footnotesize \noindent \textit{Note}: We consider
        three different levels of violation: ``Small'' ($\zeta_C =
        0.02$), ``Moderate'' ($\zeta_C = 0.10$), and ``Large''
        ($\zeta_C = 0.50$). We examine the same three different networks; the small
        world network model with the average degree of four (SW-4) and
        eight (SW-8), and the Add Health network. For the DNC estimator, we
        report the absolute mean bias, the standard error, the RMSE, and
        coverage of the 95\% confidence intervals based on the
        analytical bandwidth and the default bandwidth. The absolute mean bias, the standard error,
        and the RMSE for both estimators are standardized by the true ACPE. 
      }}
  \end{flushleft}
\end{table}

\subsection{Violation of Confounding Bridge Assumption due to Completeness}
\label{subsec:vio-comp}

\paragraph{Setup.}
In this section, we consider violations of the outcome confounding
bridge assumption (Assumption~\ref{net-assum}.3). In particular, we
consider violation of the completeness condition
(Assumption~\ref{com-wu}) we use to prove the existence of an outcome confounding bridge function. 

In particular, we modify the data generating mechanism of Section~\ref{sec:sim} as follows. For units $i = 1, \ldots, n,$
\begin{itemize}
\item[(1)] Two unobserved confounders with network dependence: 
  For $k \in \{1, 2\}$, \newline $U_{ik} = \sum_{s \geq 0} \zeta^s \sum_{j \in \cN(i; s)} \widetilde{U}_{jk}/|\cN(i; s)|$ where
  $\zeta = 0.8$ and $\widetilde{U}_{jk} \iid \textsf{Normal} (0,
  1)$. This part is the difference from the one used in Section~\ref{sec:sim}.
\item[(2)] Observed covariates with network dependence:
  $\bX_i = (X_{i1}, X_{i2}, X_{i3})$ where, for $k \in \{1, 2, 3\}$, $X_{ik} = \sum_{s \geq 0}
  \zeta^s \sum_{j \in \cN(i; s)} \widetilde{X}_{jk}/|\cN(i; s)|$,
  $\zeta = 0.8$, and $\widetilde{X}_{jk} \iid \textsf{Normal}(0,
  1)$. This part is the same as the one used in Section~\ref{sec:sim}.
\item[(3)] Observed auxiliary variable:  $C_i = U_{i1} + \beta_{UC} U_{i2} + \beta_c^\top \bX_i + \epsilon_{i0}$ where  $\epsilon_{i0} \iid
  \textsf{Normal} (0, 1)$ and $\beta_c = (0.05, 0.05, 0.05)$. The part
  of $U_{i2}$ is the difference from the one used in Section~\ref{sec:sim}.
\item[(4)] Focal behavior at the baseline: $Y_{i1} = U_{i1} +
  \beta_{UY1} U_{i2} + 0.05 C_i
  + \beta_1^\top \bX_i + \epsilon_{i1}$ where  $\epsilon_{i1} \iid
  \textsf{Normal} (0, 1)$ and $\beta_1 = (-1, -1, -1)$. The part
  of $U_{i2}$ is the difference from the one used in Section~\ref{sec:sim}.
\item[(5)] Focal behavior at the follow-up: $Y_{i2} = \tau A_i +
  0.2 Y_{i1} + 3 U_{i1} + \beta_{UY2} U_{i2} + 0.05 C_i + \beta_2^\top \bX_i + \epsilon_{i2}$ where  $\epsilon_{i2} \iid
  \textsf{Normal} (0, 1)$, and $\beta_2 = (-1, -1, -1)$. The treatment variable $A_i$ is defined as
  $A_i = \sum_{j \in \cN(i; 1)} Y_{j1} /|\cN(i; 1)|.$  The part of $U_{i2}$ is the difference from the one used in Section~\ref{sec:sim}.
\end{itemize}
The main difference is in (1) where we allow for two separate
unmeasured confounders $U_{i1}$ and $U_{i2}$. Yet, we use $W_i = C_i$ as NCO, and $Z_i = \sum_{j \in \cN(i; 1)} C_{j}
/|\cN(i; 1)|$ as NCE. Therefore, the number of unmeasured
confounders is larger than the number of NCO, and this violates the
completeness condition (Assumption~\ref{com-wu}). In this case,
an outcome confounding
bridge does not exist and Assumption~\ref{net-assum}.3 is violated.

We consider three different levels of violation using parameters
$(\beta_{UC}, \beta_{UY1}, \beta_{UY2}).$ We define ``Small'',
``Moderate'', and ``Large'' violations as follows.
\begin{itemize}
\item ``Small'': $\beta_{UC} = 0.1, \beta_{UY1} = \beta_{UY2} = 0.005$
\item ``Moderate'': $\beta_{UC} = 0.25, \beta_{UY1} = \beta_{UY2} =
  0.0125$
\item ``Large'': $\beta_{UC} = 0.5, \beta_{UY1} = \beta_{UY2} = 0.025$
\end{itemize}
We fix sample size to be 1000, and we generate 2000 simulations to evaluate estimators in terms of the
absolute mean bias, the standard error (computed as the standard
deviation of point estimates across simulations), the root mean squared error (RMSE), and coverage of
95\% confidence intervals based on the network HAC variance
estimator. We standardize the first three quantities by the true
ACPE to ease interpretation.

\paragraph{Results.}
Table~\ref{tab:sim-comp} summarizes the results of the simulation
study. Our proposed DNC estimator has small bias and has reasonable
coverage when the violation is ``small.'' However, as we expect, the
larger is the violation, the bias is larger and coverage performance becomes poorer. 

\vspace{0.2in}
\begin{table}[!h]
  \centering
  \scalebox{1}{
    \begin{tabular}{lc|ccccc}
      \hline
      \multicolumn{2}{c|}{\textbf{Simulation Design}} & \multicolumn{5}{c}{\textbf{DNC}} \\
      \hline
      \multirow{2}{*}{Network} & \multirow{2}{*}{Violation} &
                                             \multirow{2}{*}{Bias}
                               &
                                 Standard
                                           &
                                             \multirow{2}{*}{RMSE}
                               &
                                 Coverage & Coverage\\[-3pt]
                                                      & & & Error &
                               & (Analytical) & (Default) \\[3pt]
      \hline
      SW-4 & Small & 0.02 & 0.41 & 0.41 & 0.94 & 0.94 \\ 
                                                      & Moderate & 0.20 & 0.38 & 0.43 & 0.89 & 0.89 \\ 
                                                      & Large & 0.75 & 0.38 & 0.84 & 0.39 & 0.39 \\[10pt]   

      SW-8 & Small & 0.07 & 0.56 & 0.56 & 0.95 & 0.95 \\ 
                                                      & Moderate & 0.11 & 0.54 & 0.56 & 0.91 & 0.91 \\ 
                                                      & Large & 0.58 & 0.48 & 0.75 & 0.68 & 0.68 \\[10pt]    

      Add Health & Small & 0.04 & 0.44 & 0.45 & 0.95 & 0.94 \\ 
                                                      & Moderate & 0.17 & 0.44 & 0.47 & 0.89 & 0.89 \\ 
                                                      & Large & 0.70 & 0.41 & 0.81 & 0.49 & 0.49 \\ 
      \hline
    \end{tabular}}
  \caption{Operating Characteristics when the Confounding Bridge
    Assumption and the Completeness Condition are Violated.}\label{tab:sim-comp}
  \vspace{-0.1in}
  \begin{flushleft}
    \spacingset{1}{{\footnotesize \noindent \textit{Note}: We consider
        three different levels of violation: ``Small'', ``Moderate'', and ``Large''
        (see above for their definitions). We examine the same three different networks; the small
        world network model with the average degree of four (SW-4) and
        eight (SW-8), and the Add Health network. For the DNC estimator, we
        report the absolute mean bias, the standard error, the RMSE, and
        coverage of the 95\% confidence intervals based on the
        analytical bandwidth and the default bandwidth. The absolute mean bias, the standard error,
        and the RMSE for both estimators are standardized by the true ACPE. 
      }}
  \end{flushleft}
\end{table}

\end{document}